\documentclass[pre,aps,showpacs,onecolumn]{revtex4}%
\usepackage{graphicx}
\usepackage{amsmath}
\usepackage{amsfonts}
\usepackage{amssymb}
\usepackage{verbatim}
\usepackage[tight,sf]{subfigure}%
\usepackage{color}%
\usepackage{subfigure}
\usepackage{array}

\providecommand{\U}[1]{\protect\rule{.1in}{.1in}}

\begin{document}
\title{Squeezed helical elastica}
\author{Lila Bouzar$^{a,b}$, Martin Michael M\"uller$^{c,d}$, Pierre Gosselin$^{e}$, Igor M. Kuli\'c$^{d}$, Herv\'e Mohrbach$^{c,d}$ }
\affiliation{$a$\ D\'epartement de Physique Th\'eorique, Facult\'e de Physique, USTHB,
BP~32 El-Alia Bab-Ezzouar, 16111 Alger, Algeria}
\affiliation{$b$\ Laboratoire de Physique et Chimie Quantique, Universit\'e Mouloud Mammeri, BP 17, 15000 Tizi-Ouzou, Algeria}
\affiliation{$c$\ Equipe BioPhysStat, ICPMB-FR CNRS 2843, Universit\'e de Lorraine, 1
boulevard Arago, 57070 Metz, France}
\affiliation{$d$\ Institut Charles Sadron, CNRS-UdS, 23 rue du Loess, BP 84047, 67034
Strasbourg cedex 2, France}
\affiliation{$e$\ Institut Fourier, UMR 5582 CNRS-UJF, Universit\'e Grenoble I, BP74, 38402
Saint-Martin d'H\`eres, France}

\begin{abstract}
We theoretically study the conformations of a helical semi-flexible filament confined to a flat
surface. This squeezed helix exhibits a variety of unexpected shapes resembling 
circles, waves or spirals depending on the material parameters. We explore the conformation 
space in detail and show that the shapes can be understood as the mutual elastic interaction of 
conformational quasi-particles. Our theoretical results are potentially useful to 
determine the material parameters of such helical filaments in an experimental setting.

\end{abstract}

\maketitle



\section{Introduction}

Elastic objects exhibit a plethora of shapes in a confined geometry. 
Sometimes it requires a lot of imagination to deduce the original three-dimensional 
shape from the observation of its confined counterparts. 
In general, a geometrical confinement induces the breaking of a pre-existing symmetry 
of an elastic object. Soft matter physics offers many examples. For instance, a spherical 
membrane vesicle adopts an onion-like shape when confined inside a sphere of smaller size \cite{Kahraman2012}.
Elastic filaments in spherical confinement have been extensively studied, such as the morphology of a wire inside 
a cavity \cite{Najafi2011} or the shapes of semi-flexible filaments on a sphere \cite{Jemal2012}. 
A variation of the theme is the confinement of a polymer between two plates \cite{Hsu2004} 
or the morphology and dynamics of actin filaments osmotically confined to a flat surface \cite{Sanchez2010}.
Generally, one can find many more examples of rods confined to various two-dimensional surfaces in the literature \cite{Guven2014, vanderHeijden_twisted_cylinder1,vanderHeijden_twisted_cylinder2,vanderHeijden_selfcontact_cylinder}.

In this paper we consider the planar confinement of a polymer which is not straight 
but helical in its ground state. Related problems were considered before like a twisted \cite{vanderHeijden_twisted_plane} or a nonlinearly elastic \cite{Maddocks1984} rod under external loads confined on a plane. In living nature, one frequently finds helical polymers like microtubules \cite{Mohrbach2010,mt2012}, Ftsz
filaments \cite{Lu2000} and dynamin \cite{Ferguson2012}. Even whole microorganisms exhibit helicity inherited from
their constituent filaments \cite{Asakura1966}. Different helically coiled structures have also been fabricated artificially 
such as coiled carbon \cite{Volodin2000} and DNA nanotubes \cite{Douglas2007}.
To study these objects one often confines them to the focal plane of a microscope. This confinement
changes the physical properties of the underlying objects and peculiar
squeezed conformations often resembling looped waves, spirals or circles are
observed \cite{Asakura1966,Volodin2000,Douglas2007}. Here we give an explanation for 
these observations. The helical filament is modeled as a semi-flexible polymer squeezed onto a flat surface and was previously called squeelix \cite{Nam2012}. Excluded volume interactions are not taken into account in this approach, even though they are potentially relevant \cite{vanderHeijden_selfcontact_cylinder}.
The variation of the linear elastic energy of the squeelix allows to determine the shapes at zero temperature. 
Varying the material parameters allows to classify the zoo of shapes in a manner similar to Euler elastica in three-dimensional space
\cite{Nizette1999}. The results and the physical interpretation of the underlying theory are presented in the main text. 
The interested reader can find the mathematical details in the appendix.

In the following section we present the model and the fundamental equations of the squeelix elastica. 
In Sec.~\ref{sec:infinitesqueelix} we discuss the various shapes of a squeelix of infinite length. In this case these shapes are 
always ground states of the elastic energy. They can be understood qualitatively with the notion 
of conformational quasi-particles called twist-kinks \cite{Nam2012}. These twist-kinks are another 
example of a general theme that we have already encountered in the context of microtubules \cite{Kahraman2014}. 
In Sec.~\ref{sec:finitesqueelix} we will see that squeelices of finite length display a more complex behavior. 
The boundary conditions provoke the existence of metastable states which we will discuss in detail.
In Sec.~\ref{sec:exp} we suggest a procedure how experimental data can be interpreted to 
extract material parameters from the theory.


\section{A helical worm-like chain confined in two dimensions: the squeelix\label{sec:model}}

\subsection{Basic equations of the helical WLC model}

The shape of an elastic rod can be described by the spatial 
evolution of the Frenet-Serret basis $(\mathbf{n},\mathbf{b},\mathbf{t})$ attached to the
centerline of the rod. An internal twist of the rod is taken into account with the help of an 
additional local basis $(\mathbf{e}_{1},\mathbf{e}_{2},\mathbf{e}_{3})$ which rotates
with the material. This material frame or director basis $(\mathbf{e}%
_{1},\mathbf{e}_{2},\mathbf{e}_{3})$ can be written in terms of the Frenet-Serret basis 
as $\mathbf{e}_{3}%
=\mathbf{t}$, $\mathbf{e}_{1}=\mathbf{n}\cos\psi\mathbf{+b}\sin\psi$ and
$\mathbf{e}_{2}=-\mathbf{n}\sin\psi\mathbf{+b}\cos\psi$, where $\psi$ is the
twist angle. The evolution of this basis along the centerline, described by the arc length $s$, 
is given by the twist equations
$\mathbf{e}_{i}^{\prime}=\mathbf{\Omega}\times\mathbf{e}_{i}$, where
$\mathbf{\Omega}=\left(  \Omega_{1},\Omega_{2},\Omega_{3}\right)  $ is the
strain vector function and $()^{\prime}$ denotes the derivative with respect to $s$. The
components of $\mathbf{\Omega}$ are \cite{helices}: 
\begin{subequations}
\begin{align}
\Omega_{1}(s)  &  =\kappa(s)\sin\psi(s) \; ,\\
\Omega_{2}(s)  &  =\kappa(s)\cos\psi(s) \; ,\\
\Omega_{3}(s)  &  =\tau(s)+\psi^{\prime}(s)
\; ,
\end{align}
\end{subequations}
where $\kappa(s)\ge 0$ and $\tau(s)$ are the local curvature and torsion, respectively. 
The local curvature is thus $\kappa^{2}(s)=\Omega_{1}^{2}+\Omega_{2}^{2}$
and the twist density $\Omega_{3}(s)$ is the sum of the torsion and the excess
twist $\psi^{\prime}$. 

\begin{figure}
\centering
\subfigure[]{
\raisebox {6mm}{\label{fig1a}\includegraphics[scale=0.5]{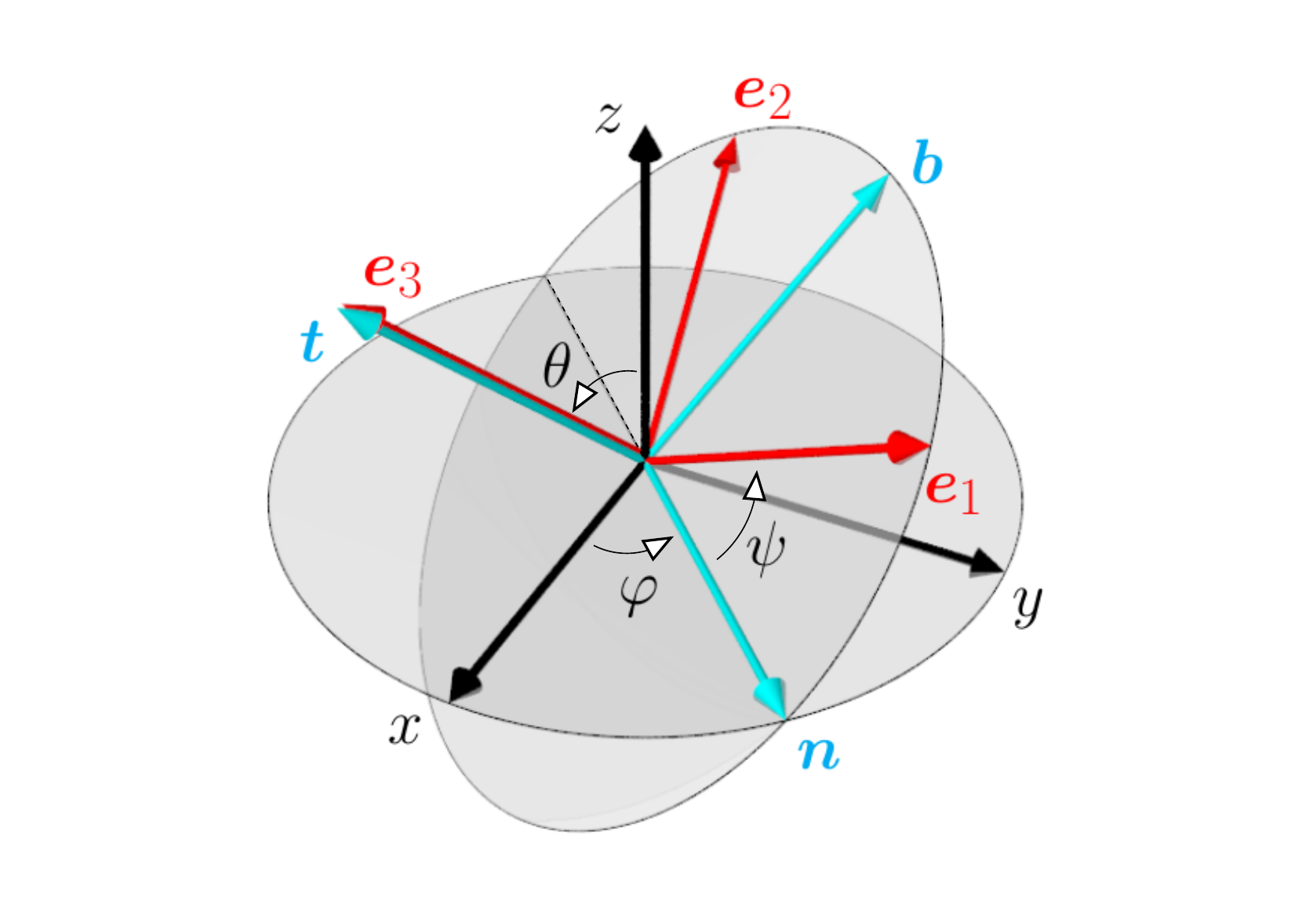}}
}
\qquad
\qquad
\subfigure[]{\label{fig1b}
\includegraphics[scale=1]{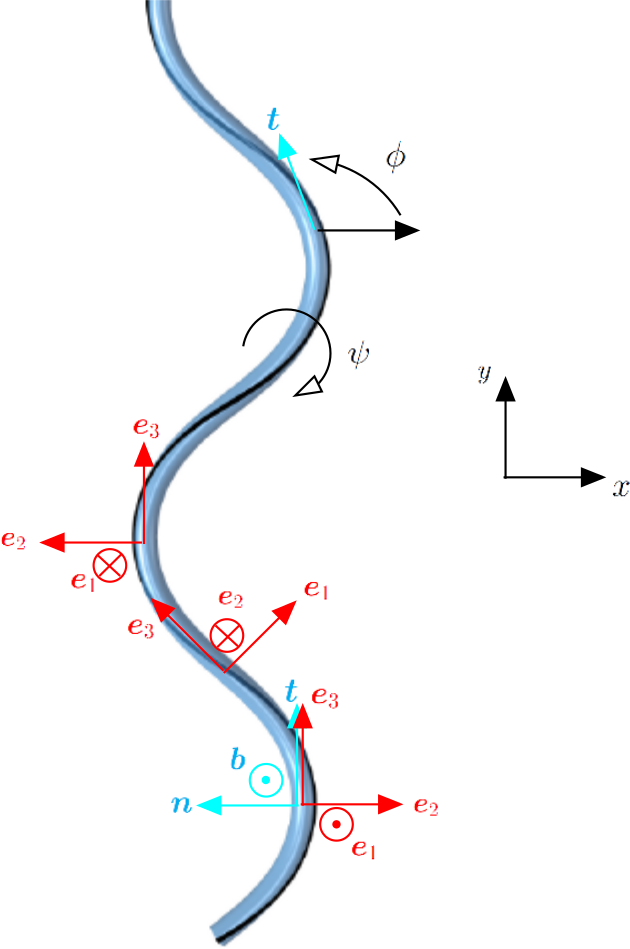}
}
\caption{(a) Representation of the Frenet-Serret basis, the director basis and the Euler angles at a position $s$ of the centerline of a helical rod in  three-dimensional space. The confinement onto the $(xy)$ plane is realized by the projection of the tangent vector $\mathbf{t}$ (and thus $\mathbf{e}_{3}$) on that plane, \textit{i.e.}, $\theta=\pi/2$. (b) Sketch of a confined helical rod with the two local bases. The Frenet-Serret vectors $\mathbf{n}$ and $\mathbf{t}$ evolve with $s$ in the $(xy)$ plane. The director vectors $\mathbf{e}_{1}$ and $\mathbf{e}_{2}$ attached to the material frame rotate in the ($\mathbf{n}\mathbf{b}$) plane with an angle $\psi$ when they evolve along the centerline. For the rest of this article the evolution of the vector $\mathbf{e}_{2}$ will be simply represented by a black ribbon drawn on the rod's surface (with $\mathbf{e}_{2}$ pointing in a direction perpendicular to the ribbon).}
\label{fig1ab}
\end{figure}

For simplification, we consider an elastic rod of circular cross-section 
with a single bending modulus whose ground state is a helix in 3D space \footnote{Note that 
a generalization of this model consists of considering two 
bending moduli which introduces an anisotropy and is, for instance, closer to DNA. This is beyond the scope of this
paper but may be an interesting starting point for future work.}.  
In linear elasticity theory, this helical worm-like chain minimizes the following energy:
\begin{equation}
E=\int\frac{B}{2}\left(  \left(  \Omega_{1}-\omega_{1}\right)  ^{2}+\left(
\Omega_{2}-\omega_{2}\right)  ^{2}\right)  +\frac{C}{2}\left(  \Omega
_{3}-\omega_{3}\right)  ^{2}ds \; ,
\label{energy3D}
\end{equation}
where $B$ and $C$ are the bending and torsional stiffness, respectively.
The positive constant parameters $\omega_{1}$ and $\omega_{2}$ are the principal intrinsic curvatures and
$\omega_{3}$ the intrinsic twist. We consider a right-handed helix in the following, \textit{i.e.}, 
$\omega_{3}>0$. One can always set $\omega_{2}=0$ by a convenient choice of the material frame. 
The bending and the twist terms of the energy given by Eq. $\left(  \ref{energy3D}\right)  $
can be minimized independently, yielding a curve of constant curvature $\kappa=\omega_{1}$
and torsion $\tau=\omega_{3}.$ In
the absence of an external torque there is no excess twist, $\psi^{\prime}%
=0$. This ground state is a helix of radius $R$ and pitch $H$ with 
\begin{equation}
  R=\frac
{\omega_{1}}{\omega_{1}^{2}+\omega_{3}^{2}} \qquad \text{and} \qquad 
H=\frac{2\pi\omega_{3}}{\omega_{1}^{2}+\omega_{3}^{2}}\; ,
\label{eq:radiuspitchofahelix}
\end{equation} 
satisfying the preferred curvature and twist everywhere.
The components of the strain vector in the director basis can be expressed with the Euler
angles $\varphi\left(  s\right)  ,$ $\theta\left(  s\right)  $ and $\psi\left(
s\right) $ (see Fig.~\ref{fig1a}): 
\begin{subequations}\label{eq:Euler}
\begin{align}
\Omega_{1} &  =\varphi^{\prime}\sin\theta\sin\psi+\theta^{\prime}\cos\psi \; , \\
\Omega_{2} &  =\varphi^{\prime}\sin\theta\cos\psi-\theta^{\prime}\sin\psi \; , \\
\Omega_{3} &  =\varphi^{\prime}\cos\theta+\psi^{\prime}
\; .
\end{align}
\end{subequations}
The curvature is then given by $\kappa^{2}(s)=\varphi^{\prime
2}\sin^{2}\theta+\theta^{\prime2}$ and the torsion is $\tau=\varphi^{\prime}%
\cos\theta.\ $

\subsection{The squeelix\label{subsec:thesqueelix}}

Confining the helical rod to the $(xy)$ plane amounts to putting $\theta=\pi/2$. 
It is convenient to introduce the angle $\phi$ between the tangent of the centerline 
and the $x$ axis defined as $\phi=\varphi-\pi /2$ (see Fig.~\ref{fig1b}). Then, Eqs.~(\ref{eq:Euler}) become
\begin{equation}
\Omega_{1}=\phi^{\prime}\sin\psi,\text{ \ \ }\Omega_{2}=\phi^{\prime}\cos
\psi,\text{ \ \ }\Omega_{3}=\psi^{\prime} \; .
\end{equation}
Hence the local curvature is simply given by $\kappa^{2}(s)=\phi^{\prime2}$ 
and the torsion $\tau=0$ as the curve is now planar (see Fig.~\ref{fig1ab}).  
The energy of a squeezed helical worm-like chain of length $L$ (the so-called \textit{squeelix}) can then be written as:
\begin{equation}
E=\frac{1}{2}\int_{-L/2}^{L/2} \left(B\left(  \phi^{\prime}-\omega_{1}\sin\psi\right)^{2}  +  C \left(  \psi^{\prime}-\omega_{3}\right)
^{2}  + B\omega_1^2 \cos^2{\psi} \right) ds \label{sqeezedenergy}
\; .
\end{equation}
Note that under confinement the curvature and twist are now coupled. Minimizing  $E$ with respect to $\phi^{\prime}$ gives
\begin{equation}
\phi^{\prime}=\omega_{1}\sin\psi  \label{EQM1} 
\; ,
\end{equation}
and Eq.~(\ref{sqeezedenergy}) reduces to a functional of $\psi(s)$ alone
\begin{equation}
E[\psi]=\frac{1}{2}\int_{-L/2}^{L/2} \left( C \left(  \psi^{\prime}-\omega_{3}\right)
^{2}  + B\omega_1^2 \cos^2{\psi} \right) ds\label{sqeezedenergyPSI}
\; ,
\end{equation}
for which the Euler-Lagrange equation is (see Appendix A):
\begin{equation}
\psi^{\prime\prime}+\frac{B\omega_{1}^{2}}{2C}\sin(2\psi)=0 \label{EQM2}
\end{equation}
with free boundary conditions, \textit{i.e.}, no torque at both ends of the filament 
\begin{equation}
\psi^{\prime}(-L/2)=\omega_{3}=\psi^{\prime}(L/2) \; . \label{BC}
\end{equation} 
Thus even in the absence of an external torque at the chain's ends, the
confinement converts the intrinsic twist into an intrinsic torque. Eq.~(\ref{EQM2}) is nothing less than the pendulum equation (with
arc length $s$ as the time and $\alpha=2\psi$ the angle of the pendulum). Its
solutions depend on the material parameters $B,C,$ $\omega_{1}$ and also
$\omega_{3}$ \textit{via} the boundary conditions Eq.~(\ref{BC}). The curvature $\kappa(s)=\phi^{\prime}(s)$ of the squeelix can
then be obtained directly from Eq.~(\ref{EQM1}) . Note that in two 
dimensions the curvature can be negative. The fact that the curvature is slaved to 
the twist is the most important consequence of the squeezing of a helical WLC.

Integrating Eq.~(\ref{EQM2}) we obtain 
\begin{equation}
\psi^{\prime}(s)=\pm\frac{1}{\lambda\sqrt{m}}\sqrt{1-m\sin^{2}\psi
}\label{psiprime}
\end{equation}
with $\lambda$ a characteristic length scale given by
\begin{equation}
\lambda=\frac{1}{\omega_{1}}\sqrt{\frac{C}{B}} \label{Lambda}
\end{equation}
and $m$ a positive real parameter. The phase plane of Eq.~(\ref{psiprime}) is well-known and the solutions 
of Eq.~(\ref{EQM2}) are particular trajectories in this plane. A detailed discussion can be found in Appendix A. 
These solutions can be determined numerically by integrating the differential 
equation~(\ref{psiprime}). Alternatively, one can use the well-known explicit solution of Eq.~(\ref{EQM2}). This will be very useful in the following since we want to compute the energy of the shapes of the squeelices, interpret them physically and discuss their stability.

The general solution of Eq.~(\ref{EQM2}) such that
$\psi(s_{0})=0$ is (see Appendix A)
\begin{equation}
\psi(s)=\pm am\left(  \frac{s-s_{0}}{\lambda\sqrt{m}}|m\right)
\; ,
\label{SOLgeneral}
\end{equation}
where $am\left(  x|m\right)  $ is the elliptic Jacobian amplitude function
whose behavior depends on the value  $m$ \cite{Abramowitz}. For the reader not familiar with elliptic functions, 
Figs.~\ref{fig3}, \ref{shapes_infini_minf1}, and \ref{fig5} show the generic characteristic behavior of $\psi(s)$.

Squeelices are stable shapes if the second variation of $E$ in Eq.~(\ref{sqeezedenergy}) with respect to $\phi^{\prime}$ and $\psi$ at the extrema of the energy is positive definite. Writing $\phi^{\prime}=\omega_{1}\sin\psi+\delta\phi^{\prime} $ we see that $\delta^2 E$ with respect to $\phi^{\prime}$ gives the contribution $\frac{1}{2}\int_{-L/2}^{L/2} B\left( \delta \phi^{\prime} \right)^2 ds$ which is positive definite. Therefore, the stability of squeelices relies on the sign of the second variation of $E[\psi]$ in Eq.~(\ref{sqeezedenergyPSI}). Note that integrals such as $E[\psi]$ never have maximisers (see, for instance, \cite{Maddocks1984} and \cite{Manning2009}). A solution (\ref{SOLgeneral}) either leads to a stable squeelix if it is a local minimizer or to an unstable one if it is a saddle point of $E[\psi]$. This whole issue is discussed in detail in Appendix~E.

An important remark is due here. The choice of the sign of $\psi(s)$ in
Eq.~(\ref{SOLgeneral}) depends on the sign of $\omega_{3}$.
In this paper we have chosen $\omega_{3}>0$. This implies that $\psi^{\prime
}(-L/2)$ is positive as well. For $m<1$ $\psi(s)$ is a monotonous function
which then has to grow with $s$. For $m>1$ $\psi(s)$ becomes a periodic
function of $s$ which has to grow in the vicinity of $-L/2$. Therefore, we
have to choose the positive sign in Eq.~(\ref{SOLgeneral}) for all $m$.

By integrating the curvature, Eq. (\ref{EQM1}), we obtain $\phi(s)$, the angle between the 
tangent vector $\mathbf{t}$ and the $x$ axis. We can then reconstruct the two-dimensional 
shapes in Cartesian coordinates with the relations:
\begin{subequations}
\begin{align}
x(s)  &  =x_{0}+\int\nolimits_{-L/2}^{s}\cos\left(  \phi(s^{\prime})\right)
ds^{\prime}\label{XYx}\; , \\
y(s)  &  =y_{0}+\int\nolimits_{-L/2}^{s}\sin\left(  \phi(s^{\prime})\right)
ds^{\prime}\; . \label{XYy}%
\end{align}
\end{subequations}
The constants of integration $m$ and $s_{0}$ have to be determined from the
boundary conditions Eq.~(\ref{BC}). For a chain of finite
length $L$ this problem turns out to be surprisingly complicated. But to grasp
a physical intuition of the squeelix we first consider a very long chain where
$L$ is much larger than any characteristic length. In this case we can neglect
the boundary conditions Eq.~(\ref{BC}).


\section{Squeelices of infinite length\label{sec:infinitesqueelix}}

A trivial solution of Eqs.~(\ref{EQM1}) and (\ref{EQM2}) is $\psi=\pm\pi/2$
which corresponds to a shape of constant curvature $\kappa=\pm\omega_{1}$,
i.e. multiple circles on top of each other. The energy density of this
configuration is $E_{0}/L=C\omega_{3}^{2}/2$. The non-trivial general solution of Eq.~(\ref{EQM2}) assuming
the condition $\psi(0)=0$ without loss of generality is thus:
\begin{equation}
\psi(s)=am\left(  \frac{s}{\lambda\sqrt{m}}|m\right)  \; . \label{psym}
\end{equation}
Therefore $\psi^{\prime}(s)=\frac{1}{\lambda\sqrt{m}}dn\left(  \frac
{s}{\lambda\sqrt{m}}|m\right)  $ and $\psi^{\prime}(0)=\frac{1}{\lambda
\sqrt{m}}$. The curvature then reads
\begin{equation}
\kappa(s)=\omega_{1}sn\left(  \frac{s}{\lambda\sqrt{m}}|m\right)
\; ,  \label{Curv}
\end{equation}
where the functions $dn$, and $sn$ are well-known elliptic Jacobian functions
with parameter $m>0$ \cite{Abramowitz}. The function $sn$ is a periodic odd function of
amplitude unity whose period is $l_{p}=4\lambda\sqrt{m}\mathcal{K}(m)$, where $\mathcal{K}(m)$
denotes the complete elliptic integral of the first kind.

From Eqs. $\left(  \ref{XYx}\right)  $, $\left(  \ref{XYy}\right)  $ and
$\left(  \ref{Curv}\right)  $ all shapes can be determined. There is an
infinite number of solutions since we do not impose the boundary conditions
Eq.\ $\left(  \ref{BC}\right)  $. This set of solutions splits into two
categories, the oscillatory and the revolving regimes of the pendulum which
correspond to $m>1$ and $m<1$, respectively. The limiting case $m=1$ is the
homoclinic pendulum that has just enough energy to make one full $\alpha=2\pi$
(or $\psi=\pi$) rotation in an infinite \textquotedblleft
time\textquotedblright\ interval. This ensemble of solutions leads to a
variety of shapes resembling loops, waves, spirals or circles that we are
going to explore.


\subsection{The energy density of a squeelix}

To each value of the parameter $m$ corresponds a different filament shape. But
a helical filament of length $L$ with material parameters $B$, $C$,
$\omega_{1}$ and $\omega_{3}$ will adopt a single ground state when squeezed
into the plane. This shape is the one minimizing the total elastic energy $E$
of the chain. In order to compare the energy of the various solutions in the
limit large $L$ we only need compute the energy per length $e=E/L$ given by
(see Appendix B)
\begin{equation}
e(m)=\frac{\omega_{1}\sqrt{BC}}{L\sqrt{m}}\left(  \mathcal{E}\left(  \psi\left(
L/2\right)  |m\right)  - \mathcal{E}\left(  \psi\left(  -L/2\right)  |m\right)  \right)
+\frac{B\omega_{1}^{2}}{2}\left(  1+\frac{1}{\mu}-\frac{1}{m}\right)
-\frac{C\omega_{3}}{L}\left(  \psi\left(  L/2\right)  -\psi\left(
-L/2\right)  \right)  \label{Ecyclegeneral}
\end{equation}
with $\mathcal{E}\left(  x|m\right)  $ the elliptic integral of the second kind \cite{Abramowitz}, and
\begin{equation}
\mu=\frac{\omega_{1}^{2}B}{\omega_{3}^{2}C}=\frac{\pi^{2}}{4}\gamma 
\; ,\label{MU}
\end{equation}
which measures the ratio of the bending energy $\propto B\omega_1^2$ over the twist energy 
$\propto C\omega_3^2$. The control parameter
\begin{equation}
\gamma=\frac{4\omega_{1}^{2}B}{\pi^{2}\omega_{3}^{2}C}\label{Gamma}%
\end{equation}
will play a crucial role in the following. 
To simplify the analysis of the energy density, we consider the two different behaviors of the 
pendulum separately.

For the oscillating pendulum ($m > 1$) we consider the energy per period since
$E/L=E_{p}/l_{p}$ with $E_{p}$ the energy of one period of oscillation and
$l_{p}= 4\lambda \mathcal{K}\left(  \frac{1}{m}\right)  $.\ Using $\psi\left(
l_{p}/2\right)  =\psi\left(  -l_{p}/2\right)  $ we obtain the energy density%
\begin{equation}
e_{m>1}(m)=\frac{B\omega_{1}^{2}}{2}\left(  1+\frac{1}{\mu}-\frac{1}{m}\right)
\; ,
\end{equation}
which is a monotonously growing function of $m$. Its minimum at $m=1$ is given
by
\begin{equation}
e_{m=1}=\underset{L\rightarrow\infty}{\lim}\frac{E}{L}=\frac{1}{2}C\omega
_{3}^{2}=E_{0}/L
\; ,
\end{equation}
which is degenerate with the energy density of the trivial solution of
constant curvature. The absence of minima for all $m>1$, implies that the
associated shapes cannot be a ground state of a squeelix of infinite length.

In the case of the revolving pendulum ($m < 1$) we compute the energy per length by
using $E/L=E_{_{cycle}}/l_{cycle}$ with $E_{_{cycle}}$ the energy for a single
cycle defined by the condition $\psi(s+l_{cycle})=\psi(s)+\pi$ with
\begin{equation}
  l_{cycle} / \lambda =2 \sqrt{m}\mathcal{K}(m)
\; .
\label{eq:lcycle}
\end{equation}
Since $\psi(l_{cycle}/2%
)=\psi(-l_{cycle}/2)+\pi$ we have $\mathcal{E}\left(  \psi\left(  l_{cycle}/2\right)  |m\right)
-\mathcal{E}\left(  \psi\left(  -l_{cycle}/2\right)  |m\right)  =2\mathcal{E}(m)$ where $\mathcal{E}\left(
m\right)  $ is the complete elliptic integral of the second kind \cite{Abramowitz}. The energy
density becomes
\begin{equation}
e_{m<1}(m)=\frac{B\omega_{1}^{2}}{m}\frac{\mathcal{E}(m)}{\mathcal{K}(m)}+\frac{1}{2}B\omega
_{1}^{2}\left(  1+\frac{1}{\mu}-\frac{1}{m}\right)  -\frac{\pi\sqrt{BC}}%
{2}\frac{\omega_{1}\omega_{3}}{\sqrt{m}\mathcal{K}(m)} \; . \label{Ecycle1}%
\end{equation}
Remarkably, $e_{m<1}(m)$ exhibits a minimum at $m=m^{\ast}$ given by the
equation
\begin{equation}
\frac{\sqrt{m^{\ast}}}{\mathcal{E}(m^{\ast})}=\sqrt{\gamma} \; . \label{msol}%
\end{equation}
Note that $\sqrt{m}/\mathcal{E}(m)\leq1$ for all $m\leq1$. Thus, this minimum only
exists for $\gamma\leq$ $1$.

\begin{figure}
\begin{center}
 \includegraphics[width=0.4\textwidth]{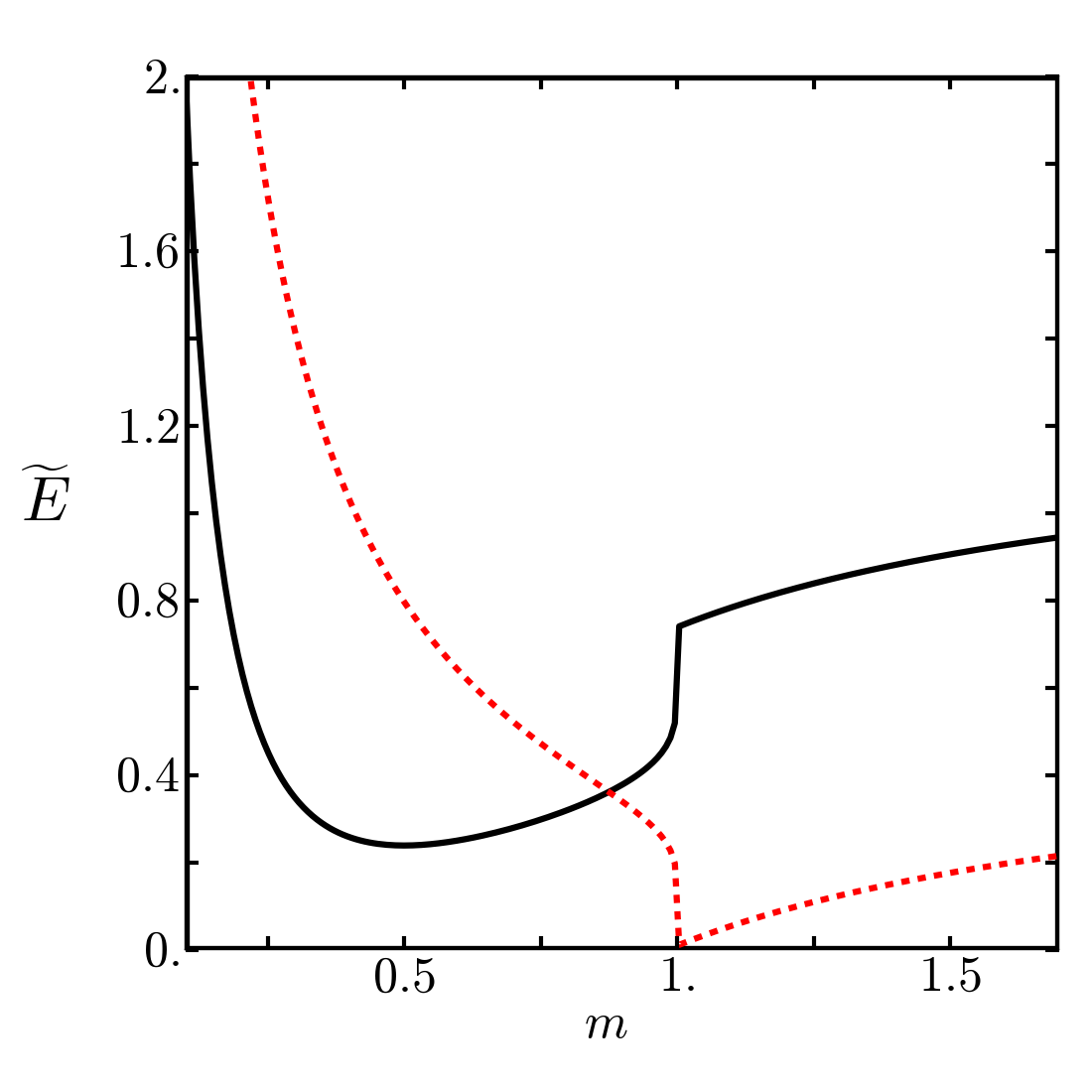}
\caption{Scaled energy density $\widetilde{E}=E/(B\omega_1^2 L)$ of a squeelix of infinite length for $\gamma=0.274$ with $C/B=1$ and $\omega_3/\omega_1 = 1.22$ (black solid curve). The minimum of this curve lies at $m^*=0.5$ as predicted by Eq.~(\ref{msol}). For $\gamma>1$ the minimum of the scaled energy is at  $m^*=1$ as 
exemplified by the red dotted curve where $\gamma=20.26$ with $C/B=1$ and $\omega_3/\omega_1 =0.14$.
\label{energyLinfini}}
\end{center}
\end{figure}

Eq.~(\ref{msol}) shows that the ground state of a squeelix of infinite length
is determined by the parameter $\gamma$. For $\gamma<1$, the ground state is
given by the set of equations Eqs. $\left(  \ref{XYx}\right)  $, $\left(
\ref{XYy}\right)  $ and $\left(  \ref{Curv}\right)  $. The parameter $m$ in
these equations is the solution of Eq.\ $\left(  \ref{msol}\right)  $ and is
therefore smaller than unity (revolving pendulum).\ For small values
$\gamma\ll 1$, we find $m^{\ast}\approx\frac{\pi^{2}}{4}\gamma=\mu.\ $When
$\gamma$ is approaching unity $\gamma\lesssim1$ we see that $m^{\ast
}\approx\gamma$. For $\gamma>1$ the minimum of $e_{m<1}(m)$ is at $m=1$ with
$e_{m<1}(1)=E_{0}/L =e_{m>1}(1)$. To illustrate our findings, Fig. \ref{energyLinfini} shows the energy density as a function
of $m$ for different values of $\gamma$. Figures \ref{fig3}-\ref{fig5}
show a variety of possible shapes depending on the material parameters. 

The parameter $\gamma$ allows to make a connection between these shapes and the three-dimensional unconfined helix. From Eqs.~(\ref{eq:radiuspitchofahelix}) and $(\ref{Gamma})$ one obtains the ratio between pitch and radius of the helix as $\frac{H}{R} = 4\sqrt{\frac{B}{C}}\frac{1}{\sqrt{\gamma}}$. In the regime $\gamma\gg 1$, the pitch of the unconfined 3D helix is much smaller than its radius, $H \ll R$, which translates to a circular squeelix after confinement onto the plane. 
In the opposite regime $\gamma\ll 1$, where the helix is extended ($H \gg R$), the confinement leads to other, nontrivial shapes with $m<1$. 
To understand these shapes in more detail we are now going to study the squeelix in terms of entities that we call twist-kinks.

\subsection{The twist-kink picture}

The solutions describing the ground state of the squeelix in terms of elliptic Jacobi functions are not very
illuminative. To gain more physical insight we will use the concept of a
twist-kink introduced in Ref.~\cite{Nam2012}. This object corresponds to a region of the filament
where the twist is highly concentrated and the curvature flips. A squeelix
can be interpreted as the result of the elastic interaction between twist-kinks along the
filament. Mathematical details are provided in Appendix C.


\subsubsection{The homoclinic pendulum $(m=1)$}

\begin{figure}
\centering
\subfigure{
\raisebox {5mm}{\includegraphics[width=4.5cm]{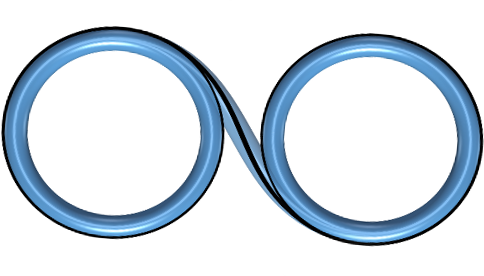}}}
\qquad
\qquad
\subfigure{
\includegraphics[width=3.5cm]{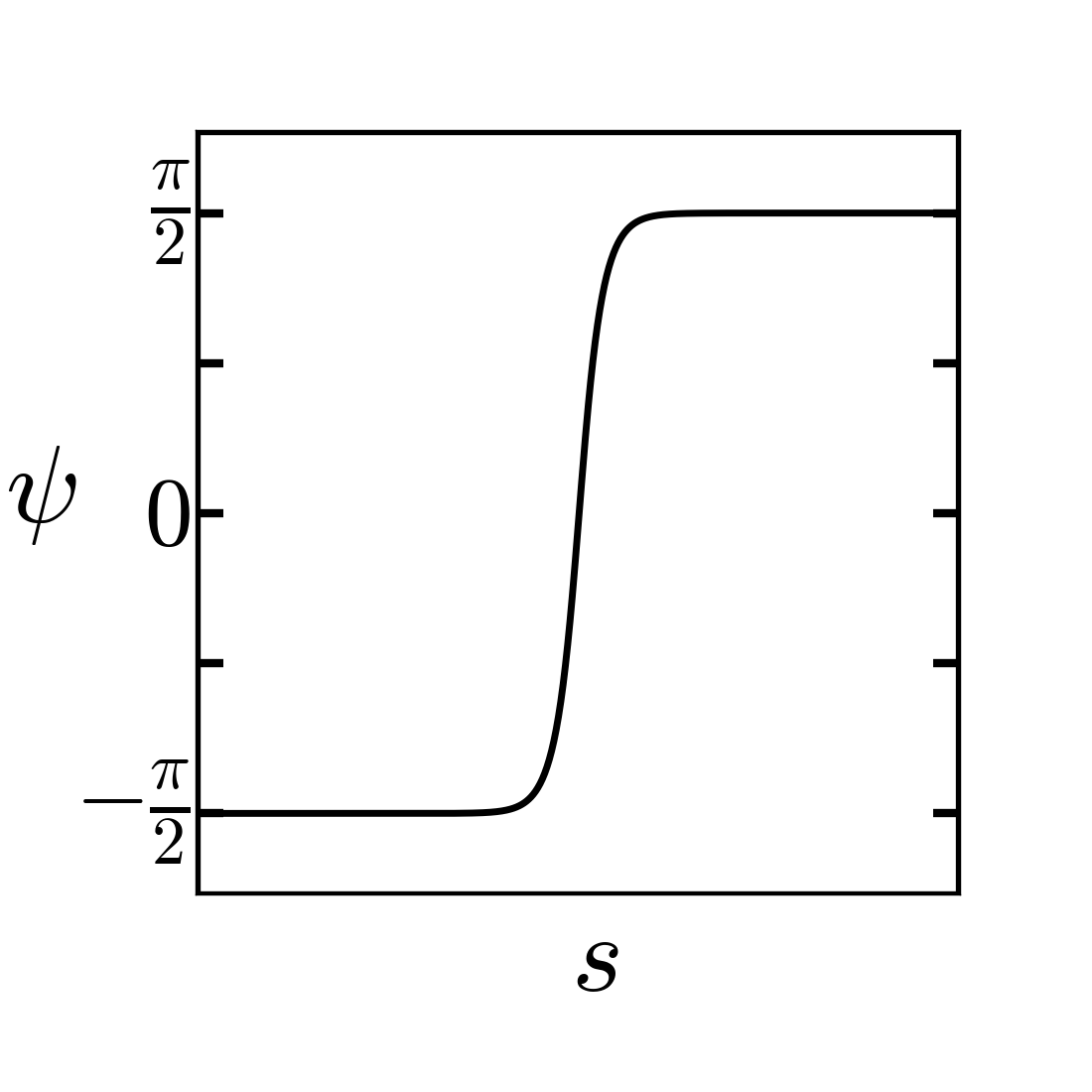}
}
\caption{Squeelix of infinite length for $m=1$ (left) and its twist $\psi(s)$ (right). The shape contains only a single twist-kink of size $\lambda$. Lengths are given in units of $\omega_1^{-1}$. In this example $\sqrt{C/B}=1$, $\omega_3=0.64\omega_1$, $\lambda=1$ and $\gamma=1$.}
\label{fig3}
\end{figure}
A filament with a \textit{single} twist-kink only exists for $m=1$. 
Eq.\ $\left(
\ref{psym}\right)  $ becomes the
Gudermann function $\psi(s)=gd\left(  s/\lambda\right)  $ which also reads%
\begin{equation}
\psi(s)=2\arctan(e^{s/\lambda})-\pi/2 \; .\label{twist}%
\end{equation}
This configuration interpolates between $\psi(-\infty)=-\pi/2$ and
$\psi(\infty)=\pi/2$ where the curvatures are opposite, \textit{i.e.}, $\kappa(\pm
\infty)=\pm\omega_{1}$. The region of the filament of size $\lambda$ where the
twist is highly concentrated and where the curvature
\begin{equation}
\kappa(s)=\omega_{1}\tanh(s/\lambda)\label{curvetwistkink}%
\end{equation}
flips (curvature inversion points) was called a \textit{twist-kink} in Ref.~\cite{Nam2012} 
in analogy to the concept of kinks in soliton physics
\cite{Currie1980}. Since $\lambda/ L \ll 1$  the filament consists of two regions of approximately constant opposite curvature $\kappa\approx\pm\omega_{1}$ separated by a region of size $\lambda$ (see Fig.~\ref{fig3}).

The energy of this chain is $E_{1}=2\sqrt{BC}\omega_{1}-\pi
C\omega_{3}+C\omega_{3}^{2}L/2.$ Therefore, the self-energy of a single
twist-kink is
\begin{equation}
\Delta E=E_{1}-E_{0}=\pi C\omega_{3}(\sqrt{\gamma}-1)
\; ,
\end{equation}
where $\gamma$ can now be interpreted as a \textit{twist-kink expulsion parameter} \cite{Nam2012}.
This terminology speaks for itself: for $\gamma>1$ the twist-kink is expelled from the filament, 
which consequently forms superimposed circles of radius $1/\omega_{1}$. For $\gamma<1$, the squeelix can by
populated by twist-kinks whose density is limited by their repulsive interactions.


\subsubsection{The revolving pendulum $(m<1)$}

\begin{figure}
\centering
\subfigure[][]{\label{fig4-a}
\raisebox {5mm}{\includegraphics[width=9cm]{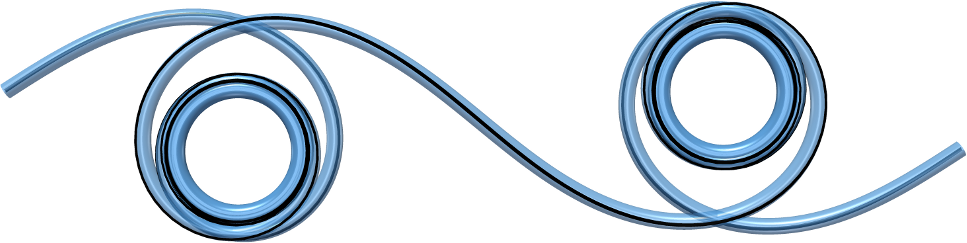}}
\qquad
\includegraphics[width=3.5cm]{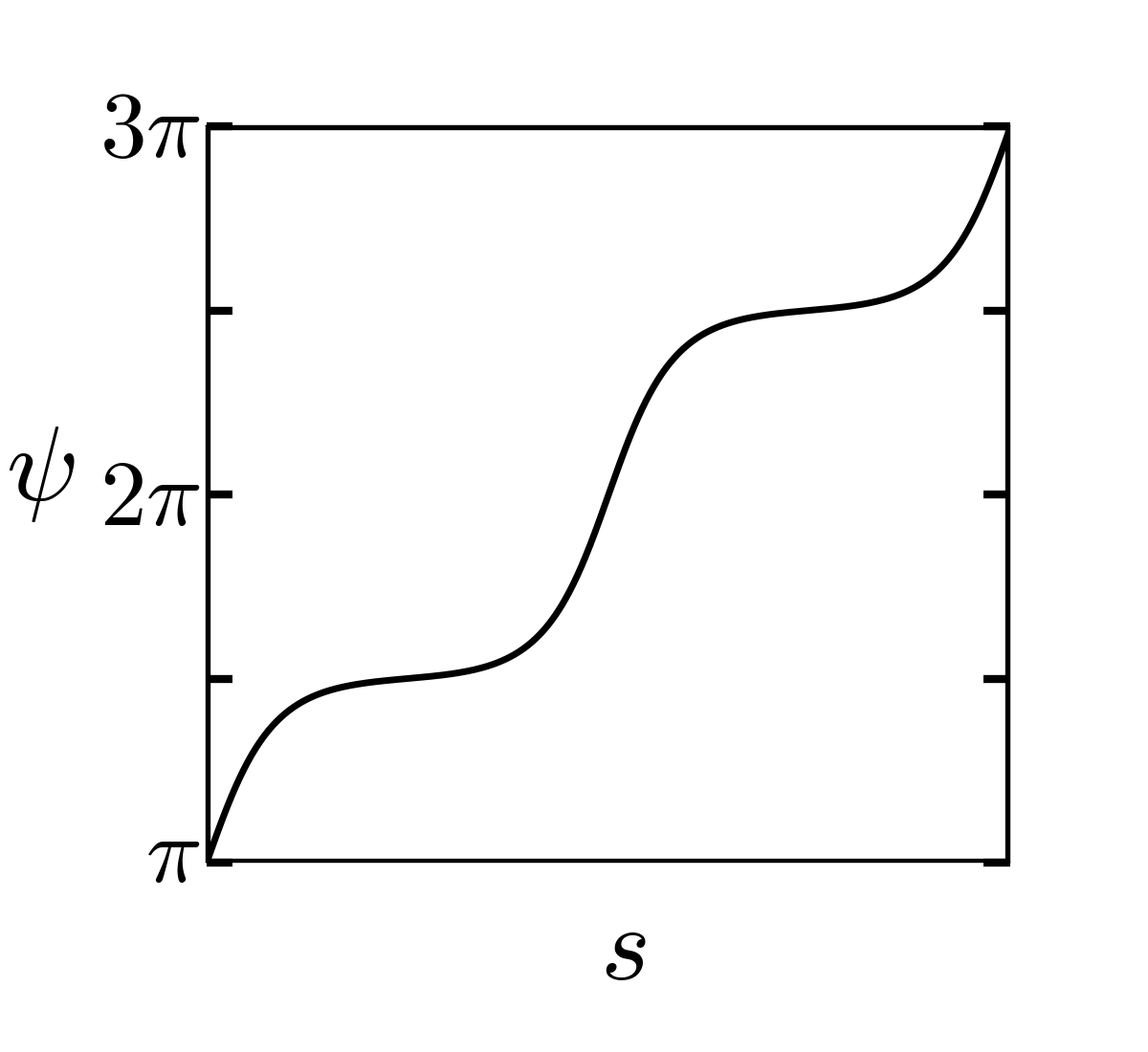}}\\
\subfigure[][]{\label{fig4-b}
\raisebox {7mm}{\includegraphics[width=8cm]{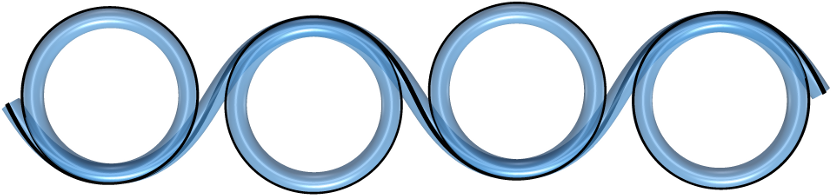}}
\qquad
\qquad
\qquad
\includegraphics[width=3.5cm]{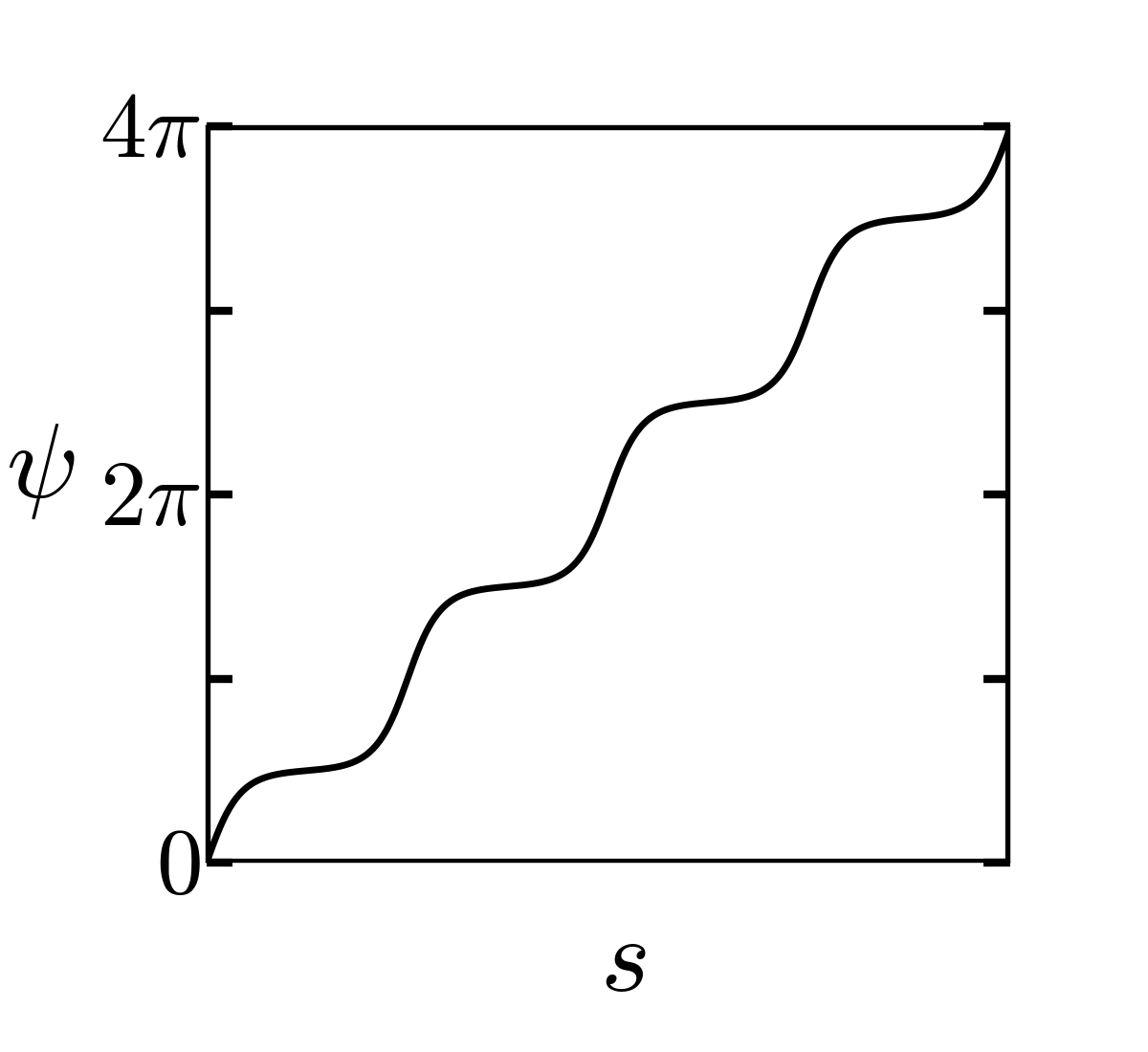}}\\
\subfigure[][]{\label{fig4-c}
\raisebox {7mm}{\includegraphics[width=8cm]{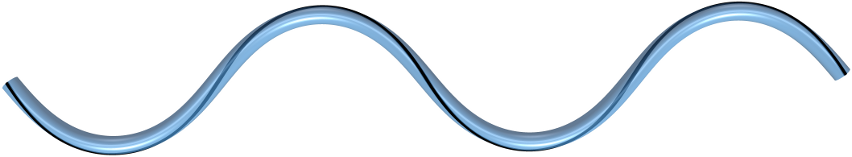}}
\qquad
\qquad
\qquad
\includegraphics[width=3.5cm]{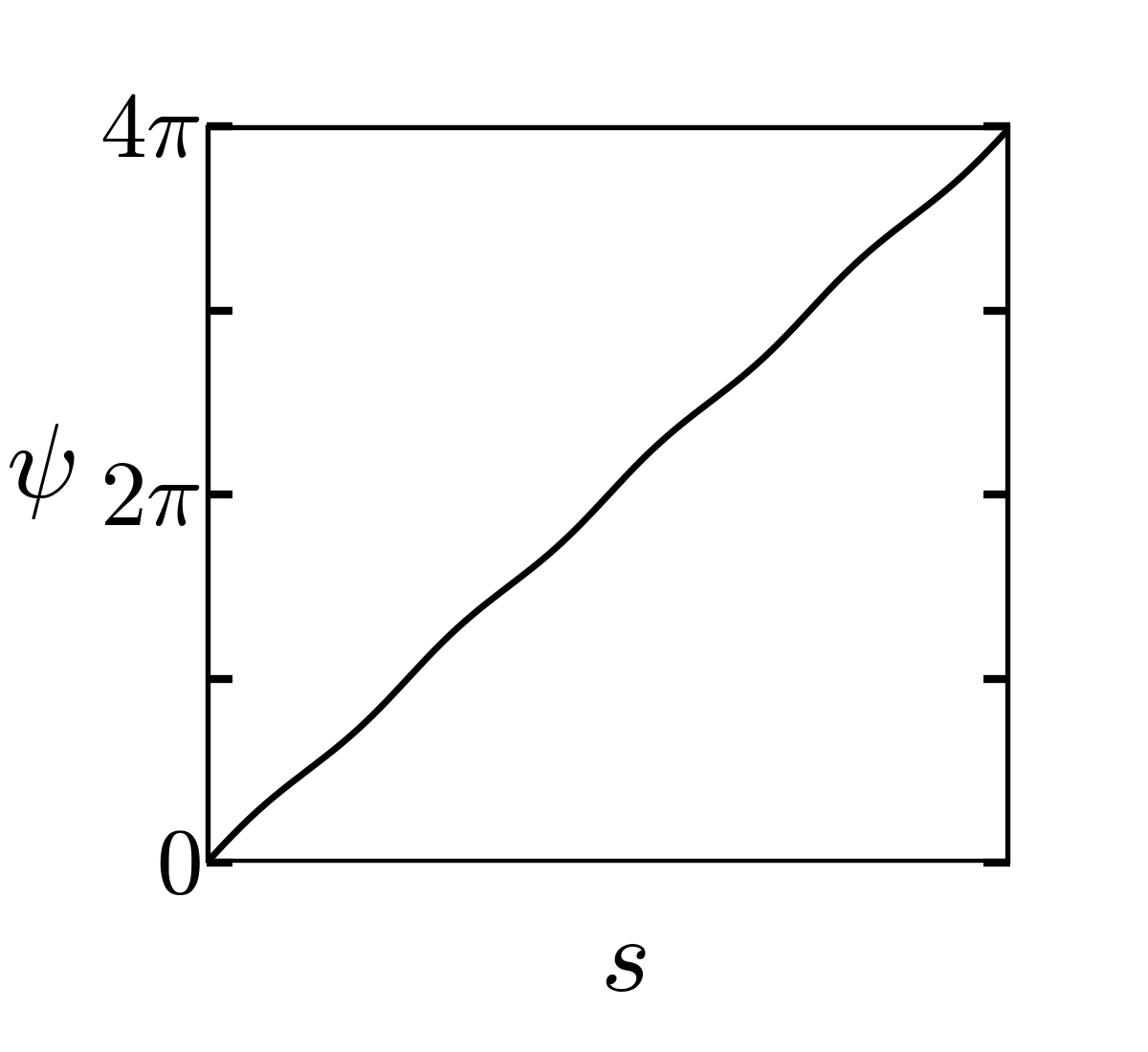}}\\
\caption{Finite sections of the shapes of squeelices of infinite length in their ground state (left) and the associated twist $\psi(s)$ (right) for $m<1$. Lengths are given in units of $\omega_1^{-1}$ and $l_{loop}=2\pi\omega_1^{-1}$. For $\gamma$ close to one typical shapes are either circular or spiral windings depending on the ratio $l_{cycle}/l_{loop}$. For instance, in (a) and (b) we have set $\gamma=0.995$ for which $m=0.999$. 
In (a) we have set $\sqrt{C/B}=10$ so that $\omega_3 \approx 0.06\omega_1$, $\lambda \approx 10.0\omega_1^{-1}$, and $l_{cycle}\approx 96.8 \omega_1^{-1}\gg l_{loop}$. In (b) we have decreased $\sqrt{C/B}$ to one so that $\omega_3\approx 0.64\omega_1$,  $\lambda \approx 1.0 \omega_1^{-1}$, and $l_{cycle}\approx 9.7\omega_1^{-1}$ which is of the order of $l_{loop}$. 
(c) For $\gamma=0.274$ one finds $m=0.5$. This corresponds to the dense twist-kink regime where the shape is sinus-like and the twist grows approximately linearly with $s$. The parameters of this example are $\sqrt{C/B}=1$ and thus  $\omega_3\approx 1.22 \omega_1$, $\lambda \approx 0.7 \omega_1^{-1}$, and $l_{cycle}\approx 2.6\omega_1^{-1}$.} 
\label{shapes_infini_minf1}
\end{figure}

In a revolving pendulum the twist angle $\psi(s)$ is a monotonously
growing function of $s$. The length $l_{cycle}$ is given by the condition
$\psi(s+l_{cycle})=\psi(s)+\pi$. As a consequence of Eq. $\left( \ref{EQM1}\right) $ 
the curvature of the filament reverses its sign every $l_{cycle}$, \textit{i.e.},
$\kappa(s+l_{cycle})=-\kappa(s)$. Depending on the ratio $l_{cycle}/\lambda$ the 
squeelix adopts different typical periodic shapes. 

When $l_{cycle}/\lambda\gg1$, we are in a regime where $m\lesssim 1$ since the expansion of Eq.~(\ref{eq:lcycle}) 
gives $l_{cycle}/\lambda\approx\ln(\frac{16}{1-m})$ at lowest order. For instance, one obtains 
$l_{cycle}/\lambda \ge10$ for $m\ge 0.999$. In this regime Eq.~(\ref{msol}) implies that $m\approx \gamma$.
The ground state of the filament is populated by a low density $\rho\simeq1/l_{cycle}$ of twist-kinks 
 of the form $\psi(s)\approx gd\left(s/\lambda\right)  +O(1-m)$.
The density is limited by the mutual kink-kink
repulsion $U_{int}\sim\pi C\omega_{3}\sqrt{\gamma}e^{-d/\lambda}$ for
$d/\lambda\gg 1$, where $d$ is the distance between the two entities (see Appendix C).

One finds three types of behavior depending on the ratio $l_{cycle}/l_{loop}$ where
$l_{loop}=\frac{2\pi}{\omega_{1}}$ is the length of a loop of constant radius
$\omega_{1}$.  
When $l_{cycle}\gg l_{loop}$ the shape of the squeelix consists of a succession of spiral windings 
with minimal radius of curvature $1/\omega_1$
and alternating sign of curvature, Fig \ref{fig4-a}. 
The twist-kinks are localized at the curvature inversion points.
When $l_{cycle}\approx l_{loop}$ each loop has only one turn Fig \ref{fig4-b}.
Finally, for $l_{cycle}<l_{loop}$ the shape consists
of a succession of flipped circular arcs (incomplete loops) separated by the twist-kinks.

When $l_{cycle}$ is of the same order or smaller than $\lambda$, no loops are formed 
and the shape of the filament is sinus-like Fig \ref{fig4-c}.
The density of twist-kinks is very high. The more $l_{cycle}/\lambda$ is decreased the more 
the twist-kinks are deformed up to the point where the notion of individual twist-kinks looses its
meaning. In the limit
$m\ll1$, the twist evolves linearly $\psi(s)\approx\frac{\pi
}{l_{cycle}}s$ with $l_{cycle}\approx\lambda\pi\sqrt{m}$ and the curvature is given by 
$\kappa(s)\approx\omega_{1}\sin(\frac{\pi}{l_{cycle}}s)$. 
This explains the sinus-like shape of the squeelix.  The number of
curvature inversion points per unit length (previously identified as the density of
twist-kinks) is given by $\rho\simeq1/l_{cycle}=\omega_{3}/\pi$ (see Appendix
D). Therefore, $m\approx\frac{\pi^{2}}{4}\gamma$, which implies that 
$\gamma$ is small in this regime.


\subsubsection{The oscillating pendulum ($m>1$)}

\begin{figure}
\subfigure[][]{\label{fig5-a}
\raisebox {4mm}{\includegraphics[width=6cm]{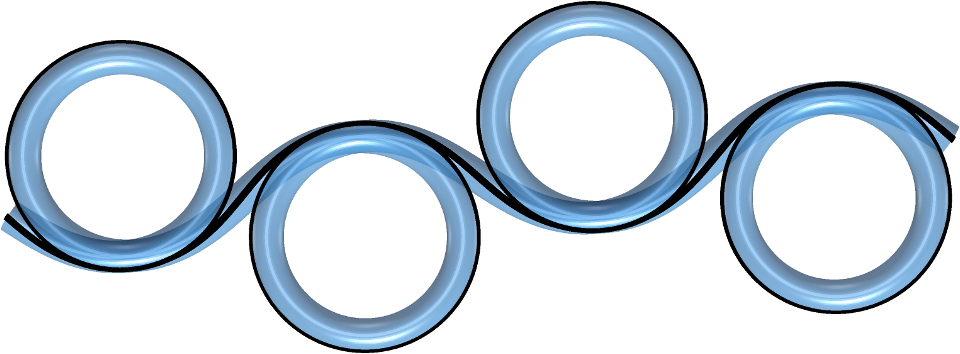}}
\qquad
\qquad
\includegraphics[width=3.5cm]{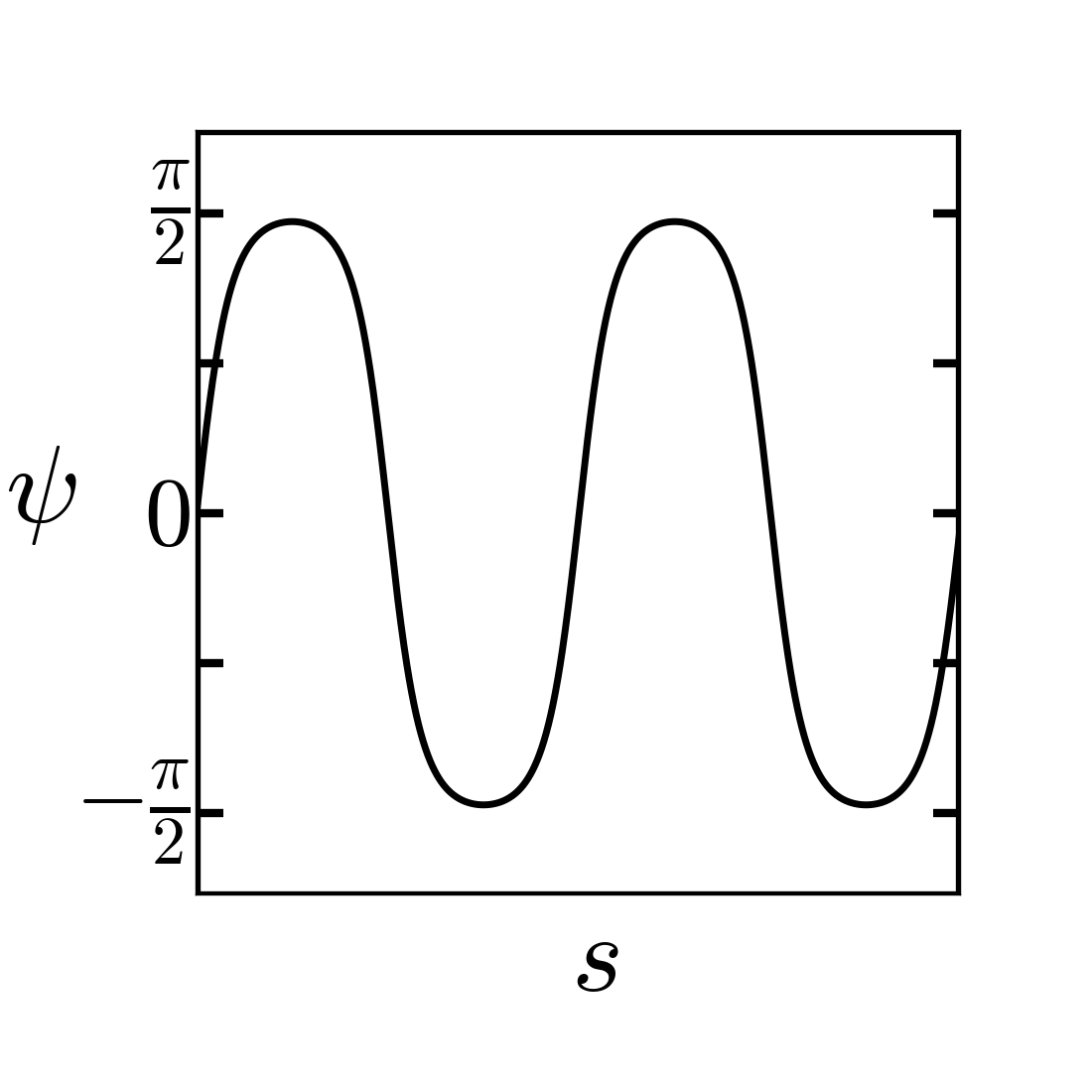}}\\
\subfigure[][]{\label{fig5-b}
\raisebox {12mm}{\includegraphics[width=7cm]{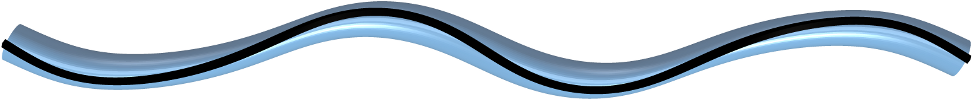}}
\qquad
\qquad
\includegraphics[width=3.5cm]{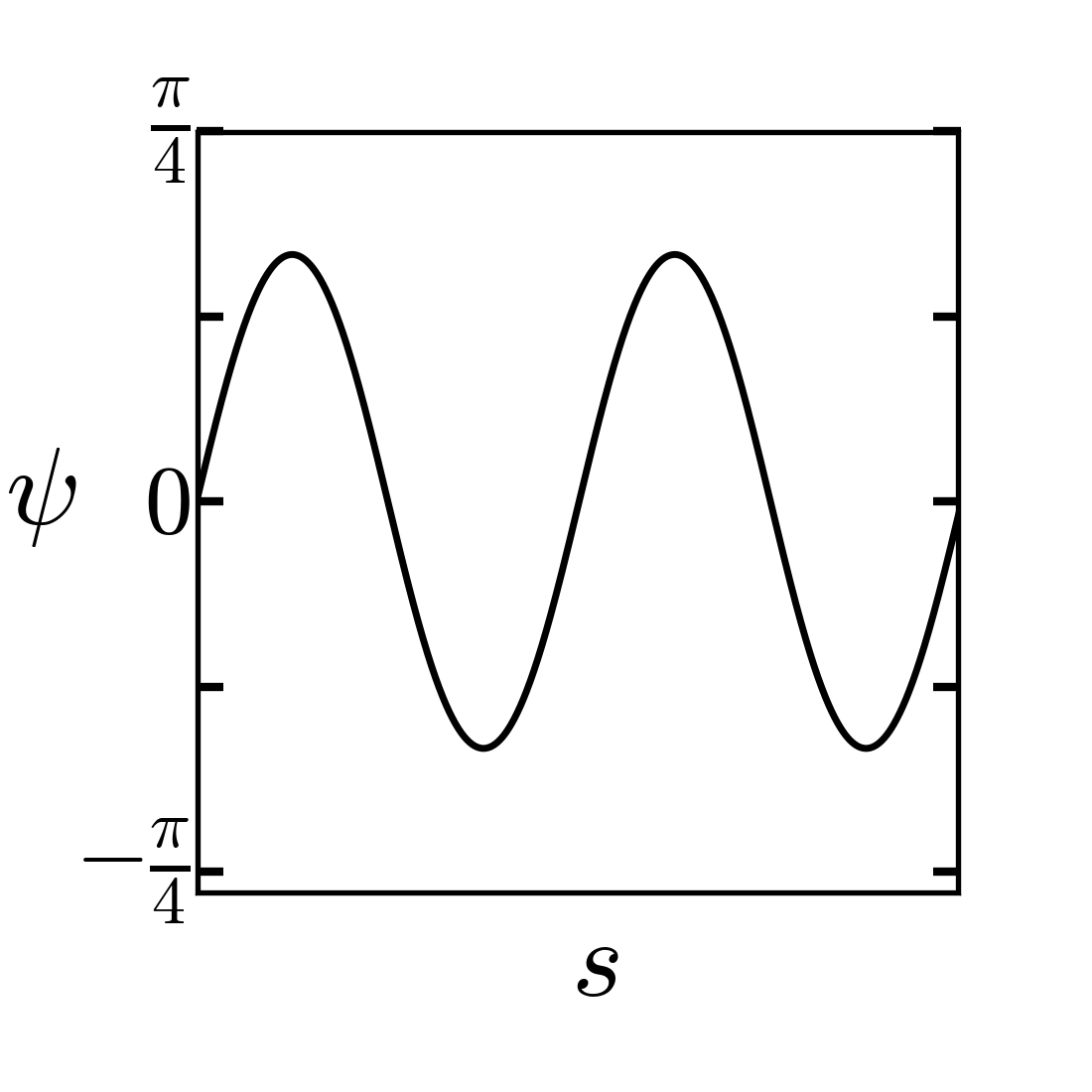}}
\caption{Finite sections of the shapes of squeelices of infinite length (left) and the associated twist $\psi(s)$ (right) for $m>1$. Lengths are given in units of $\omega_1^{-1}$ and $l_{loop}=2\pi\omega_1^{-1}$. These shapes do not minimize the energy density and are thus independent of $\gamma$. (a) Squeelix with $m=1.002$ and $\sqrt{C/B}=1$, and thus $l_{period}\approx 18\omega_1^{-1}\gg l_{loop}$. (b) Squeelix with $m=4$ and $\sqrt{C/B}=1$, and thus $l_{period}\approx 7\omega_1^{-1}$, which is of the same order of magnitude as $l_{loop}$.}
\label{fig5}
\end{figure}
Although the shapes do not minimize the energy density in this regime, we nevertheless consider
them for completeness. Moreover, we will see in the next section that similar shapes are 
local energy minima when the length of the squeelix is finite.

When $m>1$, the twist angle $\psi(s)$ oscillates periodically between two values
$[-\psi_{0},\psi_{0}]$ with $\psi_{0}=\arcsin(1/\sqrt{m})$ over an
arc length $l_{cycle}=2\lambda K\left(  \frac{1}{m}\right)$. The curvature, 
Eq. $\left(  \ref{Curv}\right)  $, can be written as
\begin{equation}
\kappa(s)=\frac{\omega_{1}}{\sqrt{m}}sn\left(  \frac{s}{\lambda}\left|\frac{1}%
{m}\right)\right.
\; .
\end{equation}
It is a periodic function of period $l_{p}=2l_{cycle}$ with maximal curvature $\kappa_{0}=\pm\omega_{1}/\sqrt{m}$.

In the regime $l_{cycle}/\lambda\gg 1$ where $m\gtrsim1$, the
formation of loops depends again on the ratio $l_{cycle}/l_{loop}$. The shape of the squeelix 
contains a low density of alternating twist-kinks and anti-twist-kinks. 
With decreasing $l_{cycle}/\lambda$ or
increasing $m$, the curvature becomes $\kappa(s)\approx\frac{\omega_{1}}%
{\sqrt{m}}\sin\left(  \frac{s}{\lambda}\right)  +O(1/m)$ with decreasing amplitude. 
The shape is sinus-like as well. The reason why this solution is not the ground
state of a squeelix is due to the fact that the anti-kink has $\psi^{\prime
}(s)<0$ around the curvature inversion point. This maximizes
the purely twist energy contribution to the total energy. Fig. \ref{fig5} show two typical shapes.


\section{Squeelices of finite length\label{sec:finitesqueelix}}

The case of a squeelix of infinite length whose structure repeats itself
allowed us to grasp a physical intuition of the system. But in the real
world, filamentous objects always have a finite size. We are now going to
study the more realistic case of a filament of finite length $L$ in
detail. The major difference is that for infinite length, the density of
twist-kinks is determined by their mutual repulsion which depends only on the
material parameters. For the finite case there is
the additional constraint that the twist-kinks must fit inside the chain.
Their density will thus depend on $L$. Our goal is to study all these solutions and give them a 
physical sense. Here we will focus only on the main
results as the mathematical details are provided in Appendices C, D and E.


\subsection{Basic equations}

For a filament of finite length $L$, the general solution of Eq. $\left(
\ref{EQM2}\right)  $ reads
\begin{equation}
\psi(s)=am\left(  \frac{s-s_{0}}{\lambda\sqrt{m}}|m\right)  \label{F1}%
\end{equation}
with the arc length $s_{0}$ such that $\psi(s_{0})=0$ (see
Section~\ref{subsec:thesqueelix}).\ Importantly, this solution must satisfy
the boundary conditions Eq.\ $\left(  \ref{BC}\right)  $
\begin{equation}
\psi^{\prime}(-L/2)=\omega_{3}=\psi^{\prime}(L/2)\label{BCLF}%
\end{equation}
with
\begin{equation}
\psi^{\prime}(s)=\frac{1}{\lambda\sqrt{m}}dn\left(  \frac{s-s_{0}}
{\lambda\sqrt{m}}|m\right)  \; .\label{F2}
\end{equation}
Plugging Eqs.~(\ref{F1}) and (\ref{F2}) into the expression of the energy, Eq.~(\ref{Ecyclegeneral}), leads to
the energy $E(m,s_{0})$ for a given length $L$ as shown in
Fig. \ref{Ems0} of Appendix D. The energy $E(m,s_{0})$ is defined on a subspace of the
functional space of the energy $E[\psi]$ in Eq.~(\ref{sqeezedenergyPSI}). 
We observe an energy landscape with many localized minima, maxima and saddle points. Integrating
Eq.~(\ref{F2}) from $s_{0}$ to $-L/2$ leads to
\begin{equation}
s_{0}=-\frac{L}{2}-\lambda\sqrt{m}\mathcal{F}\left(  \psi(-L/2)|m\right)  \; ,\label{SOM}
\end{equation}
where $\mathcal{F}\left( x|m\right)$ is the elliptic integral of the first kind \cite{Abramowitz}. As shown in Appendix A one obtains
\begin{equation}
\psi\left(  -L/2\right)  =\pm\arcsin\left(  \sqrt{\frac{1}{m}-\frac{1}{\mu}
}\right) +n\pi \label{F3}
\end{equation}
with $n=0,1$, as a consequence of Eq.~(\ref{BCLF}). These two cases are related by the transformation $\psi(s)\rightarrow \psi(s)+\pi$. 
They lead to two shapes related by the transformation $\kappa(s)\rightarrow-\kappa(s)$. These two shapes have thus the same energy. 
We therefore consider the case $n=0$ only. Thus $-\pi/2<\psi\left(  -L/2\right)  <\pi/2$ and
\begin{equation}
\frac{\mu}{1+\mu}\le m\le \mu
\end{equation}
with $\mu=\frac{\omega_{1}^{2}B}{\omega_{3}^{2}C}$.  
We still have to determine
the values of $m$ associated to the extrema of the energy. But first it is
interesting to treat $s_{0}$ as a function of $m$. This leads to two
different trajectories $s_{0,\pm}(m)$ on the energy surface $E(m,s_{0})$
depending on the sign of $\psi\left(-L/2\right)  $ in Eq.~(\ref{F3}). In
Appendix D it is shown that the trajectory $s_{0,+}(m)$ connects saddle points to minima of the energy landscape $E(m,s_{0})$ whereas $s_{0,-}(m)$ connects maxima to saddle points.  
As we will see the local maxima and saddle points of $E(m,s_{0})$ are also saddle points of the energy $E[\psi]$ (the latter having no local maximizers) and thus they lead to unstable squeelices. Although it appears possible that some of the local minima of $E(m,s_{0})$ are saddle points of $E[\psi]$, this is in fact not the case. All minima of $E(m,s_{0})$ are minima of $E[\psi]$ and their corresponding shapes are consequently stable (see Appendix E). The approach through the introduction of the energy $E(m,s_{0})$ will allow us to give a physical understanding of all these extremal shapes.

\begin{figure}
\centering
\subfigure[]{\label{Ema}\includegraphics[width=0.4\textwidth]{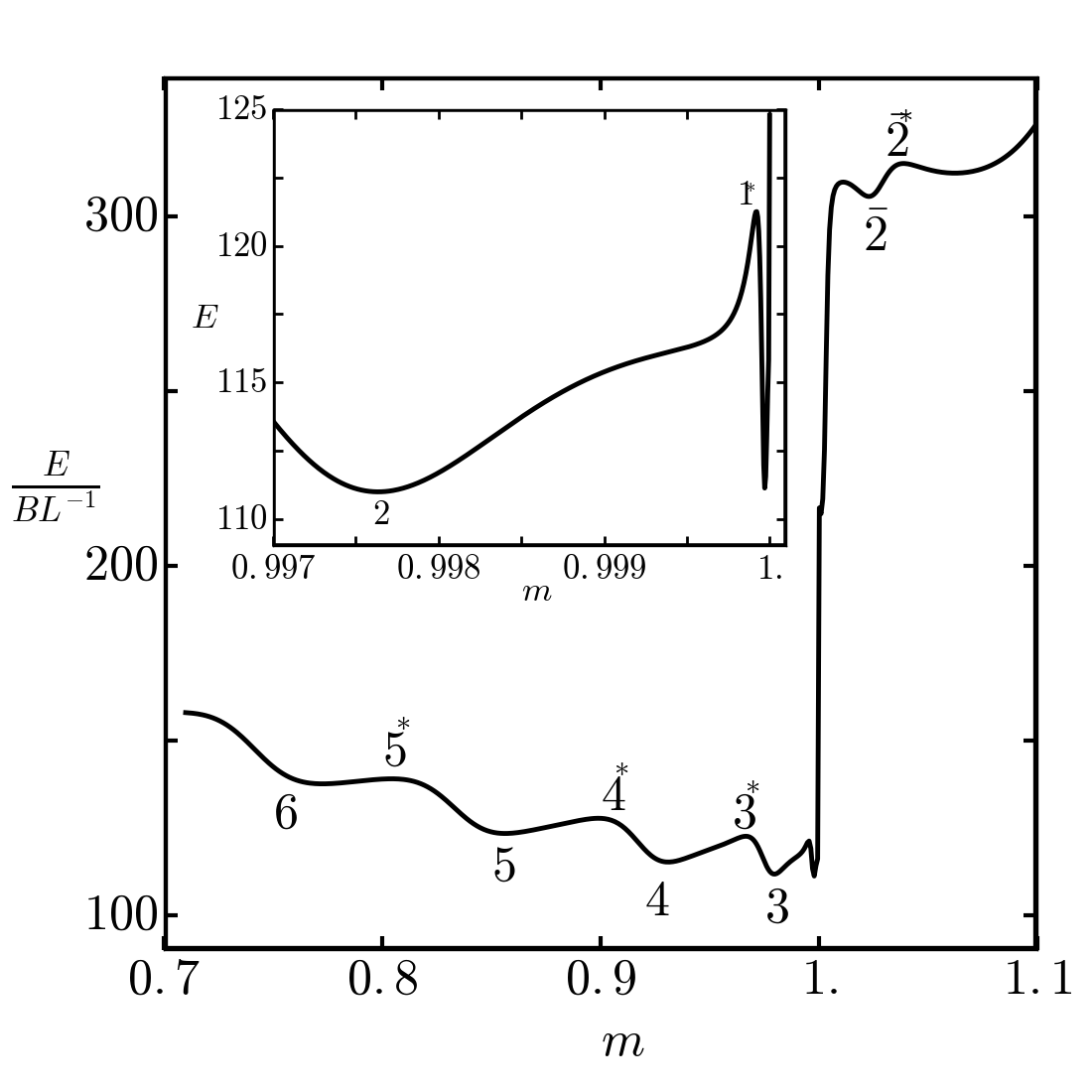}}
\qquad
\qquad
\subfigure[]{\label{Emb}\includegraphics[width=0.4\textwidth]{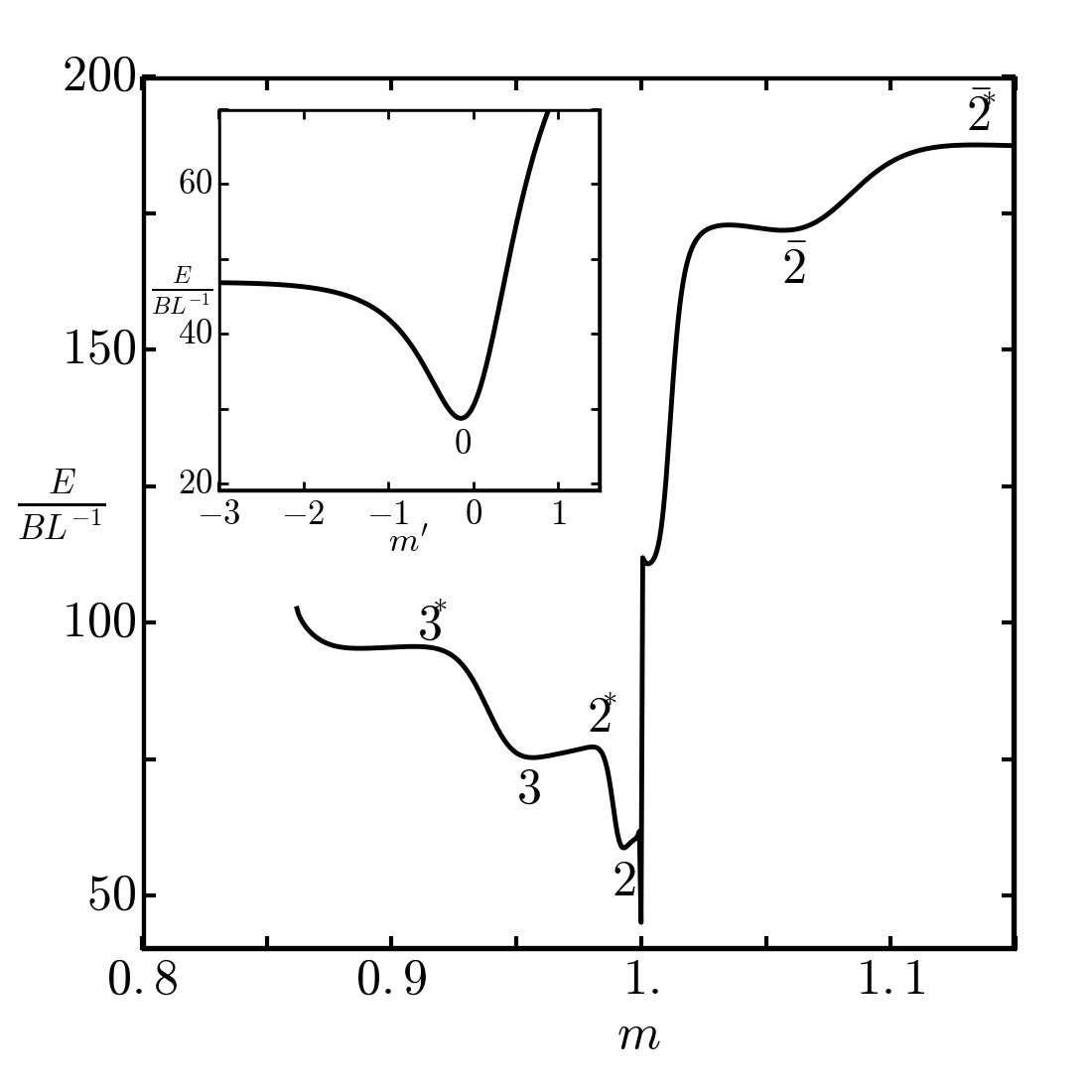}}
\caption{Energy $E(m,s_{0,+}(m))$ as a function of $m$ for different values of $\gamma$. Energies and lengths are measured in units of $BL^{-1}$ and $L$, respectively. We set $\sqrt{C/B}=1$. For $m<1$ the integer $n$ denotes the energy minima of squeelices with $n$ twist-kinks and $n^*$ denotes the energy barriers corresponding to unstable critical shapes. The notation $\bar{n}$ and $\bar{n}^*$  are used for the local minima and maxima respectively in the region $m>1$.
In this region some of the extrema are not numbered since they do not satisfy the boundary conditions, Eq.~(\ref{BC}) (see the explanation in the main text). (a) Energy $E(m,s_{0,+}(m))$ for $\gamma=0.9915$ with $\omega_1=24.4L^{-1}$ and $\omega_3=15.6L^{-1}$. The ground state corresponds to the minimum $2$ as shown in the inset. In this state the squeelix has two twist-kinks. The minima near $m=1$ (states $n=0$, $n=0^*$ and $n=1$) are too close to each other to be distinguishable at this scale (first minimum, not numbered). (b) Energy for $\gamma=2.5330$, $\omega_1=20L^{-1}$ and $\omega_3=8L^{-1}$. In this regime the ground state is the state without twist-kink (the state denoted 0) as shown in the inset where we introduced $m^{\prime}=(m-1)\cdot 10^8$ for convenience. A finite number of local minima with various numbers of twist-kinks exists. The states $0^*$, $1$ and $1^*$ are not distinguishable in the figure. In both cases (a) and (b) the states $\bar{1}$ and $\bar{1}^*$ are not distinguishable as well.}
\label{Em}
\end{figure}

To obtain the energy minima $E(m,s_{0}(m))$, we focus on the case $s_{0,+}(m)$
in the following. Fig.~\ref{Em} shows the function
$E(m,s_{0,+}(m))$ for different values of $\gamma$. One finds that the
global minimum lies in the interval $0<m<1$ for all values of $\gamma$. In the
regime of weak $\gamma$ the local minima with $m<1$ are lower than the local
minima with $m>1$. These configurations are thus less accessible at finite
temperature. For increasing $\gamma\lesssim1$, the local minima for both
$m<1$ and $m>1$ can have comparable energies, and the global minimum approaches
$m=1$ from below. Note, however, that it is only equal to one in the infinite case. In the opposite 
regime $\gamma\gg1$ local minima with $m>1$ are lower
than the local minima with $m<1$. What are the typical shapes of these extrema?

\subsection{Shapes of squeelices of finite length}

We must again separate the study in two parts. While we focus on the main
points in this section, we refer to Appendix D for more details on the
mathematical derivation.


\subsubsection{The revolving pendulum ($m < 1$)}

A revolving pendulum with the boundary conditions $\psi^{\prime}%
(-L/2)=\omega_{3}=\psi^{\prime}(L/2)$ must satisfy either
\begin{subequations}
\begin{align}
\psi\left(  L/2\right)  & =\psi\left(  -L/2\right)  +n_{a}\pi\text{
\ \ \ }(\text{case} \; \text{(a)}) \qquad \text{or} \label{psyNA} \\
\psi\left(  L/2\right)  & =-\psi\left(  -L/2\right)  +n_{b}\pi\text{
\ \ \ }(\text{case} \; \text{(b)}) \; . \label{psyNB}%
\end{align}
\end{subequations}
As discussed in the previous section we focus primarily on $\psi\left(  -L/2\right)  >0$,
\textit{i.e.}, the case $s_{0,+}$. The integers $n_{a,b}\geq1$ cannot exceed a
maximum value which depends on the chain length $L$ and the twist-kink
repulsion. The solutions of case (a) and (b) correspond to the maxima and minima of the energy $E(m,s_{0,+}(m))$, respectively, as shown in Fig.~\ref{Em}. 
Intuitively, the minima correspond to symmetric shapes (with respect to the center of the chain at
$s=0$) with $n_{b}-1$ twist-kinks disposed along the chain in an equidistant
manner. These shapes are stable, \textit{i.e.}, minimizers of $E[\psi]$ (see Appendix E). The ground state $n_{b}=n_{b}^{\ast}$ results from the optimal combination of the twist-kink self-energy (when it is negative) and the repulsive energy between them.

It is appealing to look at simple shapes associated to the states of Fig.~\ref{Em}. 
When $n_{b}=1$, the shape is a circular arc without twist-kinks, whereas $n_{a}=1$ corresponds to a curvature inversion point or a partial twist-kink localized near the end of the chain at $s=L/2$. This is the critical configuration on the top of the energy barrier of $E(m,s_{0,+}(m))$ that the system must exceed in order to reach the next minimum with $n_{b}=2$. The corresponding shape contains a single twist-kink in the middle of the chain. The procedure for higher $n$ follows the same behavior. Therefore, asymmetric shapes with a curvature inversion point, that we call a critical twist-kink, localized at $s=L/2$ correspond to case (a). They are the critical shapes on top of the energy barrier of $E(m,s_{0,+}(m))$ that must be overcome to inject (or remove) an additional twist-kink into the chain in order to reach the next local minimum (see Fig.~\ref{figminfeq1}). These critical asymmetric shapes are unstable.

To come back to the case $s_{0,-}$ note that a solution of case (a) with $s_{0,-}$ leads to a shape which is almost identical to its $s_{0,+}$ counterpart, except that the curvature inversion point that is localized at $s=L/2$ for $s_{0,+}$ is localized at $s=-L/2$ (see Appendix D for more details). In other words, the shape contains a critical twist-kink symmetrically localized at the opposite end of the chain. Since it costs the same energy to add a twist-kink from one end or the other end of the chain, both shapes have the same elastic energy. 

A solution of case (b) with $s_{0,-}$ turns out to be an energy maximum of $E(m,s_{0,-}(m))$ with a critical twist-kink at each of the two ends. For this reason its energy is even higher than the energy of the equivalent maximum of case (a) with $s_{0,+}$. The trajectory $s_{0,-}(m)$ thus passes through states which contain either one or two critical twist-kinks. 

In summary, the shapes having either one critical twist-kink near one end of the chain, \textit{i.e.}, case (a) of the trajectories $s_{0,+}$ and $s_{0,-}$ or two critical twist-kinks at both ends,  \textit{i.e.}, case (b) of the trajectory $s_{0,-}$, are all unstable. Examples are provided in Fig.~\ref{unstable shape}.


\subsubsection{The oscillating pendulum ($1<m<\mu$)}

In this regime we have the following two boundary conditions
\begin{subequations}
\begin{align}
\psi\left(  L/2\right)   &  =\psi\left(  -L/2\right)  \text{ \ \ \ }
(\text{case} \; \text{(a)}) \; ,\\
\psi\left(  L/2\right)   &  =-\psi(-L/2)\text{ \ \ \ }(\text{case} \; \text{(b)}) \; ,
\end{align}
\end{subequations}
which corresponds to equal (a) or opposite (b) curvatures at the ends of the
squeelix. The cases (a) and (b) correspond to the local maxima and minima of the energy $E(m,s_{0,+}(m))$, respectively. In Appendix E it is shown that only the solutions of case (b) with $\psi(-L/2) > 0$ (trajectory $s_{0,+}$) are stable. Therefore, as for $m<1$ we focus mainly on the trajectory $s_{0,+}$. 
In case (a) the boundary conditions Eq.~(\ref{BCLF}) imply
that $L=l_{p}n_{a}$ with the integer $n_{a}\ge1$ and $l_{p}(m)$ the period of
oscillations. 

For a given $n_{a}$ there are thus $2n_{a}$ curvature inversion
points (where $\psi\left( s\right)  =0$). However, the last one is close to
the end of the chain $s=L/2$. This shape is asymmetric with respect to the
center of the chain. Similarly to the case of the revolving pendulum it is
the critical shape on top of the energy barrier that must be overcome to reach
the next minimum of $E(m,s_{0,+}(m))$, \textit{i.e.}, a shape of type (b). In
contrast to before one now has to add or remove an anti-twist-kink at the end
of the chain. The shapes of case (b) are symmetrical with respect to $s=0$ as
their curvature inversion points are equidistant along the chain (see Fig \ref{figminfeq1}). 
They are minimizer of $E[\psi]$ and are stable.

The solutions of case (a) are again related to their counterpart with $s_{0,-}$  
and have the same energy (details in Appendix D). They are critical shapes with a curvature inversion point (close to the origin of the chain for $s_{0,-}$, and close to its end for $s_{0,+}$) on top of the energy barrier of $E(m,s_{0,+}(m))$. 

Case (b) with $s_{0,-}$ is an energy maximum as
the shapes contain two curvature inversion points near each end of the chain. These
configurations again have a much larger energy than the equivalent maximum of
case (a) with $s_{0,+}$.

The shapes having either one critical anti-twist-kink (curvature inversion point) near one end of the chain, \textit{i.e.},  case (a) of the trajectories $s_{0,+}$ and $s_{0,-}$ or two critical anti-twist-kinks at both ends,  \textit{i.e.}, case (b) of the trajectory $s_{0,-}$, are all unstable. Two examples with a single critical anti-twist-kinks can be seen in Fig.~\ref{figmsupeq1}.

Fig.~\ref{Em} shows the energy $E(m,s_{0,+}(m))$ for different values of $\gamma$. The corresponding trajectories $s_{0,+}(m)$ are solutions of the Euler-Lagrange equations with $\psi'(-L/2)=\omega_3$. They pass by the extrema we are searching for, \textit{i.e.}, which fulfill the boundary conditions at both ends, Eq. (\ref{BC}). For $m>1$ these trajectories also exhibit other extrema $m_i$ which do not satisfy $\psi'(L/2)=\omega_3$ but are nevertheless extrema $\delta E=0$ as the corresponding boundary term in Eq.~(\ref{eq:firstvariation}) vanishes due to $\delta \psi(L/2)\big|_{m_i} =\frac{\partial \psi(L/2)}{\partial m}\big|_{m_i}\delta m= 0$. 
Even though $s_{0,+}(m)$ passes by all local extrema one can find other trajectories $s(m)$ which do not satisfy the boundary conditions but have a lower energy for a given $m$ (not shown in Fig. \ref{Em}).


\subsubsection{Finding the ground state}

As we have seen, choosing the material parameters such that the \textit{twist-kink
expulsion parameter} $\gamma>1$ leads to a ground state with zero twist-kinks. The twist $\psi$ of this ground state is given by a Jacobi amplitude function
with a value $m=\underline{m}$ very close but smaller than $1$. An example is given by the state $0$ in Fig. \ref{Em} where $\underline{m}=0.9999999985$. This state satisfies the condition 
$\psi\left(  L/2\right)  =-\psi\left(
-L/2\right)  +\pi$ (cf. Eq. $\left(  \ref{psyNB}\right)  $ with $n_{b}=1$)
with $0<$ $\psi\left(  -L/2\right)  =\arcsin\left(  \sqrt{1/\underline{m}%
-1/\mu}\right)  <\pi/2$. The shape of the filament is a circular arc of curvature $\kappa=\omega_1$  in its bulk with deformed ends. This ground state is obviously degenerate with the symmetric shape $\kappa=-\omega_1$ corresponding to the transformation $\psi(s)\rightarrow \psi(s)+\pi$. 

For $\gamma<1$, the injection of twist-kinks is favored and the theory
predicts the existence of many metastable (and unstable) states with a
different number of twist-kinks within the filament. The energy of these
states consists of two terms. The negative self-energy of the twist-kinks and
their positive mutual interaction which also includes the repulsion between the
twist-kinks and the partial twist-kinks at the chain ends. The global minimum
of the energy is reached for an optimal number of twist-kinks that makes 
the best compromise between the two contributions of the energy. 

For each value of $m$ corresponding to a local minimum of the energy
$E (m,s_{0,+}(m))$ there is a state satisfying Eq. $\left(  \ref{psyNB}%
\right)  $ and having $n_{b}(m)-1$ twist-kinks. These states satisfy the
relation
\begin{equation}
L=l_{cycle}(m)n_{b}(m)-2\Delta s(m) \label{Ln}%
\end{equation}
with $\Delta s(m)=-L/2-s_{0,+}(m)=\lambda\sqrt{m}\mathcal{F}\left(  \psi(-L/2)|m\right)  $
and $l_{cycle}=2\lambda\sqrt{m}\mathcal{K}(m)$. Thus the number of twist-kinks
associated to this state is given by:
\begin{equation}
n_{b}(m)=\frac{L+2\Delta s(m)}{l_{cycle}(m)}\label{nL}%
\; .
\end{equation}
For each integer $n_{b}=1,2,...$ there is an associated value of $m$
corresponding to a local minimum of the energy. The case $n_b=0$ is relevant for very short length $L$ only (see Appendix D) and is not considered in the main text. The function $n_{b}(m)$ is
decreasing with $m\in\left[  \mu/(1+\mu),\underline{m}\right]$. The maximum
number of twist-kinks within the filament or equivalently the number of
metastable states $n_{max}$ is 
\begin{equation}
n_{max}=\left\lfloor n_{b}(\frac{\mu}{1+\mu}) \right\rfloor -1=\lfloor\frac{L}{l_{cycle}(\mu/(1+\mu))}\rfloor \; , \label{nmax}
\end{equation}
where $\left\lfloor x\right\rfloor $ denotes the largest integer less
than or equal to $x$. The value of $\underline{m}$
is given by the condition $n_{b}(\underline{m})=1$. 

It is possible to determine the global minimum of the energy $E(m,s_{0,+}(m))$ 
from the energy density of a squeelix of infinite length
$e_{m<1}(m)$ (cf. Eq. $\left(  \ref{Ecycle1}\right)  $). The energy density
$e_{m<1}(m)$  has a single minimum at $m=m^{\ast}$. Because of the cyclic
nature of $\psi\left(  s\right)  $ a portion of length $L_{n}$ of this
infinite chain which contains, say $n-1$ twist-kinks, must satisfy the
relation
\begin{equation}
L_{n}=l_{cycle}(m^{\ast})n-2\Delta s(m^{\ast})
\; .
\end{equation}
This value $L_{n}$ is the length that allows to contain $n-1$ twist-kinks in an
optimal manner, \textit{i.e.}, which minimizes $e_{m<1}(m)$. Here the twist angle
reads $\psi(s)=am\left(  \frac{s}{\lambda\sqrt{m^{\ast}}}|m^{\ast}\right)  $
if $n$ is even and $\psi(s)=am\left(  \frac{s-l_{cycle}(m^{\ast})/2}%
{\lambda\sqrt{m^{\ast}}}|m^{\ast}\right)  $ if $n$ is odd. But in both cases
$\psi^{\prime}(-L_{n}/2)=\psi^{\prime}(L_{n}/2)=\omega_{3}$. In general the
length $L$ of the filament is given and is not equal to $L_{n}$. But if we
choose $n$ such that $L_{n}<L<L_{n+1}$, the state that minimizes the
energy $E(m,s_{0,+}(m))$ is a state with $n_{b}=n$ or $n_{b}=n+1$. The
value of $n$ can be found via Eq. $\left(  \ref{nL}\right)  $ which gives
$n_{b}(m^{\ast})$. Then $n=\left\lfloor n_{b}(m^{\ast})\right\rfloor $. 
Knowing $n_{b}$ we determine $m$ from Eq. $\left(\ref{nL}\right)  $.


\subsubsection{Example}
In this paragraph we consider an example to illustrate the theory. We show some shapes associated to the energy states (minima and maxima) of Fig.~\ref{Ema}. The numbering of the shapes follows that of the figure: a shape designated by $(n)$ has $n$ twist-kinks and is a state of minimal energy, \textit{i.e.}, $n_b=n+1$. A shape $(n^*_{+})$ is a critical shape (asymmetric and unstable) with $n_a=n+1$. It is the critical shape on the top of the energy barrier between the energy minima $n$ and $n+1$. The symbol $+$ reminds us that we consider the line $s_{0,+}$. We measure all lengths in units of $L$ and the energies in units of $BL^{-1}$. We also choose $\sqrt{C/B}=1$. The shapes of the squeelix associated to the extrema of Fig. \ref{Em}(a) and (b) will be very similar (because both energy curves are defined in a comparable range of $m$). Consequently, we will only consider the shapes associated to the extrema of the energy in Fig. \ref{Ema}, where $\gamma=0.9915$, $\omega_1=24.4L^{-1}$ and $\omega_3=15.6L^{-1}$. In this case $0.7\le m\le 2.44$ and the typical size of a twist-kink is $\lambda \approx 0.04L$ which allows a maximum number $n_{max}=6$ of twist-kinks within the chain (from Eq. \ref{nmax}). We will treat the regions $m<1$ and $m>1$ separately.  
\\

\begin{figure}
\begin{center}
\centering
\begin{tabular}{cccc}
 \includegraphics[scale=0.07]{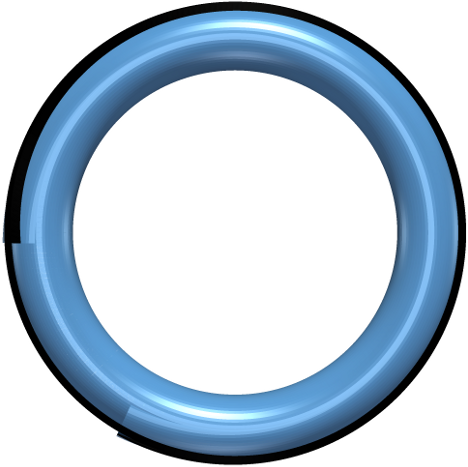} ($0$) &
\includegraphics[scale=0.07]{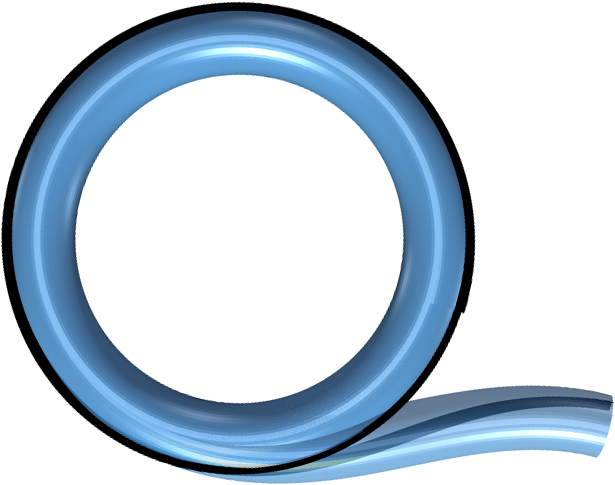} ($0^*_+$) & 
\includegraphics[scale=0.08]{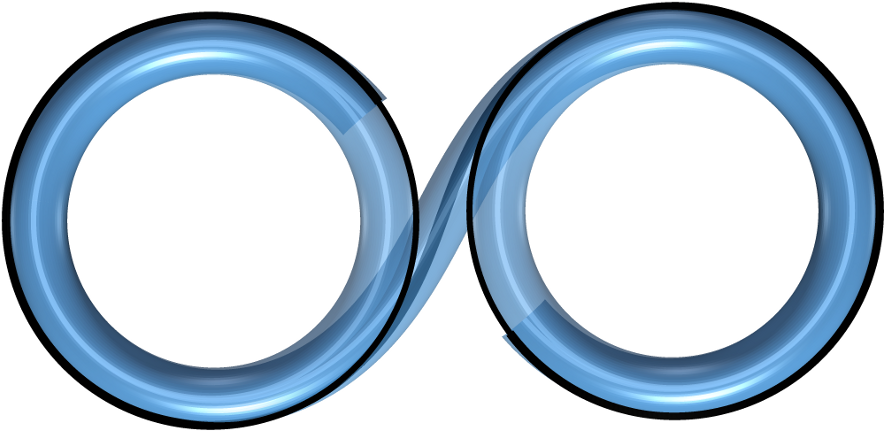} ($1$) \\
\\
\\
\includegraphics[scale=0.08]{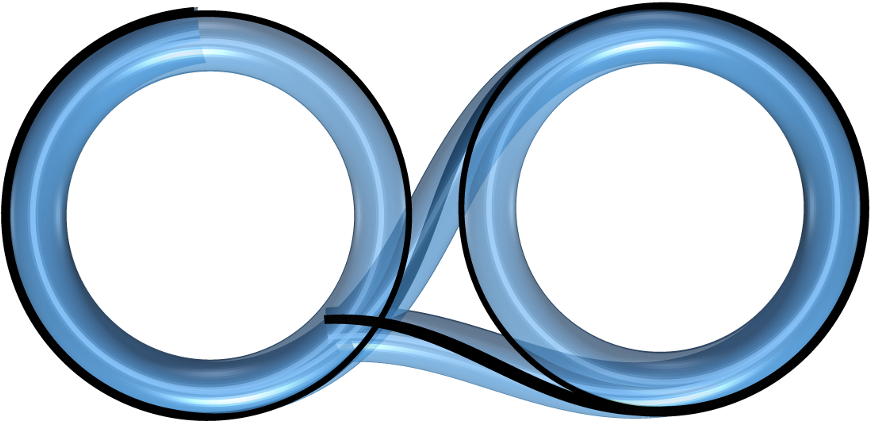} ($1^*_+$)  &
\includegraphics[scale=0.11]{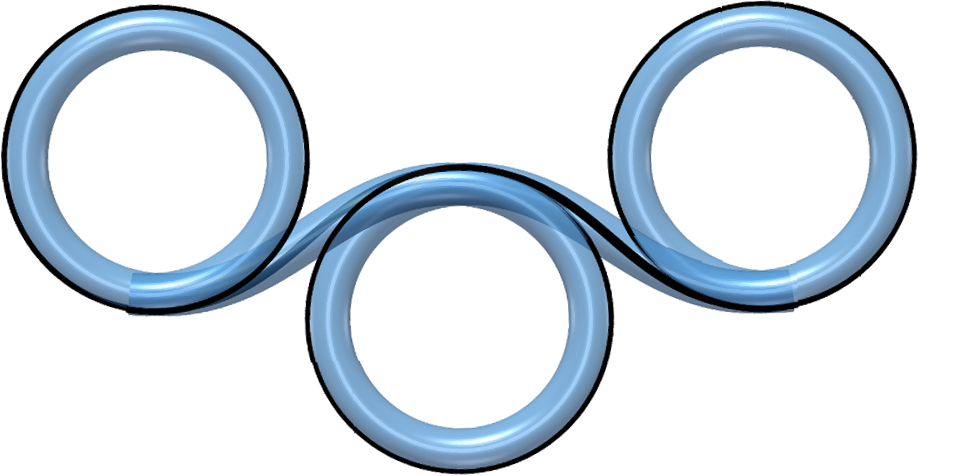} ($2$) &
\includegraphics[scale=0.11]{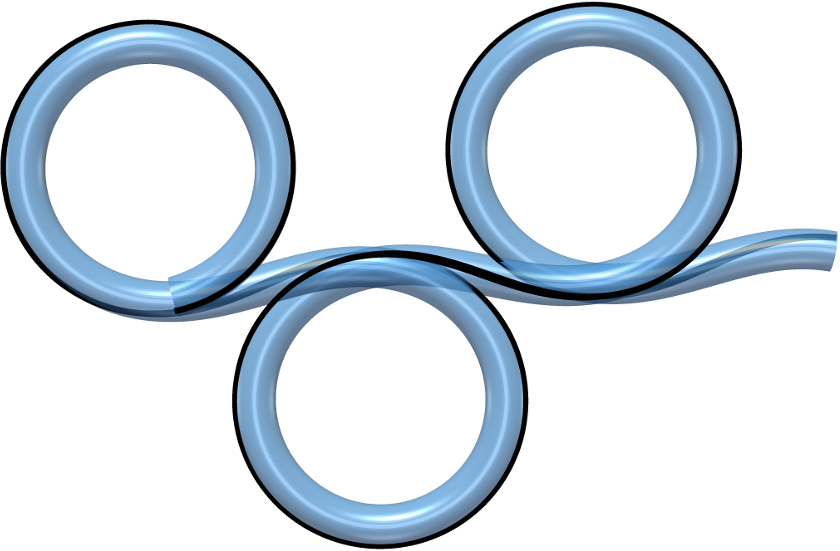} ($2^*_+$) \\
\\
\\
\includegraphics[scale=0.08]{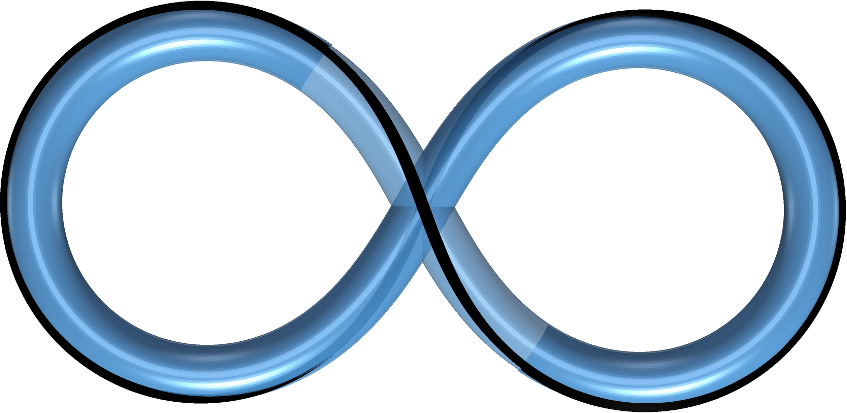} ($3$) &  
\includegraphics[scale=0.12]{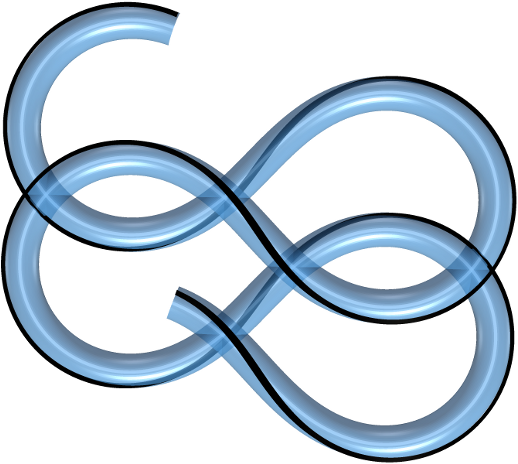} ($3^*_+$) &
\includegraphics[scale=0.11]{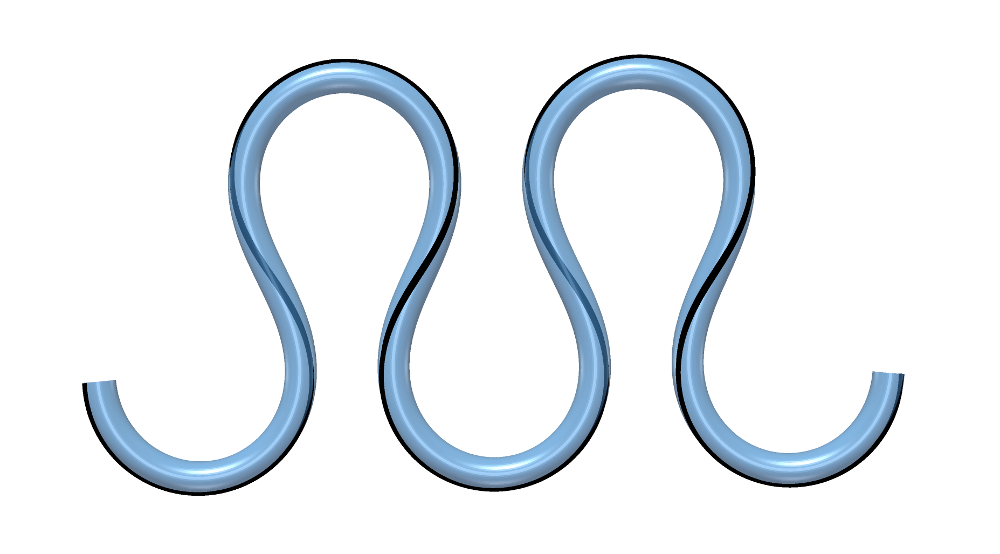} ($4$) \\
\\
\\
\includegraphics[scale=0.11]{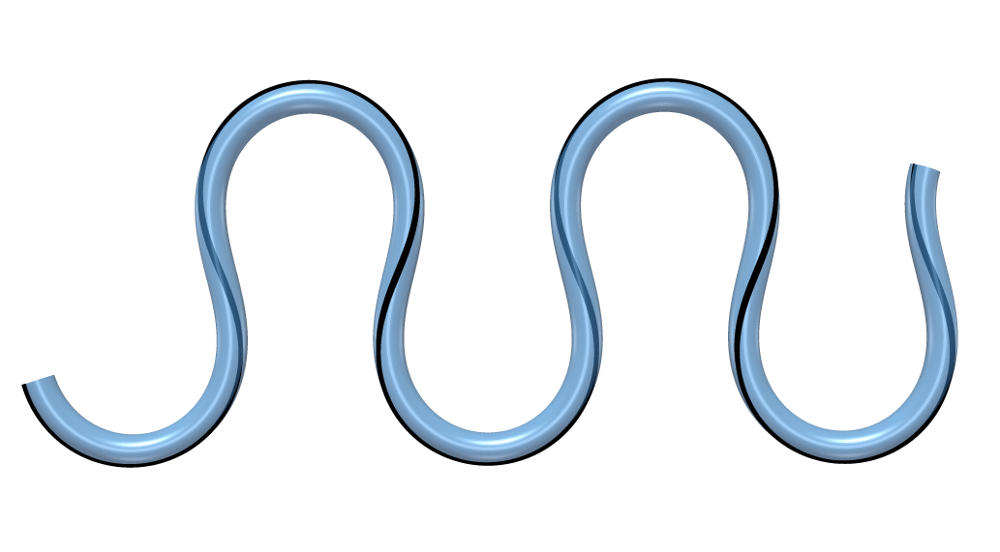} ($4^*_+$) &
\includegraphics[scale=0.14]{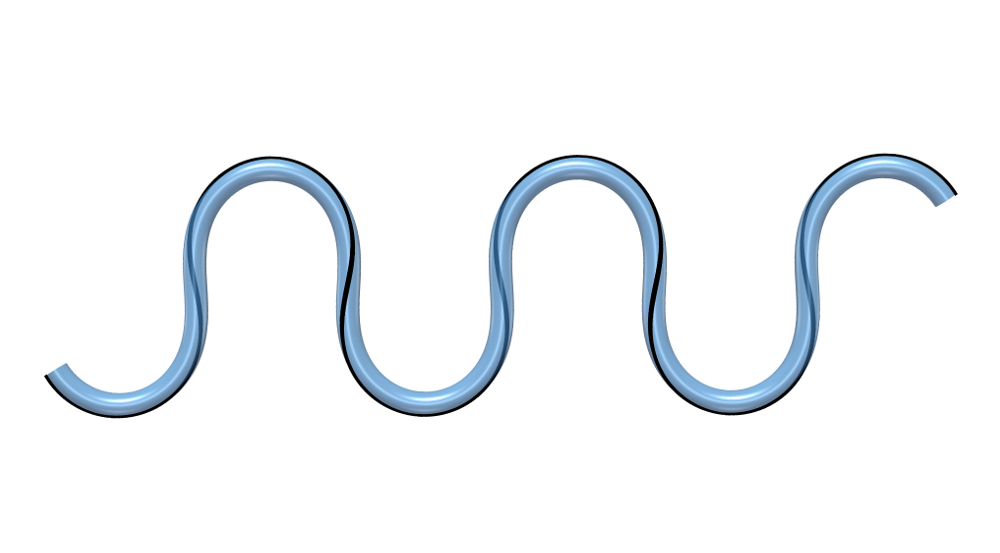} ($5$) &
\includegraphics[scale=0.16]{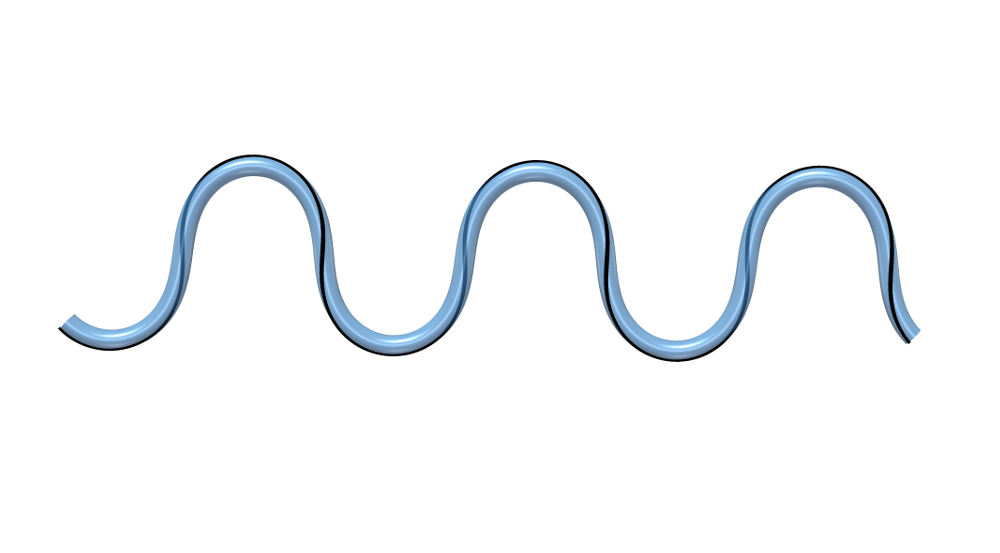} ($5^*_+$) 
\end{tabular}
\end{center}
\caption{Shapes of the squeelix in the regime $m<1$ for $\gamma=0.9915$.} 
\label{figminfeq1}
\end{figure}

\paragraph{Case $m<1$.} Table~\ref{table1} provides the values of $n_{a/b}$, $m$ and the energies of the shapes shown in Fig.~\ref{figminfeq1}. The shape $(0)$ consists of approximatively four circles on the top of each other (with deformed ends) as the perimeter of a single circle is $l_{loop}=2\pi/\omega_1\approx 0.26 L$. The ground state is given by the shape $(2)$ which has two twist-kinks, \textit{i.e.}, $n_b=3$. Squeelices of local energy minima $(n)$ have symmetric shapes. Critical configurations ($n^*$) on the top of energy barriers are asymmetric with a critical twist-kink near the end $s=L/2$ of the filament. 

\begin{table}
\begin{center}
\begin{tabular}
[c]{|c|c|c|c|}\hline
Shape & $n_{a/b}$ & $m<1$ & Energy ($BL^{-1}$)\\\hline
$(0)$ & $n_{b}=1$ & $0.9999999999471$ & $111.31$\\\hline
$(0_{+}^{\ast})$ & $n_{a}=1$ & $0.999999999595$ & $121.47$\\\hline
$(1)$ & $n_{b}=2$ & $0.99997091$ & $111.11$\\\hline
$(1_{+}^{\ast})$ & $n_{a}=2$ & $0.9999195$ & $121.26$\\\hline
$\textbf{(2)}$ & $\textbf{n}_\textbf{\textit{b}}\textbf{=3}$ & $\textbf{0.997635}$ & $\textbf{110.98}$ \\\hline
$(2_{+}^{\ast})$ & $n_{a}=3$ & $0.995359$ & $121.23$\\\hline
$(3)$ & $n_{b}=4$ & $0.97955$ & $111.72$\\\hline
$(3_{+}^{\ast})$ & $n_{a}=4$ & $0.96651$ & $122.56$\\\hline
$(4)$ & $n_{b}=5$ & $0.93093$ & $115.18$\\\hline
$(4_{+}^{\ast})$ & $n_{a}=5$ & $0.8990$ & $127.71$\\\hline
$(5)$ & $n_{b}=6$ & $0.8563$ & $123.36$\\\hline
$(5_{+}^{\ast})$ & $n_{a}=6$ & $0.8043$ & $139.00$\\\hline
\end{tabular}
\end{center}
\caption{Numerical values of the parameters of the shapes of Fig. \ref{figminfeq1}. The ground state is written in bold face.}
\label{table1}
\end{table}

\begin{figure}
\begin{center}
\begin{tabular}{ccc}
\includegraphics[scale=0.065]{fig7fig8fig15_ka1+.png} ($0^*_+$) & 
\qquad
\qquad
\includegraphics[scale=0.08]{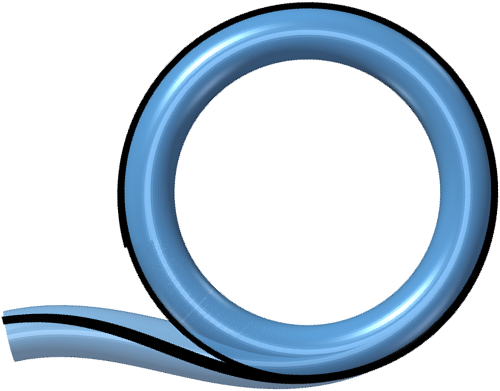} ($0^*_-$) & 
\qquad
\qquad 
\includegraphics[scale=0.066]{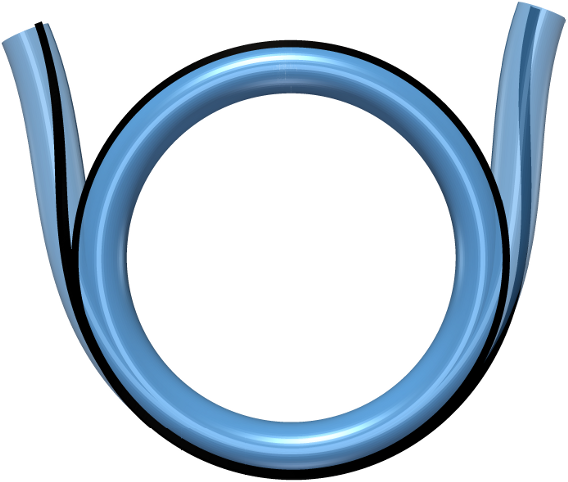} ($0^{**}_-$)
\end{tabular}
\end{center}
\caption{Examples of unstable shapes. The two shapes with a single critical twist-kink are labelled $(0^*_+)$ and $(0^*_-)$, and the shape with two critical twist-kinks $(0^{**}_-)$.}
\label{unstable shape}
\end{figure}

For completeness Fig. \ref{unstable shape} shows three unstable shapes, in particular two critical shapes on the line $s_{0,-}(m)$. The shapes $(0_-^{*})$ and $(0_+^{*})$ have a critical twist-kink symmetrically localized at the opposite ends of their chain. They have the same energy $E_{0^*_+}=E_{0^*_-}$. The shape  $(0_-^{**})$ has two critical twist-kinks localized at the two ends of the chain. Its energy is thus higher. Table~\ref{table2} provides the numerical values of the parameters characterizing these shapes. 

\begin{table}[!]
\begin{center}
\begin{tabular}
[c]{|c|c|c|c|}\hline
Shape & $n_{a/b}$ & $m$ & Energy ($BL^{-1}$)\\\hline
$(0_{+}^{\ast})$ & $n_{a}=1$ & $0.999999999595$ & $121.47$\\\hline
$(0_{-}^{\ast})$ & $n_{a}=1$ & $0.999999999595$ & $121.47$\\\hline
$(0_{-}^{\ast\ast})$ & $n_{b}=1$ & $0.9999999969$ & $131.63$\\\hline
\end{tabular}
\end{center}
\caption{Numerical values of the parameters of the shapes of Fig. \ref{unstable shape}.}
\label{table2}
\end{table}

\begin{figure}
\begin{center}
\centering
\begin{tabular}{cccc}
\includegraphics[scale=0.08]{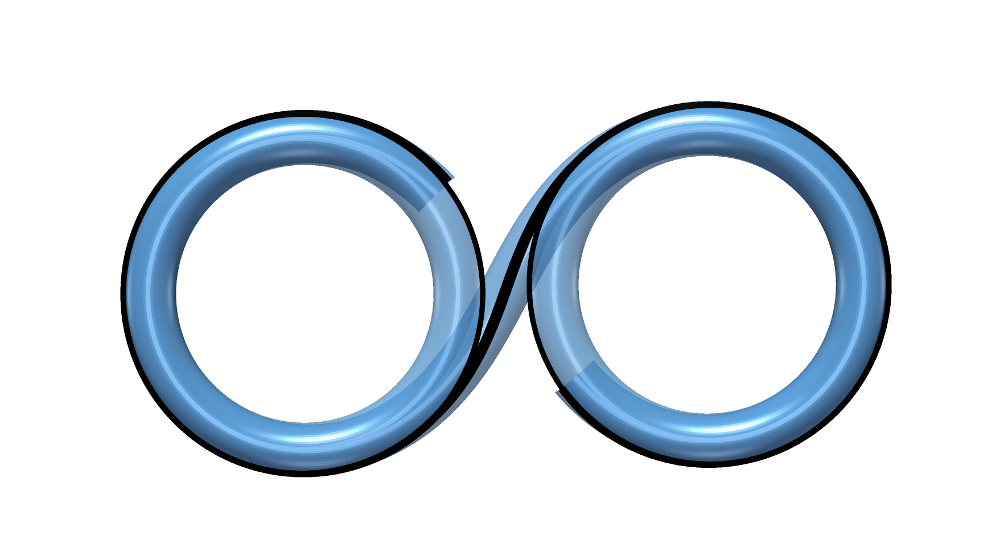} ($\bar{1}$) &
\qquad
\includegraphics[scale=0.08]{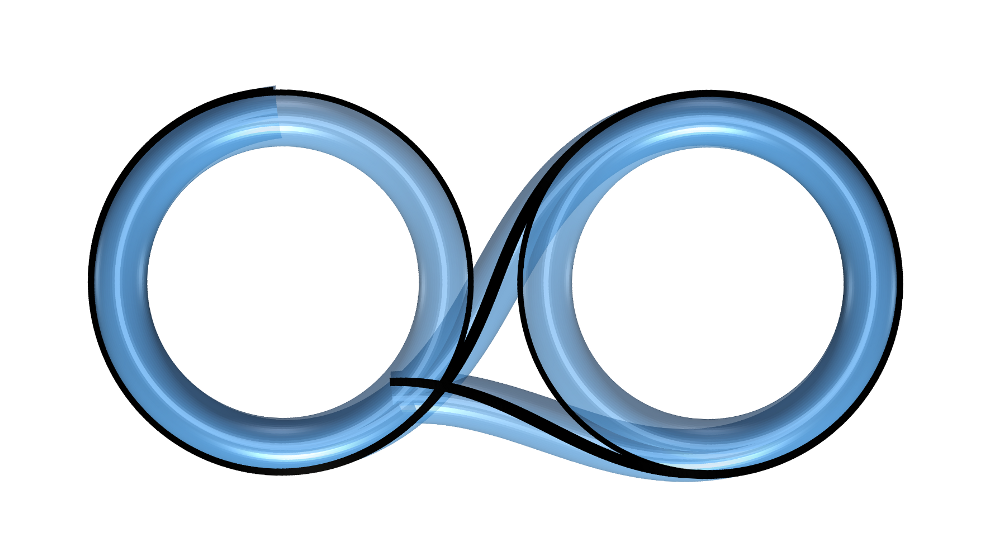} ($\bar{1}^*$) & 
\qquad 
\includegraphics[scale=0.08]{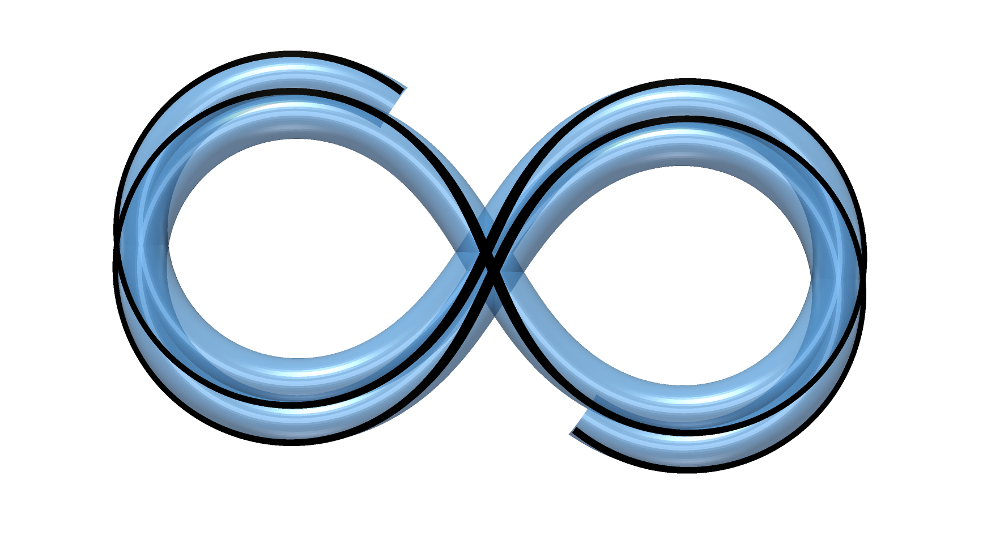} ($\bar{2}$) &
\qquad
\includegraphics[scale=0.12]{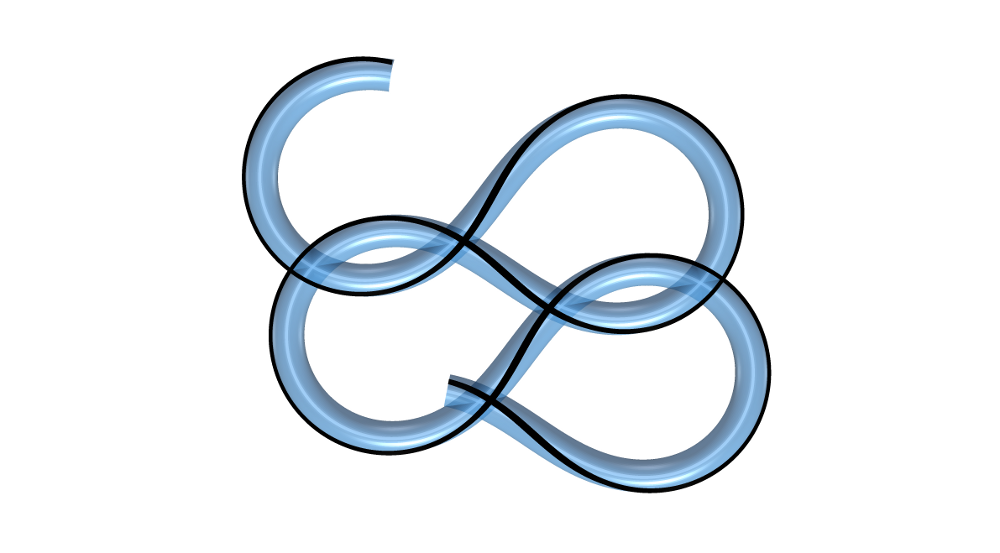} ($\bar{2}^*$)  
\end{tabular}
\end{center}
\caption{Several shapes of the squeelix in the regime $m>1$ for $\gamma=0.9915$. The two shapes  ($\bar{1}$) and ($\bar{2}$) minimize the energy of the squeelix. The shapes ($\bar{1}^*$) and ($\bar{2}^*$) have a single critical anti-twist-kink at one end of the chain and are unstable. } 
\label{figmsupeq1}
\end{figure}

\paragraph{Case $m>1$. } We consider the shapes of the squeelices associated to the four extrema in the region $m>1$ of the energy in Fig. \ref{Ema}. 
Table~\ref{table3} provides the numerical values of the parameters characterizing these shapes which are shown in Fig. \ref{figmsupeq1}. 
We observe that the shapes are similar to some of the shapes with $m<1$ but have a higher energy. This is due to the presence of 
anti-twist-kinks which maximise the pure twist energy. 

\begin{table}[!]
\begin{center}
\begin{tabular}
[c]{|c|c|c|c|}\hline
Shape & $n_{a/b}$ & $m$ & Energy ($BL^{-1}$)\\\hline
$(\overline{1})$ & $n_{b}=1$ & $1.000029$ & $209.12$\\\hline
$(\overline{1}_{+}^{\ast})$ & $n_{a}=1$ & $1.00008$ & $219.28$\\\hline
$(\overline{2})$ & $n_{b}=2$ & $1.0228$ & $305.65$\\\hline
$(\overline{2}_{+}^{\ast})$ & $n_{a}=2$ & $1.0387$ & $315.09$\\\hline
\end{tabular}
\end{center}
\caption{Numerical values of the parameters of the various shapes in Fig. \ref{figmsupeq1}. }
\label{table3}
\end{table}


\section{How to measure the material parameters\label{sec:exp}}

When confined biofilaments exhibit abnormal, wavy, spiral or circular shapes
that cannot be explained by the semi-flexible chain model, the theory of
squeelices developed here should be of some help. 
Assuming that the filament is in thermal equilibrium and does not display large thermal 
fluctuations, it is possible to have a quantitative
understanding of the experimental biofilament under study, \textit{i.e.}, a solid estimate of its material
parameters.

For a filament whose shape is wavy we know that $\gamma<1$. The general procedure to 
obtain the material parameters in this case consists of the following steps: ($i$) extract the tangent angle  
$\phi(s)$ from the experimental data and compute the curvature $\kappa(s)$. ($ii$) The maximum of $\kappa(s)$ 
gives $\omega_1$. ($iii$) The length $l_{cycle}$ corresponds to the distance between two adjacent roots of $\kappa(s)$.
($iv$) \textit{Via} Eq.~(\ref{eq:lcycle}) one obtains $\lambda$ in terms of $m$. ($v$) Plugging this result into the 
expression~(\ref{Curv}) of the curvature allows to fit the experimentally obtained curvature with a single parameter $m$.
($vi$) From $m$ one obtains $\lambda$ and thus the ratio $C/B$ using Eq.~(\ref{Lambda}). ($vii$) From Eq.~(\ref{msol}) 
one finally gets $\gamma$ from which $\omega_3$ can be deduced.

A word of caution is due here. The suggested procedure neither takes into account excluded-volume interactions 
nor the effect of a finite temperature. Both can potentially modify the resulting shapes. Self-interactions are more important in the regime 
where the theoretical squeelix forms loops that lie on top of each other. In this case the repulsion will induce a separation of the circles. A strong self-interaction might even induce the formation of additional twist-kinks extending the filament and thus changing the ground state. In the regime $\gamma\ll 1$, where the shapes are wavy, self-interactions can be safely discarded.

In this article we have scaled all lengths with the length of the filament $L$. 
In these units the shape of a squeelix is scale invariant (the maximum number of twist-kinks in the chain, $n_{max}$, is independent of $L$) but its energy decreases with $L$ (see, for instance, Tab.~\ref{table3}). At zero temperature the theory can directly be applied to any microscopic or macroscopic system. For biofilaments at finite temperature, however, scale matters, and the natural energy scale is $k_B T$. The characteristic lengths of a biofilament (like $\lambda$) are fixed and independent of $L$ and measured, for instance, in $\mu$m. The shape is not scale invariant any more. In these units the energy and $n_{max}$ grow linearly with $L$ (as one can see from Eq.~(\ref{nmax}) for $n_{max}$). 
At finite temperature the number of twist-kinks can fluctuate within the chain if the energy barrier $\Delta E$ between two adjacent minima with $n$ and $n+1$ twist-kinks, respectively, of the energy $E(m,s_{0,+}(m))$ is of order $k_B T$. 
We can estimate this barrier from the energy contribution of the deformed ends (see text below Eq.~(\ref{zeroTWA2}) in Appendix C): $\Delta E  / k_B T \approx \frac{\omega_3^2}{\omega_1} \left(\frac{C}{B}\right)^{\frac{3}{2}} l_B $, where $l_B = B / k_B T$ is the persistence length associated to the bending of the filament. We expect strong shape fluctuations when $\Delta E  / k_B T$ is of the order of one or smaller. Such strong fluctuations were already observed for a squeelix with circular ground state \cite{Nam2012}.


\section{Conclusion}
A filament confined on a flat surface is frequently encountered in experiments to 
permit its observation in the focal plane. But generally, confinement modifies the 
elastic properties. This is particular blatant for filaments that adopt a helical shape 
in free space.  The theory based on the linear elasticity of squeezed helical 
filaments is analogous to that of the two-dimensional Euler elastica. A lot of different 
shapes are found resembling circles, waves or spirals. Remarkably, a conformational quasi-particle 
called twist-kink emerges naturally from the model. In this picture the shapes of the squeelix result from the 
repulsive interaction of these quasi-particles. The extreme case of complete twist-kink
expulsion from the chain could be the explanation for the formation
of tiny actin rings confined to a flat surface \cite{Sanchez2010}. In the same manner 
wavy and circular movements of microtubules in gliding assay experiments have been explained by the active movement 
of squeelices \cite{Gosselin2016}. 

In this article we have elucidated the rich variety of shapes that can be 
found for these systems. This provides a nomenclature of squeezed helices that can potentially be 
usefull for the interpretation of experimental observations.

Confined elastic rods on a plane submitted to an additional lateral confinement were studied previously \cite{Domokos1997,Manning2005}. 
An interesting extension of the present study would be to consider the case of squeelices under such a confinement. 
This seems particularly relevant in view of experiments with double confinement of biofilaments performed by K\"oster et al \cite{Koester2012}.

\begin{acknowledgments}
The authors thank Albert Johner and Falko Ziebert for helpful discussions. They would also like to thank the referees for their 
helpful reports which helped to improve the manuscript. 
\end{acknowledgments}

\bigskip


\appendix

\section{Euler-Lagrange equations of the squeelix}

The elastic energy of the squeelix has been derived in the main text (see Eq. (\ref{sqeezedenergy})):
\begin{equation}
E=\frac{1}{2}\int_{-L/2}^{L/2} \left[B\left(  \phi^{\prime}-\omega_{1}\sin\psi\right)^{2}  +  C \left(  \psi^{\prime}-\omega_{3}\right)
^{2}  + B\omega_1^2 \cos^2{\psi} \right] ds\label{EA2}%
\; .
\end{equation}
Minimizing with respect to $\phi^{\prime}$ gives
\begin{equation}
\phi^{\prime}=\omega_{1}\sin\psi\label{curvatureA1}%
\end{equation}
and the energy becomes a function of $\psi(s)$:
\begin{equation}
E[\psi] = \frac{1}{2}\int_{-L/2}^{L/2} \left[C \left(  \psi^{\prime}-\omega_{3}\right)
^{2}  + B\omega_1^2 \cos^2{\psi} \right] ds\label{EA3}%
\; .
\end{equation}
The first variation of $E$ with respect to $\psi$ leads to
\begin{equation}
\delta E =- \int_{-L/2}^{L/2}ds\left( \left(  C\frac{d}{ds}\left(
\psi^{\prime}-\omega_{3}\right)  +\frac{1}{2} B\omega_{1}^2 \sin (2\psi) \right)
\delta\psi\right)   +\left[  C\left(  \psi^{\prime}-\omega_{3}\right)  \delta\psi\right]
_{-L/2}^{L/2}%
\label{eq:firstvariation}
\end{equation}
with free boundary conditions, \textit{i.e.}, no torque at both ends of the filament. The
condition $\delta E=0$ implies the Neumann boundary conditions:
\begin{equation}
\psi^{\prime}(-L/2)   =\psi^{\prime}(L/2)=\omega_{3}\label{BCA2}
\end{equation}
and the \textit{pendulum} equation
\begin{equation}
\psi^{\prime\prime}+\frac{1}{2\lambda^{2}}\sin(2\psi)=0\label{twistA1}
\end{equation}
with the length $\lambda=\frac{1}{\omega_{1}}\sqrt{\frac{C}{B}}$.\ Integrating
Eq. $\left(  \ref{twistA1}\right)  $\ we obtain $(\psi^{\prime})^{2}
=a_{1}-\frac{1}{2\lambda^{2}}\sin^{2}\psi$ with $a_{1}$ a positive constant of
integration. This equation can be written conveniently in the following form
\begin{equation}
\psi^{\prime}(s)=\pm\frac{1}{\lambda\sqrt{m}}\sqrt{1-m\sin^{2}\psi
}\label{psiprimeA1}%
\end{equation}
with $a_{1}=1/(\lambda^{2}m)$ and $m$ a positive real parameter.

\begin{figure}
\begin{center}
 \includegraphics[width=0.5\textwidth]{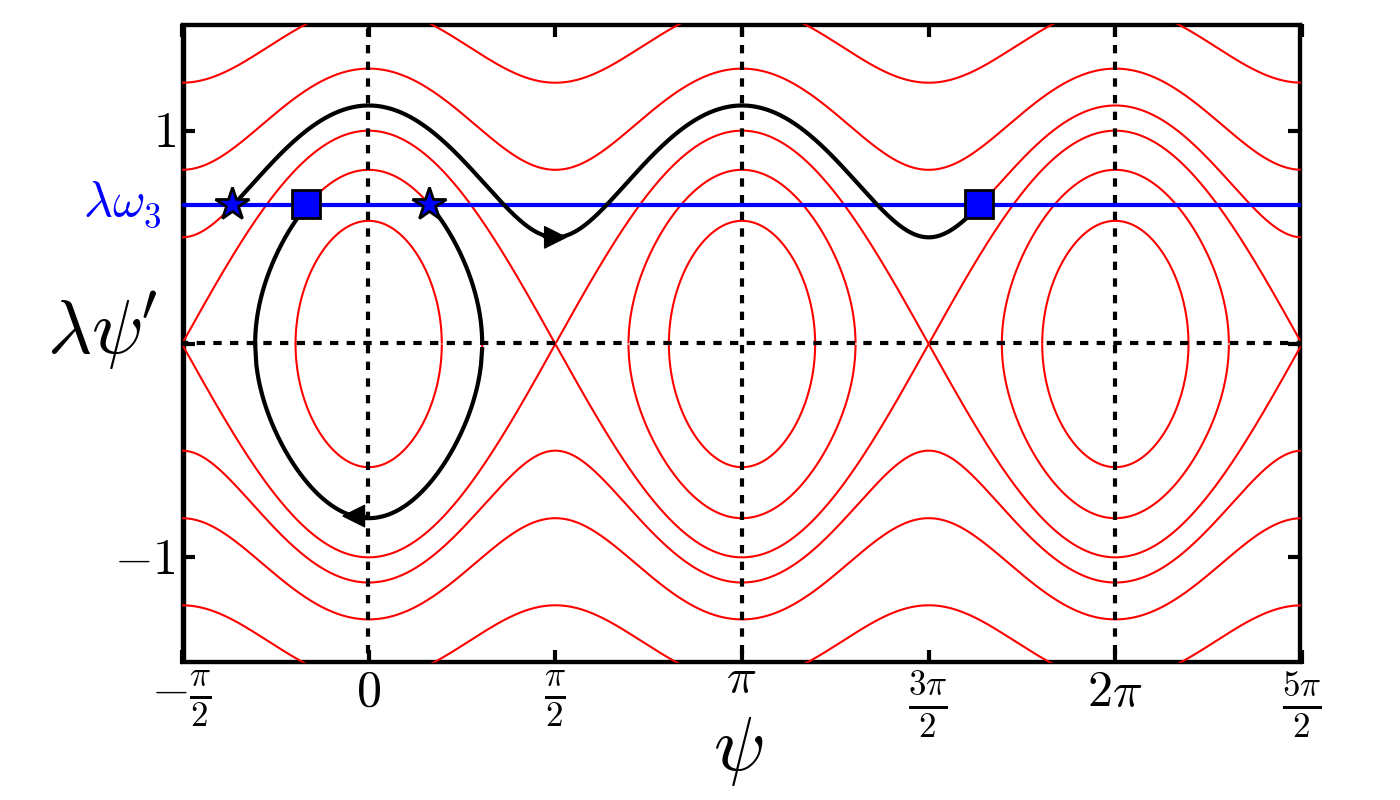}
\caption{
Phase plane of Eq.~(\ref{psiprimeA1}). Two typical trajectories are shown in black corresponding to the revolving ($m<1$) and oscillating ($m<1$)  pendulum, which are solutions of Eq.~(\ref{twistA1}). The stars represent the position $s=-L/2$, the squares $s=L/2$. These points have to lie on the line $\lambda\omega_3$ due to the boundary conditions Eq.~(\ref{BCA2}).
\label{fig:phaseplane1}}
\end{center}
\end{figure}

The phase plane of Eq.~(\ref{psiprimeA1}) is well-known and the solutions 
of Eq.~(\ref{twistA1}) are trajectories in this plane which begin and end at $\psi^{\prime}(s)=\omega_3$ as
shown in Fig.~\ref{fig:phaseplane1}. Among all these trajectories the solutions we look for are those of length $L$. 
A similar approach with Dirichlet boundary conditions is considered in Ref.~\cite{Maddocks1987} for the determination 
of the equilibrium configurations of a uniform elastic rod subject to cantilever loading. 
The stability analysis of these solutions is discussed in Appendix~E and their physical interpretation can be found in Appendix~D. 
From Fig.~\ref{fig:phaseplane1} we see that when $\lambda\omega_3>1$ the solutions correspond to a revolving pendulum ($m<1$) 
only. The oscillating pendulum ($m>1$) is a solution only in the regime $\lambda\omega_3<1$.

Eq. $\left(  \ref{psiprimeA1}\right)  $ implies $\psi\left(  -L/2\right)
=\pm\arcsin\sqrt{\frac{1}{m}-\frac{1}{\mu}}+n\pi$, with $n\in\mathbb{Z}%
$. Since $\phi^{\prime}=\omega_{1}\sin\psi$ we can limit ourself to $n=0$ and
$n=1$ without loss of generality. Since these two cases lead to two shapes related by the transformation $\kappa(s)\rightarrow-\kappa(s)$ and thus have the same energy we treat the case $n=0$ only. Consequently the twist angle at the first end can have one of the two values
\begin{equation}
\psi\left(  -L/2\right)  =\pm\arcsin\left(  \sqrt{\frac{1}{m}-\frac{1}{\mu}%
}\right)  \label{psyLA1}%
\; .
\end{equation}
This implies that $m$ lies in the interval
\begin{equation}
\frac{\mu}{1+\mu}\le m \le \mu\label{mmuA1}%
\end{equation}
with
\begin{equation}
\mu=\frac{B\omega_{1}^{2}}{C\omega_{3}^{2}} \; .
\end{equation}
As a consequence of Eq.\ $\left(  \ref{mmuA1}\right)  $ the twist is limited
to the interval $-\pi/2<\psi\left(  -L/2\right)  <\pi/2$.\ 
Integrating Eq. $\left(  \ref{psiprimeA1}\right)  $ with the positive sign
\begin{equation}
\int_{s_{0}}^{-\frac{L}{2}}ds=\lambda\sqrt{m}\int_{0}^{\psi(-\frac{L}{2}%
)}\frac{d\psi}{\sqrt{1-m\sin^{2}\psi}}
\end{equation}
yields $s_{0}$, given by the condition $\psi(s_0)=0$, in terms of $m$ for a given $L$:%
\begin{equation}
s_{0}(m)=-\frac{L}{2}-\lambda\sqrt{m}\mathcal{F}\left(  \psi(-L/2)|m\right)
\; ,\label{SOA1}%
\end{equation}
where $\mathcal{F}\left(  x|m\right)  $ is the elliptic integral of the first kind, a
growing function of $x$ \cite{Abramowitz}. Eq.\ $\left(  \ref{SOA1}\right)  $ defines 
two functions $s_{0,+}(m)$ and $s_{0,-}(m)$ depending on the sign of
$\psi(-L/2)$. These two functions $s_{0,\pm}(m)$ are symmetric with respect to the line
$s_{0}(m)=-L/2$, \textit{i.e.}, $s_{0,+}(m)=2s_{0}(\mu)-s_{0,-}(m)$ and thus meet at
the boundary $m=\mu$. Therefore $s_{0}(m)$ is defined in the interval
\begin{equation}
s_{0,-}(\frac{\mu}{1+\mu})\le s_{0}(m) \le s_{0,+}(\frac{\mu}{1+\mu}%
)\label{SOInterval}%
\end{equation}
with $s_{0,\pm}(\frac{\mu}{1+\mu})=-\frac{L}{2}\mp\lambda\frac{\pi}{2}%
\sqrt{\frac{\mu}{1+\mu}}$. 

The explicit solution of Eq. (\ref{twistA1}) is well-known:
\begin{equation}
\psi(s)=\pm am\left(  \frac{s-s_{0}}{\lambda\sqrt{m}}|m\right)  \; ,\label{psyA1}%
\end{equation}
where $am\left(  x|m\right)  $ is the elliptic Jacobian amplitude
function \cite{Abramowitz}. 
To obtain Eq.~(\ref{psyA1}) we have used the definition $\varphi=am(x|m)$ with $x=%
{\textstyle\int\nolimits_{0}^{\varphi}}
\frac{d\theta}{\sqrt{1-m\sin^{2}\theta}}$ and the relation $am\left(
x|m\right)  =-am\left(  -x|m\right)  $.\ As explained in the main text, the
boundary condition $\psi^{\prime}(-L/2)=\omega_{3}$ with $\omega_{3}>0$
imposes the positive sign in Eq.\ $\left(  \ref{psiprimeA1}\right)  $ in the
vicinity of $-L/2$, so that we must choose
\begin{equation}
\psi(s)=am\left(  \frac{s-s_{0}}{\lambda\sqrt{m}}|m\right)  \; .\label{psyplusA1}%
\end{equation}
The twist $\psi(s)$ is a growing function of $s$ for $m<1$ and periodic for $m>1$. 
The case $m=1$ is the homoclinic pendulum with a single one-half
turn, \textit{i.e.}, $\psi(s)$ changing by $\pi$ on the length $L$.

The variation of the twist is given by%
\begin{equation}
\psi^{\prime}(s)=\frac{1}{\lambda\sqrt{m}}dn\left(  \frac{s-s_{0}}%
{\lambda\sqrt{m}}|m\right) \; , \label{psyprimeA1}%
\end{equation}
where $dn\left(  x|m\right)  $ is a periodic odd elliptic Jacobian function of
period $l_{p}=2\lambda\sqrt{m}\mathcal{K}(m)$ with $\mathcal{K}(m)$ the complete elliptic integral
of the first kind \cite{Abramowitz}. The curvature $\kappa(s)=\phi^{\prime}(s)$ given by
Eq.\ $\left(  \ref{curvatureA1}\right)  $ is then%
\begin{equation}
\kappa(s)=\omega_{1}sn\left(  \frac{s-s_{0}}{\lambda\sqrt{m}}|m\right)
\label{CurvA1}%
\end{equation}
with $sn\left(  x|m\right)  $ a periodic even elliptic Jacobian function 
of period $l_{p}=4\lambda\sqrt{m}\mathcal{K}(m)$ \cite{Abramowitz}.


\section{Energy of the squeelix}
In this section we compute the elastic energy of a configuration given by Eq.~(\ref{psyplusA1}). 
From Eq.~(\ref{psyplusA1}) we have $\cos{\psi} = \mbox{cn} \left(\frac{s-s_{0}}{\lambda\sqrt{m}}|m\right)$.
Plugging this expression together with $\psi^{\prime}$ from Eq.~(\ref{psyprimeA1}) into the energy, Eq.~(\ref{EA3}),
\begin{equation}
E  =  \frac{1}{2}\int_{-L/2}^{L/2}\left(B \omega_{1}^{2}\cos^{2}\psi +C\left(\psi^{\prime2}
-2\omega_{3}\psi^{\prime}+\omega_{3}^{2}\right)\right)ds
\; ,
\label{eq:energy_integrals}
\end{equation}
we see that we have to compute three integrals:
\begin{subequations}
\begin{eqnarray}
I_1 & = & \int_{-L/2}^{L/2}\mbox{cn}^{2}\left(\frac{s-s_{0}}{\lambda\sqrt{m}}|m\right)ds = \frac{\lambda}{\sqrt{m}}\left[\mathcal{E}\left(\frac{s-s_{0}}{\lambda\sqrt{m}}|m\right)\right]_{-L/2}^{L/2}+\left(1-\frac{1}{m}\right)L
\; , \\
I_2 & = & \int_{-L/2}^{L/2}\mbox{dn}^{2}\left(\frac{s-s_{0}}{\lambda\sqrt{m}}|m\right)ds=\lambda\sqrt{m}\left[\mathcal{E}\left(\frac{s-s_{0}}{\lambda\sqrt{m}}|m\right)\right]_{-L/2}^{L/2}
\; , \\
I_3 & = & 
\int_{-L/2}^{L/2}\mbox{dn}\left(\frac{s-s_{0}}{\lambda\sqrt{m}}|m\right)ds=\lambda\sqrt{m}\left[\psi\left(s\right)\right]_{-L/2}^{L/2}
\; ,
\end{eqnarray}
\end{subequations}
where $\mathcal{E}\left(  x|m\right)  $ is the elliptic integral of the second kind \cite{Abramowitz}. To compute the integrals we have used $\int\mbox{cn}^{2}\left(s|m\right)ds=\frac{1}{m}\mathcal{E}\left(s|m\right)-\frac{\left(1-m\right)}{m}s$, $\int\mbox{dn}^{2}\left(s|m\right)ds=\mathcal{E}\left(s|m\right)$ and $\int\mbox{dn}\left(s|m\right)ds=\arcsin\left(\mbox{sn}\left(s|m\right)\right)$. 

Consequently, we obtain the elastic energy
\begin{equation}
E   =\frac{\omega_{1}\sqrt{BC}}{\sqrt{m}}\left(  \mathcal{E} \left(  \psi\left(
L/2\right)  |m\right)  - \mathcal{E}\left(  \psi\left(  -L/2\right)  |m\right)  \right)
+\frac{B\omega_{1}^{2}}{2}\left(  1+\frac{1}{\mu}-\frac{1}{m}\right)
L  \left.  -C\omega_{3}\left(  \psi\left(  L/2\right)  -\psi\left(
-L/2\right)  \right)  \right] 
 \label{EnergyA1}%
\; .
\end{equation}
This expression is correct for all $m>0$. However, when $m>1$ it is advisable for numerical reasons to transform the Jacobi elliptic functions to their analogs with a parameter $1/m$ lower than unity.
Using the relations $\mbox{cn}\left(s|m\right)=\mbox{dn}\left(\sqrt{m}s|\frac{1}{m}\right)$ and 
$\mbox{dn}\left(s|m\right)=\mbox{cn}\left(\sqrt{m}s|\frac{1}{m}\right)$ we obtain for the integrals
\begin{subequations}
\begin{eqnarray}
I_1&=&\lambda\left[\mathcal{E}\left(\frac{s-s_{0}}{\lambda}|\frac{1}{m}\right)\right]_{-L/2}^{L/2}
\; , \\
I_2&=&\lambda m\left[\mathcal{E}\left(\frac{s-s_{0}}{\lambda}|\frac{1}{m}\right)\right]_{-L/2}^{L/2}+\left(1-m\right)L
\; , \\
I_3&=&\lambda\left[\frac{\arccos\left(\mbox{dn}\left(\frac{s-s_{0}}{\lambda}|\frac{1}{m}\right)\right) \mbox{sn}\left(\frac{s-s_{0}}{\lambda}|\frac{1}{m}\right)}{\sqrt{1 - \mbox{dn}^2\left(\frac{s-s_{0}}{\lambda}|\frac{1}{m}\right)}}\right]_{-L/2}^{L/2}
\; .
\end{eqnarray}
\end{subequations}
Therefore, the energy formula~(\ref{EnergyA1}) can also be conveniently written
\begin{equation}
E_{m>1}= \omega_{1} \sqrt{BC} \left[\mathcal{E}\left(\frac{s-s_{0}}{\lambda}|\frac{1}{m}\right)\right]_{-L/2}^{L/2}+\frac{B\omega_{1}^{2}}{2}\left(-1+\frac{1}{\mu}+\frac{1}{m}\right)L-C\omega_{3}\left[ \frac{\arccos\left(\mbox{dn}\left(\frac{s-s_{0}}{\lambda}|\frac{1}{m}\right)\right) \mbox{sn}\left(\frac{s-s_{0}}{\lambda}|\frac{1}{m}\right)}{\sqrt{1 - \mbox{dn}^2\left(\frac{s-s_{0}}{\lambda}|\frac{1}{m}\right)}} \right]_{-L/2}^{L/2}\label{eq:energy_msup1}
\end{equation}
for $m>1$.

\begin{figure}
\centering
\subfigure[]{
\includegraphics[width=0.48\textwidth]{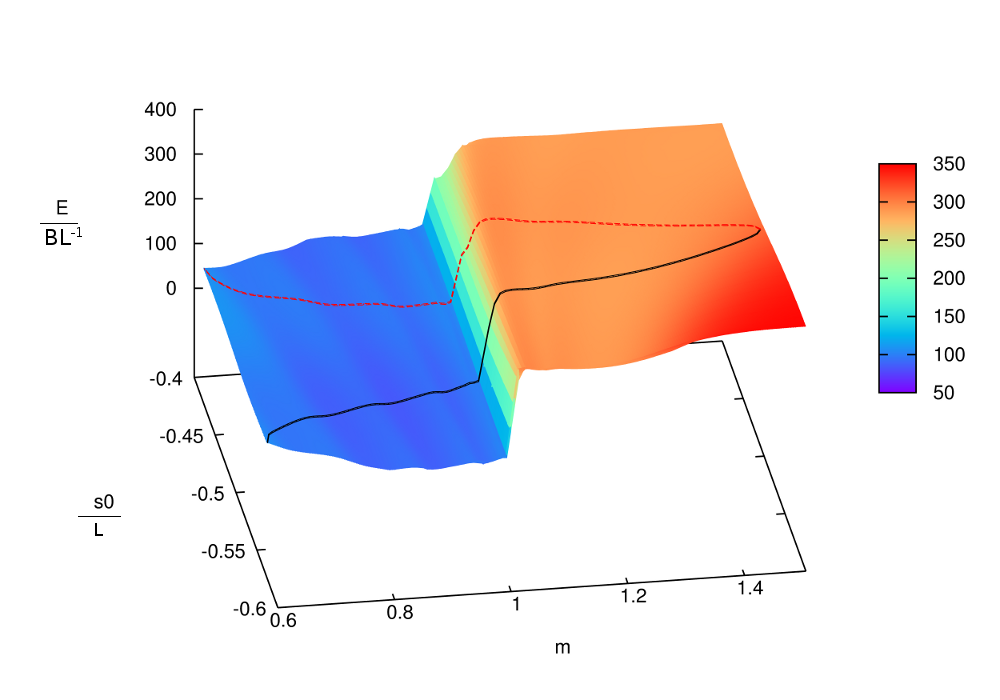}
}
\hfill
\subfigure[]{
\includegraphics[width=0.48\textwidth]{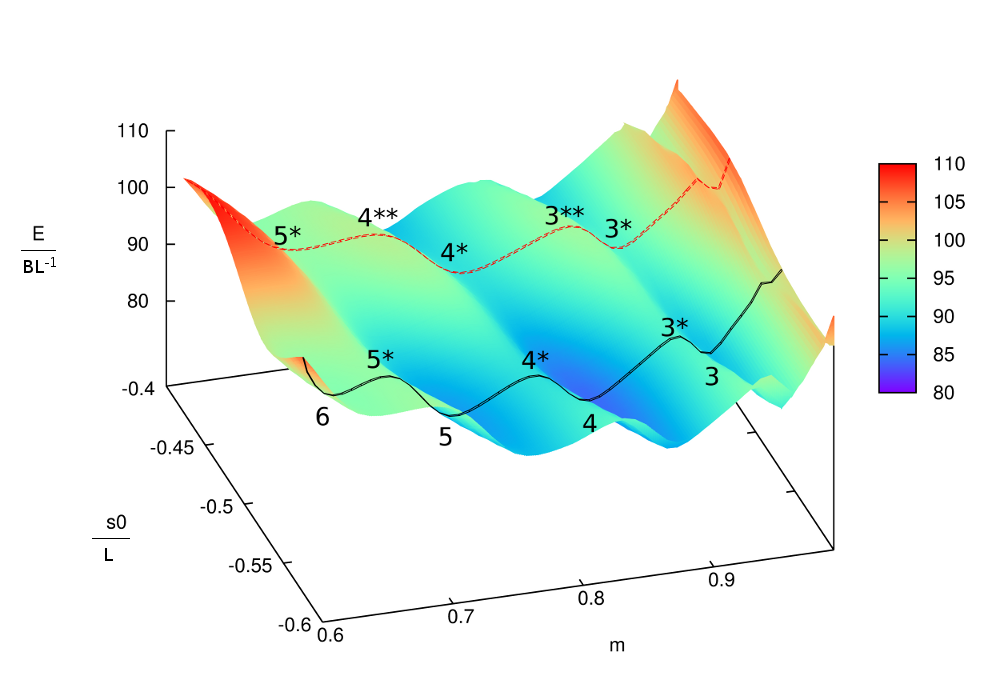}
}
\caption{Energy landscape $E(m,s_0)$ for $\gamma=0.6079$, $\omega_1=20L^{-1}$, $\omega_3=16.33L^{-1}$ and $\sqrt{C/B}=1$. Energies and lengths are measured in units of $BL^{-1}$ and $L$,  respectively. (a) The curve $s_{0,+}(m)$ in black passes by all minima and the curve $s_{0,-}(m)$ in red passes by all maxima. Both curves also pass by saddle points of same energy. (b) Zoom of the energy in the regime $m<1$. The integers $n$, $n^*$ and $n^{**}$ denote the minima, the saddle points and the maxima, respectively.}
\label{Ems0}
\end{figure}

Consider the elastic energy, Eq.~(\ref{EnergyA1}), as a function of $m$ and $s_{0}$. 
The domain of admissible values for the variables is defined by Eqs.~(\ref{mmuA1}) 
and (\ref{SOInterval}). Fig.~\ref{Ems0} shows an example: One observes a complex energy landscape with a lot of
extrema, that are minima, maxima and saddle points. These extrema correspond
to the stable or unstable shapes of the squeelices verifying the
boundary conditions Eqs.~(\ref{BCA2}). The function $E(m,s_{0}(m))$ where $s_{0}(m)$ is given by Eq.~(\ref{SOA1})  
represents a trajectory on the energy landscape $E(m,s_{0})$
that passes by all its extrema. We also observe a sharp distinction in the
behavior of the $E(m,s_{0})$ for the revolving ($m<1$) and oscillating pendulum ($m>1$), respectively (see Fig.~\ref{Ems0}). 
In the main text, we discuss the shapes of the squeelices that are associated to all these extrema.

\section{Life without Jacobi elliptic functions---the twist-kink picture}

The concept of a twist-kink was introduced in Ref.~\cite{Nam2012} by an
analogy between the energy of the squeelix and the energy of a semi-flexible chain
under tension which contains sliding loops \cite{Kulic2005,Kulic2007}. We
will not use this analogy here. Instead we will exploit the results of Appendix A to
determine the shape of a single twist-kink. In this maybe more physical approach 
we will not need Jacobi elliptic functions 
for the description of the shapes. The drawback is that we are limited to the regimes $m\approx1$ and
$m\approx0$ as we will see in the following.

Neglecting the boundary conditions Eq.\ $\left(  \ref{BCA2}\right)  $ one directly obtains 
the trivial solution $\psi(s)\approx\pm\pi/2.\ $The shape is circular
with constant radius of curvature $1/\omega_{1}$ and energy\ $E_{0}%
=C\omega_{3}^{2}L/2$.\ For a squeelix of finite length $L$, the boundary conditions
Eq.\ $\left(  \ref{BCA2}\right)  $ impose deformations at both ends of the
chain.\ For small deformations $\psi(s)\pm\pi/2\ll1$ at the chain's ends, and the
solution can be written approximately as \cite{Nam2012}
\begin{equation}
\psi(s)\approx\frac{\lambda\omega_{3}\sinh(s/\lambda)}{\cosh(L/(2\lambda))}%
\pm\pi/2  \; ,\label{zeroTWA2}%
\end{equation}
which is valid in the regime $\lambda\omega_{3}\ll1$. The energy of this
configuration is $\widetilde{E}_{0}\approx(2\lambda/L+1)E_{0}$. Thus, $2\lambda
E_{0}/L$ is the energy contribution of the deformed ends.\ Note that for large deformations at the boundary, the exact solution Eq.\ $\left(
\ref{psyplusA1}\right)  $ must be considered because deformations at the ends
are actually pieces of a twist-kink.\ This case is treated in Appendix D. For
now we assume that $\lambda\omega_{3}/L \ll1/L$.\ 

We now consider a twist angle $\psi(s)$ that increases by $\pi$ along the chain.\ The solution
is given by Eq.\ $\left(  \ref{psyplusA1}\right)  $ with $m=1$ and reads%
\begin{equation}
\psi(s)=2\arctan(e^{s/\lambda})-\pi/2\label{psyA2}%
\end{equation}
with $\psi(-\infty)=-\pi/2$ and $\psi(\infty)=\pi/2$.\ The curvature is thus
\begin{equation}
\phi^{\prime}=\omega_{1}\tan\left(  s/\lambda\right)
\; .
\end{equation}
For $\lambda/L\ll1$ the shape is made of two circular arcs of inverse
curvature separated by a region of curvature inversion of size $\lambda$ that
we have named a twist-kink \cite{Nam2012}. The energy of this configuration is
\begin{equation}
E_{1TK}=\pi C\omega_{3}\left(  \gamma-1\right)  +\frac{C\omega_{3}^{2}}{2}L
\end{equation}
with $\gamma=\frac{4\omega_{1}^{2}B}{\pi^{2}\omega_{3}^{2}C}$. We could take
into account the effect of the boundary conditions by simply adding their
elastic energy $\approx2\lambda E_{0}/L$ to $E_{1TK}$. Comparing with the zero
twist-kink case $E_{0}=C\omega_{3}^{2}L/2$ we find an expression for the self-energy of a
twist-kink:
\begin{equation}
\Delta E_{1}=E_{1TK}-E_{0}=\pi C\omega_{3}\left(  \sqrt{\gamma}-1\right)
\; .
\end{equation}
Therefore, $\gamma=1$ separates the regimes of positive and negative self-energy. 
For $\gamma>1$ the ground state is a circular arc.\ Decreasing 
$\gamma$, twist-kinks will pop-up within the chain with a density limited by
their mutual repulsion. This corresponds to the regime $m\lesssim 1$ (the revolving pendulum). To give a more quantitative foundation to this
argument let us compute the interaction energy of two twist-kinks separated by a
distance $d$ such that $d/\lambda\gg 1$. In this dilute regime $\gamma
\lesssim1$.\ 

With the following two twist-kinks ansatz%
\begin{equation}
\psi(s)=2\arctan\left(  e^{\left(  s+d/2\right)  /\lambda}\right)
+2\arctan\left(  e^{(s-d/2)/\lambda}\right)  -\frac{\pi}{2} \; ,
\end{equation}
such that $\psi(-\infty)=-\pi/2$ and $\psi(\infty)=3\pi/2$, the energy
reads%
\begin{equation}
E_{2TK}-E_{0}=2\Delta E_{1}+4\sqrt{CB}\omega_{1}\exp(-d/\lambda
))\label{repulsion}%
\; .
\end{equation}
The repulsive interaction between two twist-kinks scales as $\sim
\exp(-d/\lambda)$ in the dilute regime $\lambda\ll d$. Generalizing for $n$
twist-kinks, with a mutual separation $d=L/(n+1)\gg\lambda$ we obtain the relation%
\begin{equation}
E^{dilute}_{nTK}-E_{0}\approx n\Delta E_{1}+2n\sqrt{CB}\omega_{1}\exp(-\frac{L}%
{\lambda(n+1)}) \; .\label{Nrepulsion}%
\end{equation}
For $n\gg1$ the optimal value $n^{\ast}$ will correspond to the largest
integer such that $E^{dilute}_{nTK}-E^{dilute}_{(n-1)TK}<0$, which leads to%
\begin{equation}
n^{\ast}=\left\lfloor \frac{L}{\lambda\ln\left(  \sqrt{\gamma}/\left(
1-\sqrt{\gamma}\right)  \right)  }\right\rfloor
\; ,
\label{eq:nast}
\end{equation}
where $\left\lfloor x\right\rfloor $ is the notation for the largest integer
less than or equal to $x$. For $\gamma\lesssim1$, the density $\rho
=n^{\ast}/L$ is very small and the shapes consist of a succession of circular
arcs (or spirals) with opposite curvatures and separated by twist-kinks (see
figures in the main text). It is interesting to compare this density with the exact density $\rho = 1/l_{cycle} = (2\lambda \sqrt{m} K(m))^{-1}$ which is approximately $(\lambda \ln(\frac{16}{1-m}))^{-1}$ for $m\lesssim 1$. Comparing the two expressions of the density we find the relation $m\approx17-16/\sqrt{\gamma}$ between $m$ and $\gamma$.

Note that the oscillating pendulum ($m>1$) can be treated similarly in
this dilute regime. But now Eq. $\left(  \ref{repulsion}\right)  $ must be
interpreted in terms of the mutual repulsion between twist-kinks and
anti-twist-kinks. Corresponding shapes are thus similar to the revolving
pendulum case. However,  in the regions where the curvature changes sign, $\psi$ goes
to zero instead of repeatedly increasing by $\pi.$ In the main
text it is shown that the oscillating pendulum is never the
ground state of the squeelix. This is easy to understand.\ An oscillating
twist $\psi$ implies that its derivative $\psi^{\prime}(s)$ changes sign periodically. A negative derivative 
increases the pure twist energy around the curvature inversion point by $\frac
{C}{2}\int\left(  \psi^{\prime}-\omega_{3}\right)  ^{2}ds\sim\lambda
C\omega_{3}^{2}/2=\lambda E_{0}/L$.\ Nevertheless, this contribution is small for $\lambda/L\ll1$.

The right-hand side of Eq.~(\ref{eq:nast}) diverges as $\sqrt{\gamma}$ approaches $1/2$, 
which would mean that one can pack an arbitrary large number of TKs in a chain of length
$L$. This divergence is non-physical and is due to the fact that the long-range repulsive interaction 
$e^{-d/\lambda}$ is not strong enough to stabilize
the gas of twist-kinks against collapsing. Therefore, we must look at the opposite
regime of high twist-kink density and compute the short-range repulsion between
them. 

When $\gamma\approx0$ is decreased, the twist-kinks get more confined and their
density becomes high. They are also very deformed compared to their free
state Eq.\ $\left(  \ref{psyA2}\right)  $ and the notion of individual
twist-kinks actually looses its meaning.\ Nevertheless, we keep using the
terminology for convenience.\ When the separation between adjacent  twist-kinks is
$d\ll\lambda,$ the chain is being forced to overtwist so that $\psi\left(
s\right)  $ tends to become linear with $s$. A good ansatz in this regime is
\begin{equation}
\psi(s)=\frac{\pi}{d}s-\frac{\pi}{2}\text{ \ for }0\ll s\ll
L=nd\label{psydenseA2} 
\; ,%
\end{equation}
where $n$ the number of twist-kinks along the chain (as $\psi\left(  L\right)
-\psi\left(  0\right)  =\pi n)$ is assumed very large. In this regime $m \approx 0$.
The curvature becomes
\begin{equation}
\kappa(s)\approx\omega_{1}\cos\frac{\pi}{d}s \; .%
\end{equation}
Pluging Eq.\ $\left(  \ref{psydenseA2}\right) $ in Eq.~(\ref{EA2}) the total energy in this dense regime is%
\begin{equation}
E^{dense}_{nTK} \approx\frac{\omega_{1}^{2}B}{4}L+\frac{\pi^{2}}{2}\frac{Cn^{2}%
}{L}-\pi C\omega_{3}n + E_0
\; ,
\end{equation}
so that the optimum number $n^{\ast}$ minimizing this energy is
\begin{equation}
n^{\ast}\approx\frac{\omega_{3}}{\pi}L\label{highdensityA2}%
\; .
\end{equation}
Therefore, $\psi(s)=\omega_{3}s$ and $\psi^{\prime}(s)=\omega_{3}$ which
satisfies the boundary conditions automatically. The tangent angle to the
filament is then given by%
\begin{equation}
\phi(s)=\phi(-L/2)+\omega_{1}\sin(\omega_{3}s)\label{phiA2}
\; .
\end{equation}
Note that in the case of the oscillating pendulum, when $\gamma\approx0$, the
twist-kink-anti-twist-kink couple is very dense and the energetic cost of the
twist contribution is very high when $\psi(s)$ oscillates very
fast.\ Therefore, it is not necessary to treat this case further. 

\begin{table}
\begin{center}
\begin{tabular}{|c|c|c|c|}
\hline 
$\lambda/d$ & $E^{dense}_{nTK}$ & $E^{dilute}_{nTK}$ & $E$\tabularnewline
\hline 
\hline 
$10^{2}$ & $1600.$ & $-3\centerdot10^{8}$ & $1580.$\tabularnewline
\hline 
$10$ & $1600.$ & $-2\centerdot10^{6}$ & $1580.$\tabularnewline
\hline 
$1$ & $1600.$ & $-1\centerdot10^{4}$ & $1570.$\tabularnewline
\hline 
$10^{-1}$ & $1600.$ & $962.$ & $954$$.$\tabularnewline
\hline 
$3.10^{-2}$ & $1600.$ & $370.$ & $369.$\tabularnewline
\hline 
\end{tabular}
\end{center}
\caption{Comparison of the energies in the twist-kink picture with the exact result. 
The values are obtained with $\omega_{1}=\omega_{3}=80L^{-1}$, $n=26$, and $\lambda \omega_3 \approx 0.3$. All energies are given in units of $B/L$. \label{numerics}}
\end{table}

Taking a large value of $n$ and varying $\lambda$
one obtains Table~\ref{numerics} which shows a quantitative comparison between the energies in the twist-kink picture and the exact expression, Eq.~(\ref{EnergyA1}). We observe a good agreement in both asymptotic regimes $\lambda/d \ll 1$ and $\lambda/d \gg 1$.

The approach of this section can be used to explain the different
shapes discussed in the main text in a physical manner. In particular, it is valid in the regime of
very low and very high density of twist-kinks for the case of infinite long
chains, but also for finite chains as long as the boundary effects can be neglected.


\section{Squeelices of finite length}

In this section we will treat the problem exactly. We will nevertheless refer to 
the twist-kink nomenclature whenever it is convenient. We will consider  
the cases of the revolving and the oscillating pendulum separately.

\subsection{The revolving pendulum ($m<1$)}

\begin{figure}
\begin{center}
 \includegraphics[width=0.5\textwidth]{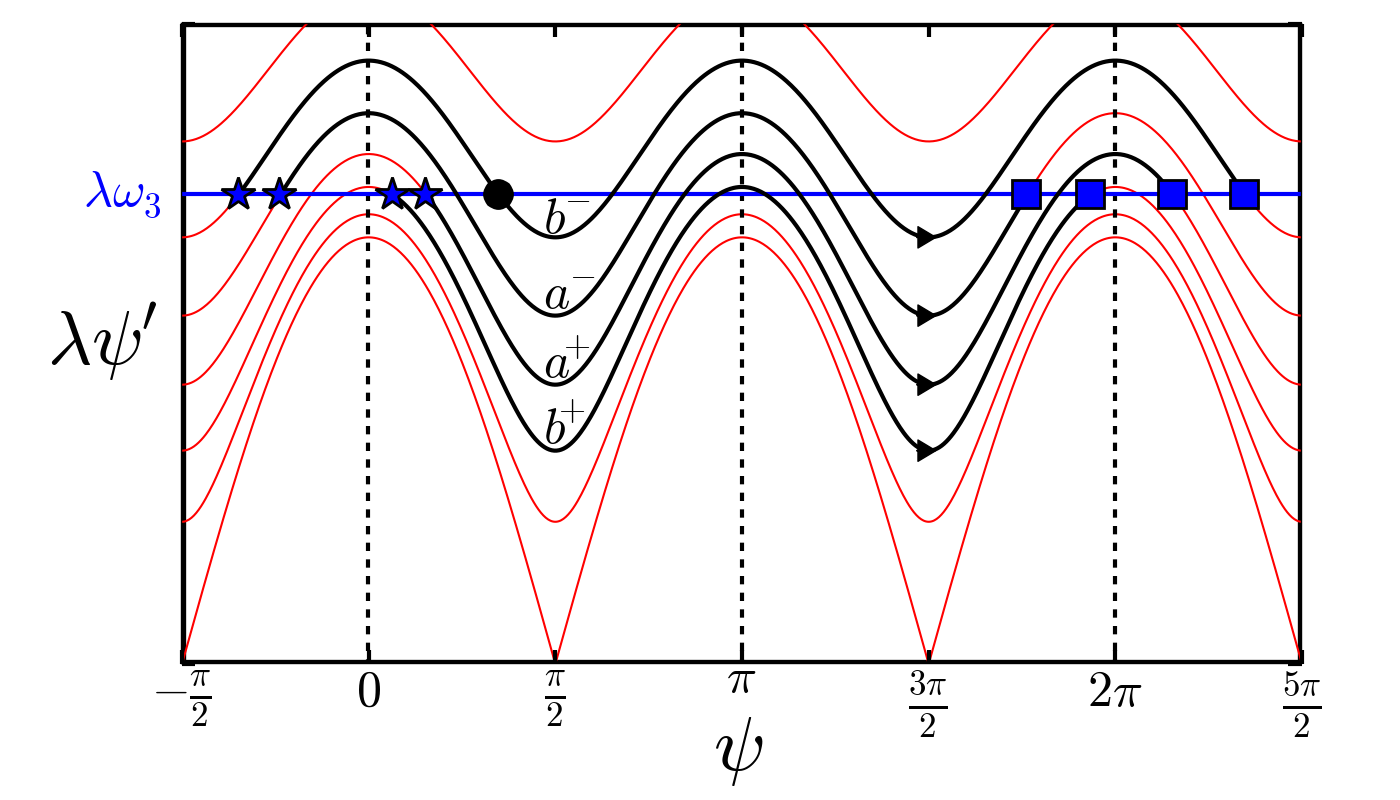}
\caption{
Phase plane of the revolving pendulum with four typical trajectories in black associated to the cases (a) and (b) (see Eqs.~(\ref{eq:aveck})). The stars represent the position $s=-L/2$, the squares $s=L/2$ for these trajectories. The subscript $\pm$ refers to the sign of $\psi(-L/2)$, which is either positive or negative. The shown trajectories 
correspond to shapes with a single twist-kink in the bulk of the chain, \textit{i.e.}, $n_a=n_b=2$. 
They are shown for illustrative reasons but do not have the same length. Nevertheless, with this restriction in mind we see that $a^+$ has a shape of type $(1_+^*)$, $a^-$  of type $(1_-^{*})$, $b^-$  of type $(1_-^{**})$, and $b^+$ of type $(1)$ as depicted in Fig.~\ref{shape_appendix}.   
In Appendix E we show that only the trajectories of type $b^+$ lead to stable shapes. The black circle indicates the position $s=L/2$ of the case $n_b=0$ (see text).
\label{fig:phaseplane2}}
\end{center}
\end{figure}

In this regime the twist, given by Eq.~(\ref{psyplusA1}), 
is a growing function of $s$. Eq.~(\ref{psiprimeA1}) 
reads with the positive sign%
\begin{equation}
\psi^{\prime}(s)=\frac{1}{\lambda\sqrt{m}}\sqrt{1-m\sin^{2}\psi}=\frac
{1}{\lambda\sqrt{m}}dn\left(  \frac{s-s_{0}}{\lambda\sqrt{m}}|m\right)  >0
\; ,
\end{equation}
where $dn\left(  x|m\right)  $ is a periodic odd elliptic Jacobian function of
period $l_{p}=2\lambda\sqrt{m}K(m)$ which is positive in this regime, \textit{i.e.},
$1\le dn\left(  x|m\right) \le \sqrt{1-m}$ \cite{Abramowitz}. The boundary conditions $\psi^{\prime
}(-L/2)=\psi^{\prime}(L/2)=\omega_{3}$ imply
\begin{equation}
\sin\psi\left(  -L/2\right)  =\pm\sin\psi\left(  L/2\right)
\; ,
\label{sinL}%
\end{equation}
which is equivalent to the two cases
\begin{subequations}
\label{eq:aveck}
\begin{align}
\psi\left(  L/2\right)  & =\psi\left(  -L/2\right)  +n_{a}%
\pi\text{ \ \ }\left(  \text{case (a)}\right) \qquad \text{and} \label{avecka}%
\\
\psi\left(  L/2\right)  & =-\psi\left(  -L/2\right)  +n_{b}%
\pi\text{ \ \ }\left(  \text{case (b)}\right) \; , \label{aveckb}%
\end{align}
\end{subequations}
where $n_{a,}n_{b}\in\mathbb{Z}$. In both cases $\psi\left(  -L/2\right)$ equals
$\pm\arcsin\left(  \sqrt{\frac{1}{m}-\frac{1}{\mu}}\right)$ according to Eq.~(\ref{psyLA1}). This leads to four typical 
trajectories in the phase plane (see Fig.~\ref{fig:phaseplane2}). Our goal is to identify those trajectories that satisfy the length constraint, \textit{i.e.}, to find all possible values of $m$ for a given $L$.


\subsubsection*{Case (a)}
An illustration of this case is given by the trajectories $a^+$ and $a^-$ in Fig.~\ref{fig:phaseplane2}.
Using the relation $am\left(  x|m\right)  +k\pi=am\left(  x+2k \mathcal{K}(m)|m\right)
$, Eq. $\left(  \ref{avecka}\right)  $ leads to

\begin{equation}
L=2n_{a}\lambda\sqrt{m}\mathcal{K}(m)\label{Lka}
\end{equation}
with $n_{a}=+1,+2,$... Here $n_{a}$ is the number of curvature inversion points within the filament (when $\psi=k\pi$ with 
$k$ a natural number). For each allowed value of $n_{a}$ there is a single solution of Eq.~(\ref{Lka}) with a particular $m$.

Since $m$ lies in the interval $\frac{\mu}{1+\mu}\le m \le \mu$ for $\mu<1$ and $\frac{\mu
}{1+\mu}\le m\le 1$ for $\mu\ge 1$ the length $L$ is bounded by
\begin{equation}
L_{\min}\leq\frac{L}{n_{a}}\leq L_{\max}\label{Lcondition}%
\end{equation}
with $L_{\min}=2\lambda\sqrt{\frac{\mu}{1+\mu}}\mathcal{K}(\frac{\mu}{1+\mu})$ and
$L_{\max}=2\lambda\sqrt{\mu}\mathcal{K}(\mu)$ for $\mu<1$ or $L_{\max}\rightarrow\infty$
for $\mu\ge 1$.\ 

In order to satisfy the condition of acceptable values of $L$, Eq.\ $\left(  \ref{Lcondition}\right)$, the number of
twist-kinks is limited, \textit{i.e.}, $n_{a}\in\left[  n_{a,\min},n_{a,\max}\right]  $
with $n_{a,\min}=\left\lfloor L/L_{\max}\right\rfloor $ and $n_{a,\max
}=\left\lfloor L/L_{\min}\right\rfloor $. The notation $\left\lfloor
x\right\rfloor $ again denotes the largest integer less than or equal to $x$. 

For $n_{a}=1$, the chain contains a curvature inversion point or a \textit{single} partial twist-kink which is localized close to one end of the chain. This requires a minimum length $L_{\min}$. 

From Eq.~$\left(\ref{SOA1}\right)$ we obtain:
\begin{equation}
s_{0,\pm}=-\frac{L}{2}\mp\lambda\sqrt{m}\mathcal{F}\left(  \arcsin\left(  \sqrt{\frac
{1}{m}-\frac{1}{\mu}}\right)  |m\right)  \; . \label{soplusmoins}
\end{equation}
We have thus two different solutions $\psi(s)$ and consequently two
different shapes of the squeelix. The stability analysis of Appendix E shows that all these shapes are \textit{unstable}, \textit{i.e.}, saddle points of the elastic energy $E[\psi]$ (see Eq.~(\ref{EA3})) of the squeelix.

\begin{figure}
\centering
\includegraphics[width=9cm]{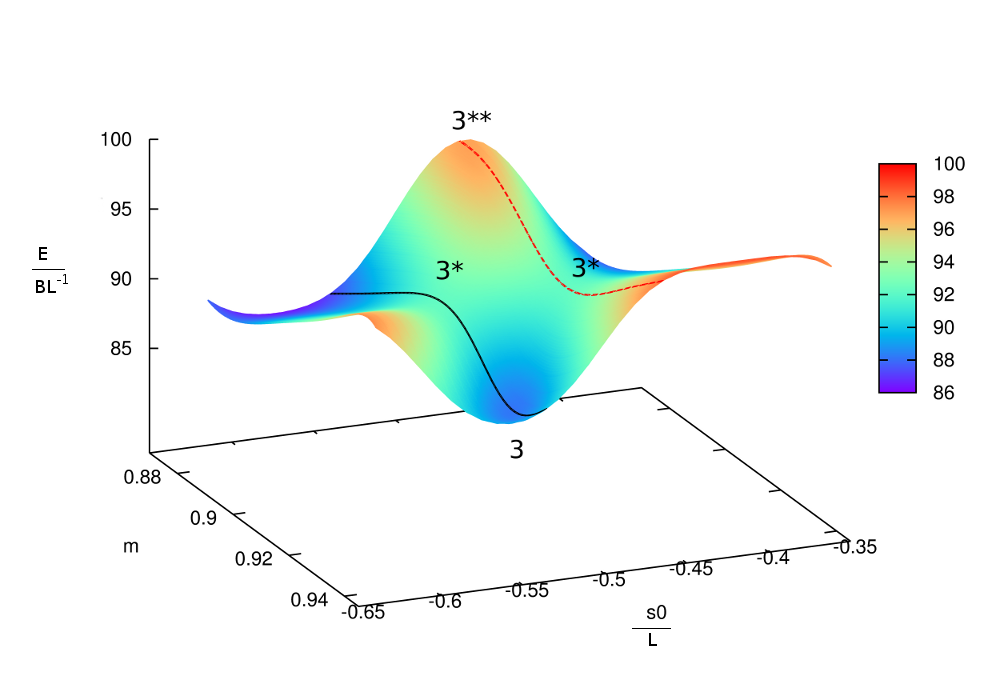}
\caption{Zoom of the energy landscape $E(m,s_0)$ for $\gamma=0.6079$, $\omega_1=20L^{-1}$, $\omega_3=16.33L^{-1}$ and $\sqrt{C/B}=1$. The curve $s_{0,+}(m)$ passes by the minimum $3$ (containing three twist-kinks, \textit{i.e.}, $n_{b}$=4), then the saddle point $3^{*}$ ($n_{a}$=4) and goes to the next minimum $4$ (with $n_{b}$=5, not shown). The curve $s_{0,-}(m)$ comes from the maximum  $2^{**}$ ($n_{b}$=3, not shown), then passes by the saddle point $3^{*}$  and reaches the next maximum $3^{**}$ ($n_{b}$=4). Energies and lengths are measured in units of $BL^{-1}$ and $L$,  respectively.}
\label{Ems0zoom}
\end{figure}

\begin{figure}
\centering
\subfigure[]{\label{Em_appendixa}
\includegraphics[width=0.4\textwidth]{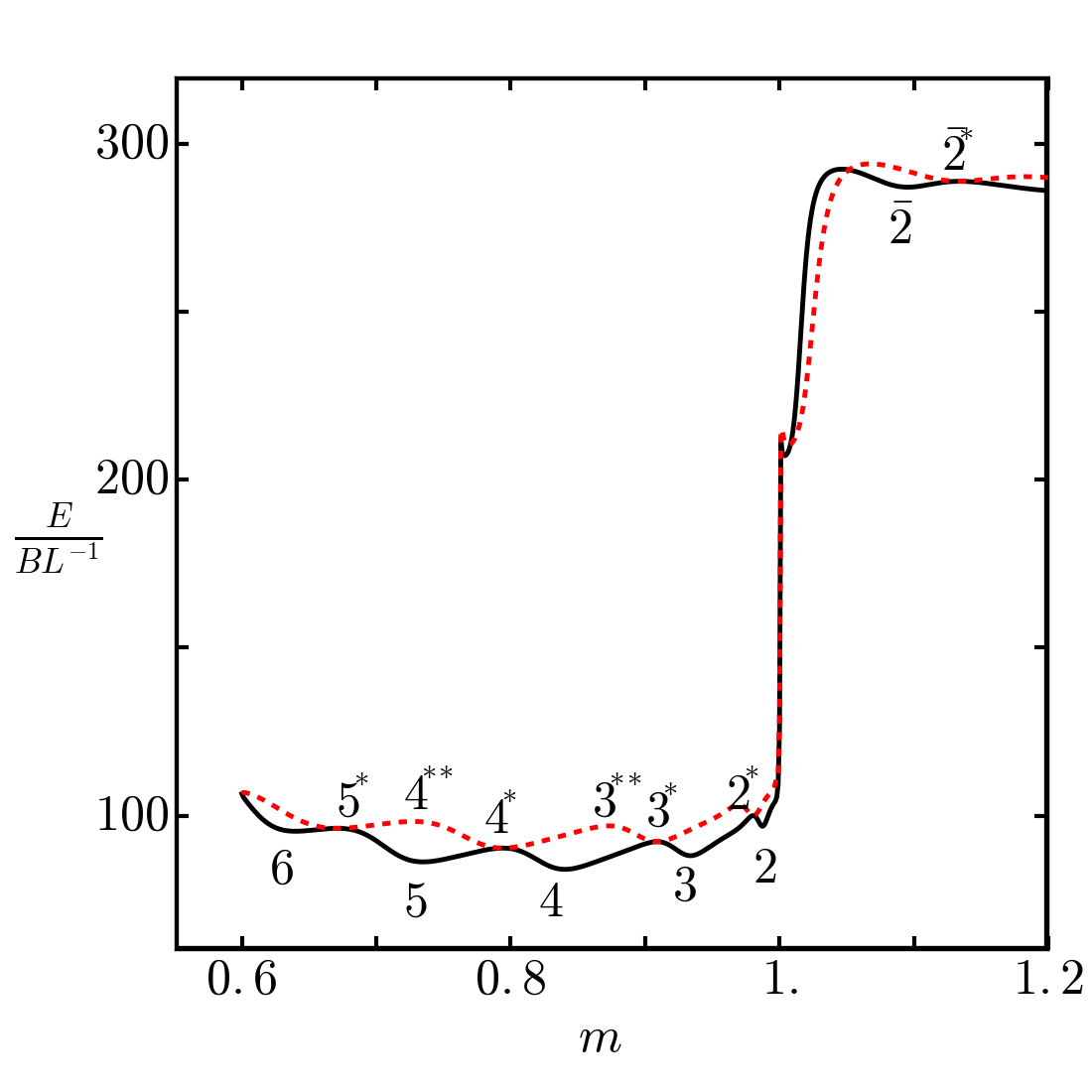}
}
\qquad
\qquad
\subfigure[]{\label{Em_appendixb}
\includegraphics[width=0.4\textwidth]{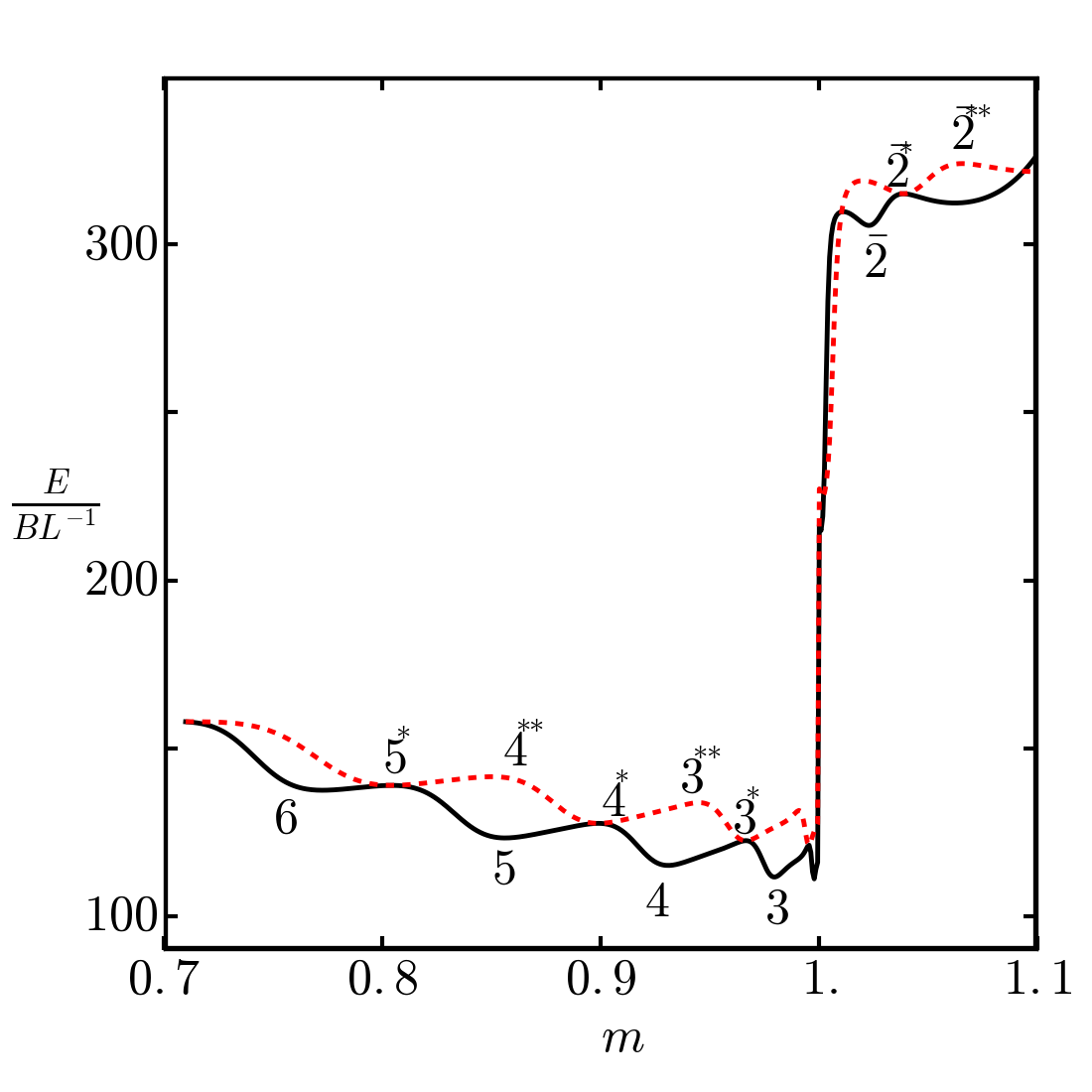}
}
\caption{The energies $E(m,s_{0,+}(m))$ (black solid curve) and $E(m,s_{0,-}(m))$ (red dotted curve), respectively for (a) $\gamma=0.6079$, $\omega_1=20L^{-1}$, $\omega_3=16.33L^{-1}$ and $\sqrt{C/B}=1$. (b) $\gamma=0.9915$, $\omega_1=24.4L^{-1}$, $\omega_3=15.6L^{-1}$ and $\sqrt{C/B}=1$. The case (b) is discussed in detail in the main text.  Energies and lengths are measured in units of $BL^{-1}$ and $L$,  respectively.}
\label{Em_appendix}
\end{figure}

For $s_{0,+}$ the shape has $n_{a}-1$ twist-kinks
in the bulk and a curvature inversion point, called critical twist-kink, near $s=L/2$. These shapes are the maxima of the curve $E(m,s_{0,+}(m))$. The case $s_{0,-}$ is very similar except that the critical twist-kink is situated near $s=-L/2$. Although they are minima of $E(m,s_{0,-}(m))$ they have the same energy than their $s_{0,+}$ counterpart. These two shapes labelled ($n^*$)  in Figs.~\ref{Ems0zoom} and \ref{Em_appendix} are saddle points of the energy landscape $E(m,s_0)$ with the same energy. They are the critical shapes on the top of the energy barrier of $E(m,s_{0,+}(m))$ that must be overcome to inject or expel a single critical twist-kink (localized at one of the ends of the squeelix) within the chain (see Fig.~\ref{Em_appendix}).

\subsubsection*{Case (b)}
An illustration of this case is given by the trajectories $b^+$ and $b^-$ in Fig.~\ref{fig:phaseplane2}.
Using the properties $am\left(  x|m\right)  +k\pi=am\left(  x+2k \mathcal{K}(m)|m\right)
$ and $am\left(  x|m\right)  =-am\left(  -x|m\right)  $ Eq. $\left(
\ref{aveckb}\right)  $ implies
\begin{equation}
s_{0}=-n_{b}\lambda\sqrt{m}\mathcal{K}(m) \; .
\end{equation}
Plugging this result into Eq.~(\ref{SOA1}) we obtain:
\begin{equation}
L=2\lambda\sqrt{m}\left[  n_{b}\mathcal{K}(m)\mp \mathcal{F}\left(  \arcsin\sqrt{\frac{1}{m}%
-\frac{1}{\mu}}|m\right)  \right] \; . \label{Lmb}
\end{equation}
Thus for each value of $n_{b}$, we have two values for $m$ that satisfy
Eq.~(\ref{Lmb}) and thus two values $s_{0,\pm}$. The two
associated shapes are very different as explained in the main text. 

(i) The case with the plus sign in Eq.~(\ref{Lmb}) corresponds
to the initial condition $\psi\left(  -L/2\right)  =-\arcsin\left(  \sqrt{\frac
{1}{m}-\frac{1}{\mu}}\right) $, \textit{i.e.}, the trajectory $s_{0,-}$. In Appendix E it is shown that all these extrema on the trajectory $s_{0,-}$ are \textit{unstable}.

Consider first $n_b=0$ which corresponds to the portion of trajectory of $b^-$ between the star and the circle in Fig.~\ref{fig:phaseplane2}. The $\psi$ is an incomplete twist-kink centered in the middle of the chain, \textit{i.e.}, $s_0=0$. One can show that this symmetric solution is a minimum of the curve $E(m,s_{0,-}(m))$ but a saddle point of the landscape $E(m,s_0)$.  It turns out that this solution does not exist for $L>L_{min}$. 

For $n_{b}=1,2...$ similar to case (a) the length $L$ is bounded by
\begin{equation}
\left(  n_{b}+1\right)  L_{\min}\le L \le n_{b}L_{\max}
\end{equation}

\begin{figure}
\begin{center}
\begin{tabular}{llll}
 \includegraphics[scale=0.065]{fig7fig15_kb1+.png} ($0$) &  \includegraphics[scale=0.065]{fig7fig8fig15_ka1+.png} ($0^*_+$) &  \includegraphics[scale=0.08]{fig8fig15_ka1-.png} ($0^*_-$) &  \includegraphics[scale=0.066]{fig8fig15_kb1-.png} ($0^{**}_-$) \\
 \\
 \\
 \includegraphics[scale=0.075]{fig7fig15_kb2+.png} ($1$) &  \includegraphics[scale=0.075]{fig7fig15_ka2+.png} ($1^*_+$) & \includegraphics[scale=0.073]{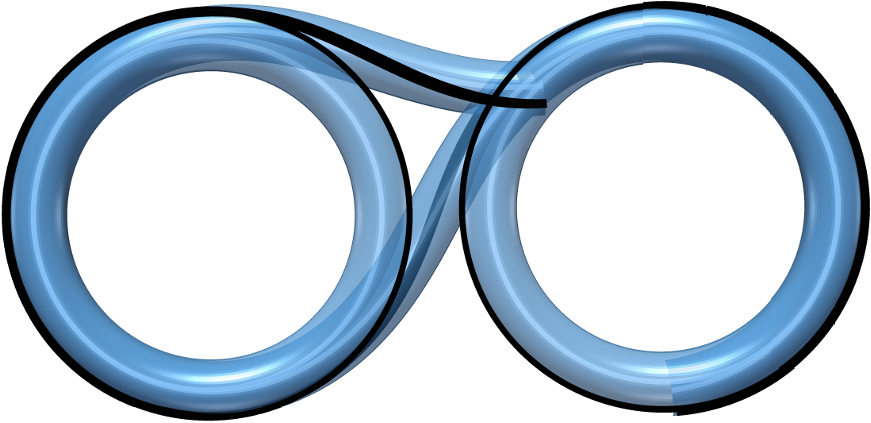} ($1^*_-$) & \includegraphics[scale=0.095]{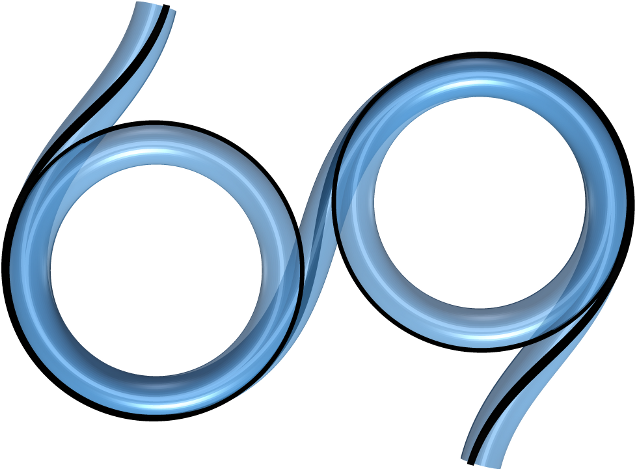}($1^{**}_-$)  \\
 \\
 \\
 \includegraphics[scale=0.1]{fig7fig15_kb3+.png} ($2$)&  \includegraphics[scale=0.102]{fig7fig15_ka3+.png} ($2^*_+$) &  \includegraphics[scale=0.102]{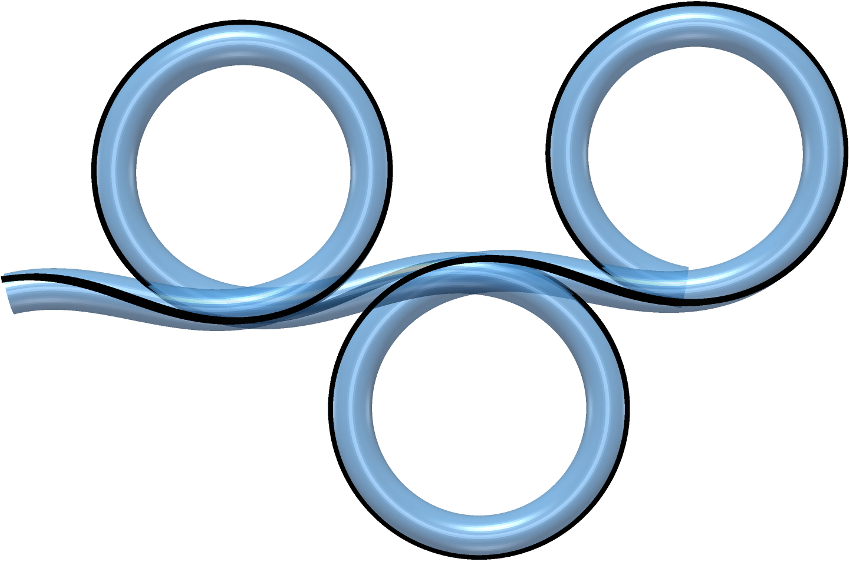} ($2^*_-$) &  \includegraphics[scale=0.11]{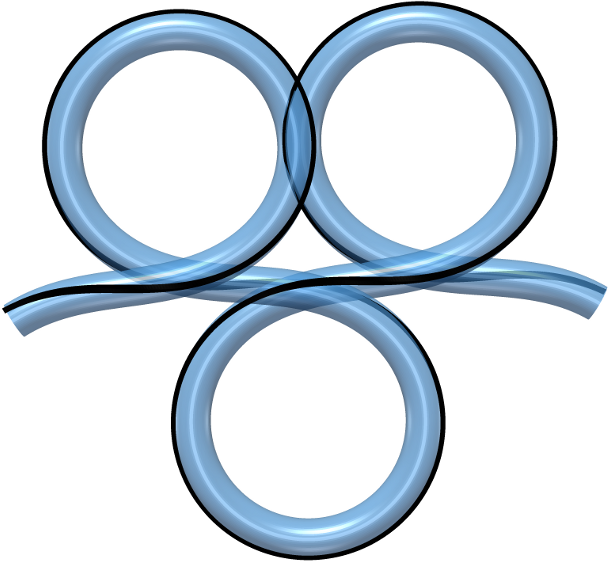} ($2^{**}_-$) \\
 \\
 \\
 \includegraphics[scale=0.07]{fig7fig15_kb4+.png} ($3$) &  \includegraphics[scale=0.11]{fig7fig15_ka4+.png} ($3^*_+$) &  \includegraphics[scale=0.11]{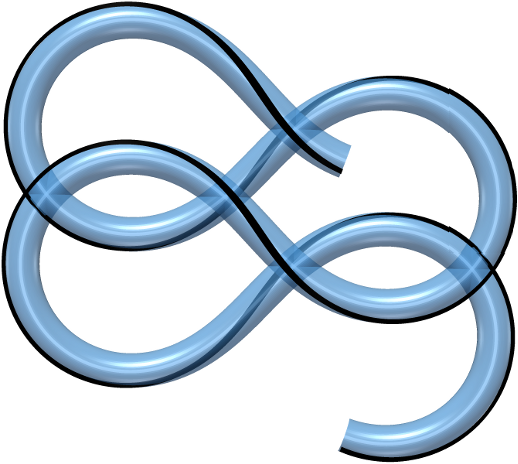} ($3^*_-$) &  \includegraphics[scale=0.11]{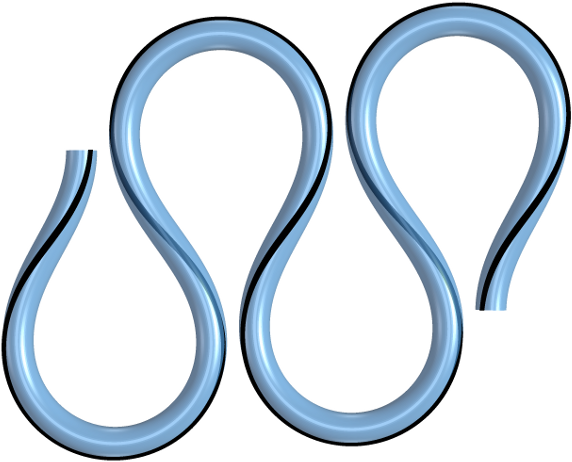} ($3^{**}_-$)  \\

\end{tabular}
\end{center}
\caption{Shapes of the squeelix in the regime $m<1$ for $\gamma=0.9915$, $\omega_1=24.4L^{-1}$, $\omega_3=15.6L^{-1}$ and $\sqrt{C/B}=1$. These shapes correspond to the first extrema of the energy in Fig.~\ref{Em_appendixb}.} 
\label{shape_appendix}
\end{figure}

In order to satisfy the condition above, $n_{b}\in\left[
n_{b,\min},n_{b,\max}\right]  $ with $n_{b,\min}=\left\lfloor L/L_{\max
}\right\rfloor $ and $n_{b,\max}=\left\lfloor L/L_{\min}\right\rfloor -1$.
Note that for $n_{b}=1$ there is a minimum length, \textit{i.e.}, $2L_{\min}\le L\le L_{\max
}$. The reason for this is that there are \textit{two} critical twist-kinks within
the chain for $n_{b}=1$, localized at both ends (see Fig.~\ref{shape_appendix}).

These symmetric solutions, labelled ($n^{**}$),  are maxima of the curve $E(m,s_{0,-}(m))$ and maxima of the landscape $E(m,s_0)$. They are the critical shapes with two critical twist-kinks
(localized at both ends of the squeelix) on the top of the energy barrier of $E(m,s_{0,-}(m))$ that
must be overcome to inject or expel one or both twist-kinks (see Figs.~\ref{Ems0zoom} and \ref{Em_appendix}).

(ii) The case with the minus sign in Eq.\ $\left(  \ref{Lmb}\right)$ corresponds
to the initial condition $\psi\left(  -L/2\right)  =\arcsin\left(  \sqrt{\frac
{1}{m}-\frac{1}{\mu}}\right) $, \textit{i.e.}, the trajectory $s_{0,+}$. In this case one obtains
\begin{equation}
\left(  n_{b}-1\right)  L_{\min}\le L \le n_{b}L_{\max}
\end{equation}
with $n_{b}=1,2...$ In order to satisfy this condition $n_{b}\in\left[
n_{b,\min},n_{b,\max}\right]  $ with $n_{b,\min}=\left\lfloor L/L_{\max
}\right\rfloor $ and $n_{b,\max}=1+\left\lfloor L/L_{\min}\right\rfloor $.
Note that for $n_{b}=1$ the length $L$ can be arbitrarily small, \textit{i.e.},
$0<L\le L_{\max}$. This is the situation with\textit{ zero} twist-kinks within the
chain. It is thus the ground state for $\gamma>1$ and corresponds to
Eq. $\left(\ref{zeroTWA2}\right)  $. When $L>L_{\min}$, a twist-kink can be
injected within the chain. It will be localized in the center of the
squeelix. More twist-kinks will be disposed in an equidistant manner (see Fig. \ref{shape_appendix}). 
Here $n_{b}-1$ gives the number of twist-kinks in the squeelix. These solutions are local minima of the energy $E(m,s_{0,+}(m))$ (see Figs. \ref{Ems0zoom} and \ref{Em_appendix}). As shown in the
stability analysis of Appendix E, these solutions lead to \textit{stable
shapes} that are also minima of the full elastic energy $E[\psi]$ of the squeelix.

Note that the case $m=1$ gives 
\begin{equation}
\psi(s)=2\arctan \left( \frac{s-s_{0,+}(1)}{\lambda} \right)-\pi/2\label{psym1}
\end{equation}
with $s_{0,+}(1)$ given by Eq.~(\ref{soplusmoins}). 
This function satisfies the boundary condition at $s=-L/2$ by construction but $\psi'(L/2)<\omega_3$. Therefore, it is never a solution of the equation of motion of a squeelix of finite length. 

Figure~\ref{shape_appendix} illustrates the theory developped in this appendix. It shows the shapes that are associated to the first extrema in 
Fig.~\ref{Em_appendixb} in the regime $m<1$. For further shapes of this example with $m<1$ see Fig.~\ref{figminfeq1} in the main text.

\begin{table}
\begin{center}
\begin{tabular}
[c]{|c|c|c|c|}\hline
Shape & $n_{a/b}$ & $m$ & Energy ($BL^{-1}$)\\\hline
$(0)$ & $n_{b}=1$ & $0.9999999999471$ & $111.31$\\\hline
$(0_{+}^{\ast})$ & $n_{a}=1$ & $0.999999999595$ & $121.47$\\\hline
$(0_{-}^{\ast})$ & $n_{a}=1$ & $0.999999999595$ & $121.47$\\\hline
$(0_{-}^{\ast\ast})$ & $n_{b}=1$ & $0.9999999969$ & $131.63$\\\hline
$(1)$ & $n_{b}=2$ & $0.99997091$ & $111.11$\\\hline
$(1_{+}^{\ast})$ & $n_{a}=2$ & $0.9999195$ & $121.26$\\\hline
$(1_{-}^{\ast})$ & $n_{a}=2$ & $0.9999195$ & $121.26$\\\hline
$(1_{-}^{\ast\ast})$ & $n_{b}=2$ & $0.9997774$ & $131.42$\\\hline
$\textbf{(2)}$ & $\textbf{n}_\textbf{\textit{b}}\textbf{=3}$ & $\textbf{0.997635}$ & $\textbf{110.98}$\\\hline
$(2_{+}^{\ast})$ & $n_{a}=3$ & $0.995359$ & $121.23$\\\hline
$(2_{-}^{\ast})$ & $n_{a}=3$ & $0.995359$ & $121.23$\\\hline
$(2_{-}^{\ast\ast})$ & $n_{b}=3$ & $0.99089$ & $131.54$\\\hline
$(3)$ & $n_{b}=4$ & $0.97955$ & $111.72$\\\hline
$(3_{+}^{\ast})$ & $n_{a}=4$ & $0.96651$ & $122.56$\\\hline
$(3_{-}^{\ast})$ & $n_{a}=4$ & $0.96651$ & $122.56$\\\hline
$(3_{-}^{\ast\ast})$ & $n_{b}=4$ & $0.94501$ & $133.87$\\\hline
\end{tabular}
\end{center}
\caption{Numerical values of the parameters of the shapes of Fig. \ref{shape_appendix}.}
\label{table_appendixD}
\end{table}


\subsection{The oscillating pendulum ($1<m$)}

For $m>1$ the twist, Eq.\ $\left(  \ref{psyplusA1}\right)$, 
is a periodic function of period $l_{p}=4\lambda \mathcal{K}(\frac{1}{m})$.\ We used the
fact that $\mathcal{K}(m)=\frac{1}{\sqrt{m}}\mathcal{K}(\frac{1}{m})$ for $m>1$.\ The amplitude of
the oscillations is $\psi_{0}=\arcsin\sqrt{1/m}<\pi/2$.\ The twist variation
Eq.\ $\left(  \ref{psiprimeA1}\right)  $ is then very different from its
$m<1$ counterpart
\begin{equation}
\psi^{\prime}(s)=\frac{1}{\lambda\sqrt{m}}cn\left(  \frac{s-s_{0}}{\lambda
}|\frac{1}{m}\right)
\; ,
\end{equation}
where $cn\left(  x|m\right)  $ is an odd periodic elliptic Jacobian function
that oscillates between $-1$ and $1$ and thus changes its sign \cite{Abramowitz}. In this
regime the periodic curvature reads
\begin{equation}
\kappa(s)=\frac{\omega_{1}}{\sqrt{m}}sn\left(  \frac{s-s_{0}}{\lambda}%
|\frac{1}{m}\right)
\end{equation}
with the period $l_{p}$.\ The boundary condition $\psi^{\prime}(-L/2)=\psi
^{\prime}(L/2)=\omega_{3}$ now implies the following cases%
\begin{subequations}
\label{eq:LM1}
\begin{align}
\psi\left(  L/2\right)   &  =\psi\left(  -L/2\right)  \text{ \ \ \ (case (a))} \; ,\label{LM1a}\\
\psi\left(  L/2\right)   &  =-\psi(-L/2)\text{ \ \ \ (case (b))}%
\; , \label{LM1b}%
\end{align}
\end{subequations}
which corresponds to equal (a) or opposite (b) curvatures at the ends of the squeelix.
\begin{figure}
\begin{center}
 \includegraphics[width=0.5\textwidth]{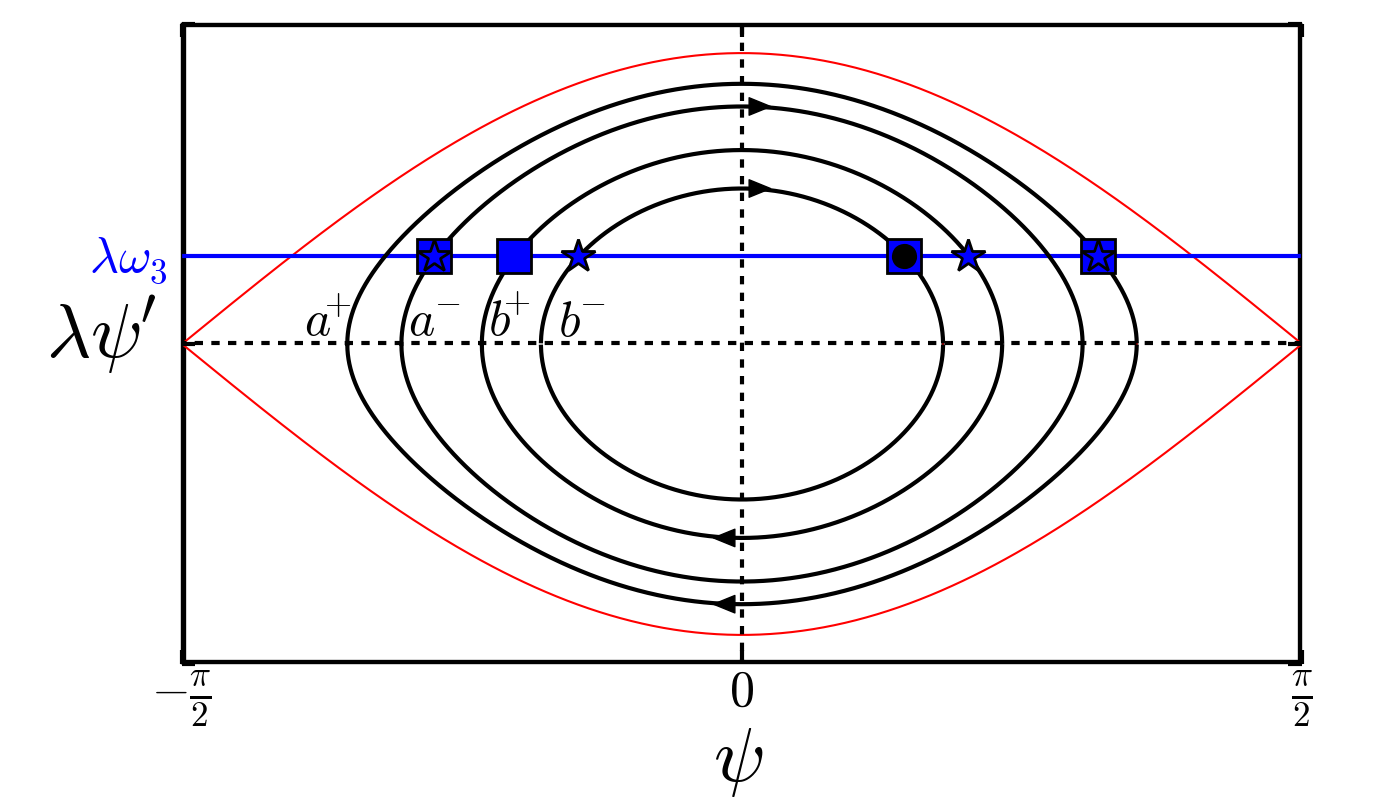}
\caption{
Phase plane of the oscillating pendulum with four typical trajectories in black associated to the cases $(a)$ and $(b)$ (see Eqs.~(\ref{eq:LM1})). The stars represent the position $s=-L/2$, the squares $s=L/2$ for these trajectories. The subscript $\pm$ refers to the sign of $\psi(-L/2)$, which is either positive or negative. 
In Appendix E we show that only the trajectories of type $b^+$ lead to stable shapes. The black circle indicates the position $s=L/2$ of the case $n_b=0$ (see text).
\label{fig:phaseplane3}}
\end{center}
\end{figure}
Eq. $\left(  \ref{psyLA1}\right)  $ is still valid, \textit{i.e.}, $\psi\left(
-L/2\right)  =\pm\arcsin\left(  \sqrt{\frac{1}{m}-\frac{1}{\mu}}\right)  $. 
Thus $-\psi_{0}\le \psi\left(  -L/2\right) \le \psi_{0}$ which implies $1\le m\le \mu
$. The equation for $s_{0}$ becomes
\begin{equation}
s_{0,\pm}=-\frac{L}{2}\mp\lambda \mathcal{F}\left(  \arcsin\left(  \sqrt{1-\frac{m}{\mu
}}\right)  |\frac{1}{m}\right) \; , \label{SOMsup1}%
\end{equation}
where $s_{0,+}$ corresponds to $0<\psi\left(  -L/2\right)  <\pi/2$ and
$s_{0,-}$ to $-\pi/2<\psi\left(  -L/2\right)  <0$. We used the relation
$\mathcal{F}\left(  \varphi|m\right)  =\frac{1}{\sqrt{m}}\mathcal{F}\left(  \arcsin(\sqrt{m}%
\sin\varphi)|\frac{1}{m}\right)  $.


\subsubsection*{Case (a)}

An illustration of this case is given by the trajectories $a^+$ and $a^-$ in Fig.~\ref{fig:phaseplane3}. 
Eq.\ $\left(  \ref{LM1a}\right)  $ implies
\begin{equation}
L=n_{a}l_{p}(m) \; , \label{LMsup1}%
\end{equation}
where $n_{a}=+1,+2,...$ is the number of periods in the chain. We have the
constraint
\begin{equation}
l_{p}(\mu)\leq\frac{L}{n_{a}}%
\end{equation}
as $K\left(  \frac{1}{m}\right)  \rightarrow\infty$ with $m\rightarrow1$.
Therefore $1\le n_{a}\le n_{a,\max}$ with $n_{a,\max}=\left\lfloor L/l_{p}%
(\mu)\right\rfloor $. For a given length there is a maximal number of
oscillations. For each allowed value of $n_{a}$ there is one solution of Eq. $\left(  \ref{LMsup1}\right)  $ with a given $m$. The arc length $s_{0}$ is
related to $L$ by Eq. $\left(  \ref{SOMsup1}\right)$.

The associated shapes turn out to be \textit{unstable} (see Appendix E). The two types of shapes
obtained with $s_{0,\pm}$ have the same energy. They correspond to the critical shapes on the top of the energy barrier that must be overcome to inject or expel a single anti-twist-kink (localized at the chain end $s=L/2$ for $s_{0,+}$ or at $s=-L/2$ for $s_{0,-}$) within the chain (see main text for some shapes).


\subsubsection*{Case (b)}
An illustration of this case is given by the trajectories $b^+$ and $b^-$ in Fig.~\ref{fig:phaseplane3}. 
Equation~(\ref{LM1b})  implies $am\left(  \frac{L/2-s_{0}%
}{\lambda\sqrt{m}}|m\right)  =-am\left(  \frac{-L/2-s_{0}+l_{p}}{\lambda
\sqrt{m}}|m\right) $ which leads to%
\begin{equation}
s_{0}=-\frac{1}{2}n_{b}l_{p}(m)
\end{equation}
with $n_{b}=0,+1,+2,...$
Plugging this result into Eq. $\left(  \ref{SOMsup1}\right)  $ we have
\begin{equation}
L=2\lambda\left(  n_{b}2 \mathcal{K}\left(  \frac{1}{m}\right)  \mp \mathcal{F}\left(
\arcsin\left(  \sqrt{1-\frac{m}{\mu}}\right)  |\frac{1}{m}\right)  \right)
\; .
\label{Lmsup1b}%
\end{equation}
Therefore, for a given value $n_{b}$ we have two different solutions for $m$
depending on the sign $\mp$. 
The case $n_b=0$ corresponds to the portion of trajectory of $b^-$ between the star and the circle in Fig.~\ref{fig:phaseplane3}. One can make the same analysis as in the case $m<1$. The solution is
symmetric with $s_0=0$ but is unstable. It is a minimum of the curve $E(m,s_{0,-}(m))$ but a saddle point of the landscape $E(m,s_0)$. This solution does not exist for $L> L_{min}$.

For $n_{b}=+1,+2,...$ Eq.\ $\left(  \ref{Lmsup1b}\right)  $ implies in
both cases
\begin{equation}
l_{p}(\mu)\le\frac{L}{n_{b}} \; ,%
\end{equation}
which leads to $1<n_{b}<$ $n_{b,\max}$ with $n_{b,\max}=\left\lfloor
L/l_{p}(\mu)\right\rfloor $. The two solutions for a given $n_{b}$ are not
equivalent. The shapes with the positive sign in Eq.$\left(  \ \ref{Lmsup1b}%
\right)  $ corresponds to the case $-\pi/2<\psi\left(  -L/2\right)  <0$ and
are unstable (see Appendix E). They are the critical shapes with two
anti-twist-kinks (localized at both ends of the squeelix) on the top of the
energy barrier that must be overcome to inject or expel one or both
anti-twist-kinks.

Only the solution with the negative sign in Eq.~(\ref{Lmsup1b}) 
corresponding to $0<\psi\left(  -L/2\right)  <\pi/2$  leads to \textit{stable}
shapes that are therefore minima of the elastic energy. The shapes are such
that the curvature inversion points are distributed in an equidistant manner
within the squeelix (see main text for some shapes). 

In Fig.~\ref{Em_appendix} we 
observe that some extrema are not numbered. These extrema at position, say $m_i$, originate from our choice of the functional space. They do do not fulfill the boundary conditions Eq.~(\ref{BC}) but instead satisfy $\delta \psi(L/2)\big|_{m_i} =\frac{\partial \psi(L/2)}{\partial m}\big|_{m_i}\delta m= 0$ so that $\delta E=0$ in Eq.~(\ref{eq:firstvariation}) is still satisfied.


\section{Stability analysis}

To study the stability of the solutions~(\ref{psyplusA1}) which satisfy the Neumann boundary conditions~(\ref{BCA2}), 
we perform the second variation of the energy Eq.~(\ref{EA3})
\begin{equation}
\delta^{2}E=C\int_{-L/2}^{L/2}\left(  \tilde{\psi}(s)^{\prime2}+V(s)\tilde
{\psi}(s)^{2}\right)  ds \label{SecondvariationAD}
\end{equation}
with $V(s)=\frac{1}{\lambda^{2}}\left(  2\sin^{2}\psi-1\right)  $ and
$\tilde{\psi}=\delta\psi$ a small variation around a solution~(\ref{psyplusA1}). 
Integrating Eq. (\ref{SecondvariationAD}) by parts gives
\begin{equation}
\delta^{2}E=C\left[  \tilde{\psi}\frac{d}{ds}\tilde{\psi}\right]
_{-L/2}^{L/2}+C\int_{-L/2}^{L/2}\tilde{\psi}\mathcal{L}\tilde{\psi} \, ds
\; , \label{deltaE2AD}
\end{equation}
where $\mathcal{L}=-\frac{d^{2}}{ds^{2}}+V(s)$ is a Lam\'e operator. We now
distinguish the cases $m<1$ and $m>1$. 


\subsection{The revolving pendulum ($m<1$)}
Stability problems with fluctuations that satisfy the Neumann boundary conditions are discussed in Ref.~\cite{Manning2009}. 
In our case the Neumann boundary condition is a natural condition for the extremal energy configuration only.
There is no external torque that would impose the Neumann conditions to the full twist $\psi(s)$. At finite temperature all fluctuations $\tilde{\psi}(s)$ around an extremal energy configuration are physically allowed and not only those with $\tilde{\psi}^{\prime}(s)=0$.
Therefore, in our analysis we consider a general fluctuation which does not necessarily respect the Neumann boundary conditions. Doing so we will find the fluctuation~$\tilde{\psi}(s)$ (around a particular solution of the Euler-Lagrange equation) that minimizes the second derivative of the energy $\delta^{2}E$. When this minimum of $\delta^{2}E$ is positive, the solution is stable with respect to any kind of fluctuation (in particular those respecting the Neumann conditions). When it is negative, the considered solution is unstable.

Starting with a fluctuation of the form $\tilde{\psi}\rightarrow\tilde{\psi
}+\sigma s+\rho$ with $\tilde{\psi}=0$ at the chain ends, Eq.~(\ref{deltaE2AD}) becomes%
\begin{equation}
\delta^{2}E=C\left\{  A_{1}\sigma^{2}+2A_{2}\sigma\rho+A_{3}\rho^{2}
+\int_{-L/2}^{L/2}\left(  2\sigma sV(s)\tilde{\psi}+2\rho V(s)\tilde{\psi
}+\tilde{\psi}\mathcal{L}\tilde{\psi}\right)  ds\right\} 
\label{eq:deltaenergy_minf1}
\end{equation}
with
\begin{subequations}
\begin{align}
A_{1} &  =\int_{-L/2}^{L/2}s^{2}V(s)ds+L \; , \\
A_{2} &  =\int_{-L/2}^{L/2}sV(s)ds \; , \\
A_{3} &  =\int_{-L/2}^{L/2}V(s)ds \; .
\end{align}
\end{subequations}
For given $\sigma$ and $\rho$, we aim to minimize $\delta^{2}E$.  This
amounts to solve the equation%
\begin{equation}
\sigma sV\left( s\right)  +\rho V\left(  s\right)  +\mathcal{L}\tilde{\psi
}=0\label{eq:psitilde}%
\end{equation}
with $\tilde{\psi}=0$ at the chain's ends. The solution of Eq.\ $\left(
\ref{eq:psitilde}\right)  $ is a minimum if its second order fluctuation is
positive. One thus has to find the sign of $\int_{-L/2}^{L/2} \bar{\psi} \mathcal{L} \bar{\psi} \, ds$, 
where $\bar{\psi}$ is the fluctuation of the fluctuation $\tilde{\psi}$. The Lam\'e operator 
has the eigenvalues 0, $\frac{1-m}{m}$, and $\frac{1}{m}$. Thus, the second order fluctuation is
positive definite. The implicit dependence
of this solution on the parameters $\sigma$ and $\rho$ is linear. It is thus
a combination%
\begin{equation}
\tilde{\psi}\left(  s\right)  =\alpha\tilde{\psi}^{\left(  0\right)
}(s)+\sigma\tilde{\psi}^{\left(  1\right)  }\left(  s\right)  +\rho
\tilde{\psi}^{\left(  2\right)  }\left(  s\right)
\; .\label{eq:psitildedecomposition}%
\end{equation}
Then from Eq. (\ref{eq:psitilde}) we obtain
\begin{align}
\mathcal{L}\tilde{\psi}^{\left(  0\right)  }\left(  s\right)   &
=0 \; ,\label{eq:psi0tilde}\\
sV\left(  s\right)  +\mathcal{L}\tilde{\psi}^{\left(  1\right)  }\left(
s\right)   &  =0\label{eq:psi1tilde} \; ,\\
V\left(  s\right)  +\mathcal{L}\tilde{\psi}^{\left(  2\right)  }\left(
s\right)   &  =0\label{eq:psi2tilde}%
\end{align}
with $\tilde{\psi}^{\left(  i\right)  }$ equal to zero at the boundaries. The first
equation shows that $\tilde{\psi}^{\left(  0\right)  }\left(  s\right)
=dn\left(  \frac{s-s_{0}}{\lambda\sqrt{m}}|m\right)  $ is the zero mode 
of $\mathcal{L}$. Since $dn\left(  \frac{s-s_{0}}{\lambda
\sqrt{m}}|m\right)  $ does not satisfy the boundary condition, we have
$\alpha=0$.
To build the two other components $\tilde{\psi}^{\left(  1\right)  }$ and
$\tilde{\psi}^{\left(  2\right)  }$we first set%
\begin{align}
\varphi\left(  \frac{s-s_{0}}{\lambda\sqrt{m}},m\right)   &  =\frac{1}%
{1-m}\left[  dn\left(  \frac{s-s_{0}}{\lambda\sqrt{m}}|m\right) \mathcal{E}\left(
am\left(  \frac{s-s_{0}}{\lambda\sqrt{m}}|m\right)  |m\right)  \right.  \\
&  \left.  -mcn\left(  \frac{s-s_{0}}{\lambda\sqrt{m}}|m\right)  sn\left(
\frac{s-s_{0}}{\lambda\sqrt{m}}|m\right)  \right]  .
\end{align}
The general solution of the second and the third equations (\ref{eq:psi1tilde} -\ref{eq:psi2tilde}) are
\begin{align}
\tilde{\psi}^{\left(  1\right)  } &  \left(  s\right)  =C_{1}dn\left(
\frac{s-s_{0}}{\lambda\sqrt{m}}|m\right)  +C_{2}\varphi\left(  \frac{s-s_{0}%
}{\lambda\sqrt{m}}|m\right)  -s  \; ,\\
\tilde{\psi}^{\left(  2\right)  } &  \left(  s\right)  =C_{3}dn\left(
\frac{s-s_{0}}{\lambda\sqrt{m}}|m\right)  +C_{4}\varphi\left(  \frac{s-s_{0}%
}{\lambda\sqrt{m}}|m\right)  -1 \; .
\end{align}
To fix the coefficients $C_{i}$ we take into account the boundary conditions
$\tilde{\psi}^{\left(  i\right)  }\left(  \pm L/2\right)  =0$. Using
\begin{align}
\psi\left(  -L/2\right)   &  =am\left(  \frac{-L/2-s_{0}}{\lambda\sqrt{m}%
}|m\right)  =\pm\arcsin\sqrt{\frac{1}{m}-\frac{1}{\mu}} \; ,\\
\psi\left(  L/2\right)   &  =am\left(  \frac{L/2-s_{0}}{\lambda\sqrt{m}%
}|m\right)  =%
\begin{cases}
\psi\left(  -L/2\right)  +n_{a}\pi\text{ \ \ \ (case}(a)\text{)}\\
-\psi\left(  -L/2\right)  +n_{b}\pi\text{ \ \ \ (case}(b)\text{)}
\end{cases}
\end{align}
(see Appendix D), we get for the case (a)%
\begin{align}
\tilde{\psi}_{a}^{\left(  1\right)  }\left(  s\right)   &  =-L\sqrt{\frac{\mu
}{m}}\frac{\mathcal{E}\left(  \psi\left(  -L/2\right)  |m\right)  -\frac{1}{2}\sqrt{\mu
m}\sin\left(  2\psi\left(  -L/2\right)  \right)  }{\mathcal{E}\left(  \psi\left(
L/2\right)  |m\right)  -\mathcal{E}\left(  \psi\left(  -L/2\right)  |m\right)
}dn\left(  \frac{s-s_{0}}{\lambda\sqrt{m}}|m\right)  \; , \\
&  +L\sqrt{\frac{\mu}{m}}\frac{1-m}{\mathcal{E}\left(  \psi\left(  L/2\right)
|m\right)  -\mathcal{E}\left(  \psi\left(  -L/2\right)  |m\right)  }\varphi\left(
\frac{s-s_{0}}{\lambda\sqrt{m}},m\right)  -s \nonumber\\
\tilde{\psi}_{a}^{\left(  2\right)  }\left(  s\right)   &  =\sqrt{\frac{\mu
}{m}}dn\left(  \frac{s-s_{0}}{\lambda\sqrt{m}}|m\right)  -1 \; ,
\end{align}
and for the case (b)
\begin{align}
\tilde{\psi}_{b}^{\left(  1\right)  }\left(  s\right)   &  =-L\sqrt{\frac{\mu
}{m}}\frac{\mathcal{E}\left(  \psi\left(  -L/2\right)  |m\right)  -\frac{1}{2}\sqrt{\mu
m}\sin\left(  2\psi\left(  -L/2\right)  \right)  }{\mathcal{E}\left(  \psi\left(
L/2\right)  |m\right)  -\mathcal{E}\left(  \psi\left(  -L/2\right)  |m\right)
+\sqrt{\mu m}\sin\left(  2\psi\left(  -L/2\right)  \right)  }dn\left(
\frac{s-s_{0}}{\lambda\sqrt{m}}|m\right)  \; ,\\
&  +L\sqrt{\frac{\mu}{m}}\frac{1-m}{\mathcal{E}\left(  \psi\left(  L/2\right)
|m\right)  -\mathcal{E}\left(  \psi\left(  -L/2\right)  |m\right)  +\sqrt{\mu m}%
\sin\left(  2\psi\left(  -L/2\right)  \right)  }\varphi\left(
\frac{s-s_{0}}{\lambda\sqrt{m}}|m\right)  -s \nonumber\\
\tilde{\psi}_{b}^{\left(  2\right)  }\left(  s\right)   &  =\sqrt{\frac{\mu
}{m}}dn\left(  \frac{s-s_{0}}{\lambda\sqrt{m}}|m\right)  -1 \; .
\end{align}
In order to determine the sign of $\delta^{2}E(\tilde{\psi})$ we
use the decomposition (\ref{eq:psitildedecomposition}) to write
\begin{align}
2\sigma sV\left(  s\right)  \tilde{\psi}+2\rho V\left(  s\right)  \tilde
{\psi}+\tilde{\psi}\mathcal{L}\tilde{\psi} &  =\nonumber\\
&  \sigma^{2}sV\left(  s\right)  \tilde{\psi}^{\left(  1\right)  }+\rho
^{2}V\left(  s\right)  \tilde{\psi}^{\left(  2\right)  }+\sigma\rho\left(
sV\left(  s\right)  \tilde{\psi}^{\left(  2\right)  }+V\left(  s\right)
\tilde{\psi}^{\left(  1\right)  }\right) \; .
\end{align}
Replacing this equality in Eq. (\ref{eq:deltaenergy_minf1}), we get
\begin{equation}
\delta^{2}E(\tilde{\psi})=C\left\{  \sigma^{2}\tilde{A_{1}%
}+2\sigma\rho\tilde{A_{2}}+\rho^{2}\tilde{A_{3}}\right\} \; .
\end{equation}
Then the stability conditions can be written%
\begin{subequations}
\begin{align}
\tilde{A_{1}}  &  >0\label{eq:stabcond1} \; ,\\
\tilde{A_{3}}  &  >0\label{eq:stabcond2}\; ,\\
\tilde{A_{1}}\tilde{A_{3}}-\tilde{A_{2}}^{2}  &  >0 \; , \label{eq:stabcond3}%
\end{align}
\end{subequations}
where we have defined%
\begin{subequations}
\begin{align}
\tilde{A_{1}} &  =A_{1}+\int_{-L/2}^{L/2}sV\left(  s\right)  \tilde{\psi
}^{\left(  1\right)  }ds \; ,\\
\tilde{A_{2}} &  =A_{2}+\frac{1}{2}\int_{-L/2}^{L/2}\left(  sV\left(
s\right)  \tilde{\psi}^{\left(  2\right)  }+V\left(  s\right)  \tilde{\psi
}^{\left(  1\right)  }\right)  ds \; ,\\
\tilde{A_{3}} &  =A_{3}+\int_{-L/2}^{L/2}V\left(  s\right)  \tilde{\psi
}^{\left(  2\right)  }ds \; .
\end{align}
\end{subequations}
Using Eqs. (\ref{eq:psi1tilde}-\ref{eq:psi2tilde}) and the boundary conditions
we obtain
\begin{align}
\tilde{A_{1}}  &  =A_{1}+\int_{-L/2}^{L/2}sV\left(  s\right)  \tilde{\psi
}^{\left(  1\right)  }\left(  s\right)  ds \nonumber\\
&  =\int_{-L/2}^{L/2}s^{2}V(s)ds+\int_{-L/2}^{L/2}s\left(  \mathcal{L}%
+\frac{d^{2}}{ds^{2}}\right)  \tilde{\psi}^{\left(  1\right)  }\left(
s\right)  ds+L \nonumber\\
&  =\int_{-L/2}^{L/2}s\left(  \frac{d^{2}}{ds^{2}}\tilde{\psi}^{\left(
1\right)  }\right)  ds+L \nonumber\\
&  =\frac{L}{2}\left(  \frac{d}{ds}\tilde{\psi}^{\left(  1\right)  }\left(
-L/2\right)  +\frac{d}{ds}\tilde{\psi}^{\left(  1\right)  }\left(  L/2\right)
\right)  +L
\end{align}
and in the same manner
\begin{align}
\tilde{A_{2}}  &  =A_{2}+\frac{1}{2}\int_{-L/2}^{L/2}\left(  sV\left(
s\right)  \tilde{\psi}^{\left(  2\right)  }+V\left(  s\right)  \tilde{\psi
}^{\left(  1\right)  }\right)  ds \nonumber\\
&  =\int_{-L/2}^{L/2}V\left(  s\right)  sds+\frac{1}{2}\int_{-L/2}%
^{L/2}\left(  s\left(  \mathcal{L}+\frac{d^{2}}{ds^{2}}\right)  \tilde{\psi
}^{\left(  2\right)  }+\left(  \mathcal{L}+\frac{d^{2}}{ds^{2}}\right)
\tilde{\psi}^{\left(  1\right)  }\right)  ds \nonumber\\
&  =\frac{1}{2}\int_{-L/2}^{L/2}\left(  s\frac{d^{2}}{ds^{2}}\tilde{\psi
}^{\left(  2\right)  }+\frac{d^{2}}{ds^{2}}\tilde{\psi}^{\left(  1\right)
}\right)  ds \nonumber\\
&  =\frac{1}{2}\left(  \frac{L}{2}\left(  \frac{d}{ds}\tilde{\psi}^{\left(
2\right)  }\left(  -L/2\right)  +\frac{d}{ds}\tilde{\psi}^{\left(  2\right)
}\left(  L/2\right)  \right)  \right)  +\frac{1}{2}\left[  \frac{d}{ds}%
\tilde{\psi}^{\left(  1\right)  }\left(  s\right)  \right]  _{-L/2}^{L/2}%
\; ,
\end{align}
as well as
\begin{align}
\tilde{A_{3}}  &  =A_{3}-\int_{-L/2}^{L/2}V\left(  s\right)  \tilde{\psi
}^{\left(  2\right)  }ds \nonumber\\
&  =\int_{-L/2}^{L/2}V\left(  s\right)  ds+\int_{-L/2}^{L/2}\left(
\mathcal{L}+\frac{d^{2}}{ds^{2}}\right)  \tilde{\psi}^{\left(  2\right)
}ds=\left[  \frac{d}{ds}\tilde{\psi}^{\left(  2\right)  }\left(  s\right)
\right]  _{-L/2}^{L/2} \; .
\end{align}
Now we have to distinguish between the cases (a) and (b) to check the
stability conditions Eqs. (\ref{eq:stabcond1}-\ref{eq:stabcond3}). After
calculating the different terms in $\tilde{A_{i}}$ and using the expression of
$\tilde{\psi}_{a,b}^{\left(  i\right)  }\left(  s\right)  $ we arrive at%
\begin{subequations}
\begin{align}
\tilde{A_{1}}_{a}  &  =LC_{2a}\frac{\sqrt{\mu}}{\lambda m} \; ,\\
\tilde{A_{2}}_{a}  &  =-\frac{L}{2}\frac{\sqrt{\mu}}{\lambda}\sin\left(
2\psi\left(  -L/2\right)  \right)  \; ,\\
\tilde{A_{3}}_{a}  &  =0 \; ,
\end{align}
\end{subequations}
and%
\begin{subequations}
\begin{align}
\tilde{A_{1}}_{b}  &  =\frac{L}{\lambda}\left(  \frac{L}{4}\frac{\mu}{\sqrt
{m}}\sin\left(  2\psi\left(  -L/2\right)  \right)  +\frac{\sqrt{\mu}}{m}%
C_{2b}\right)  \; ,\\
\tilde{A_{2}}_{b}  &  =0 \; ,\\
\tilde{A_{3}}_{b}  &  =\frac{\sqrt{\mu}}{\lambda}\sin\left(  2\psi\left(
-L/2\right)  \right)  \; .
\end{align}
\end{subequations}
The conditions Eqs.\ (\ref{eq:stabcond1}-\ref{eq:stabcond3}) are never satisfied
for the case (a) which thus corresponds to unstable solutions. This follows directly from Eq. (\ref{eq:stabcond3}) with $\tilde{A_{3}}_{a}=0$. 

For the stability conditions of case (b) to be satisfied, we must have
$\tilde{A_{1}}_{b}>0$ and $\tilde{A_{3}}_{b}>0$.\ For $\tilde{A_{3}}_{b}$ this
is true only if $\psi\left(  -L/2\right)  \in\left]  0,\frac{\pi}{2}\right]
$. For $\tilde{A_{1}}_{b}$, we have to check the sign of $C_{2b}$ which is the
sign of $\mathcal{E}\left(  \psi\left(  L/2\right)  |m\right)  - \mathcal{E}\left(  \psi\left(
-L/2\right)  |m\right)  $. It is always positive because $\mathcal{E}\left(  x,m\right)
$ is a monotonically increasing function. Then $\psi^{\prime}\left(  s\right)  >0$
(as we have chosen $\omega_{3}>0$) leads to $\psi\left(  \frac{L}{2}\right)
>\psi\left(  -\frac{L}{2}\right)  $. Therefore, case (b) with $\psi(  -L/2 ) >0 $ corresponds 
to stable solutions whereas all other cases are unstable.


\subsection{The oscillating pendulum ($m>1$)}

In the case where $m>1$ the twist $\psi$ is a periodic function of $s$. We thus
distinguish two cases.

\subsubsection*{Case (a)}

The shape of the squeelix exhibits a certain number of waves or oscillations. In the
following we will consider a chain of length $L$ with one period $l_p=L$, and thus we
choose $s=-L/2$ to $s=-L/2+l_{p}=L/2$. We start with a perturbation of the form
\begin{equation}
\tilde{\psi}(s)=\cos\psi\left(  s\right)  \; .
\end{equation}
The second order variation of the energy, Eq. (\ref{deltaE2AD}), follows as
\begin{align}
\delta^{2}E  &  =\left[  -\frac{C}{\lambda\sqrt{m}}cn\left(  \frac{s-s_{0}
}{\lambda\sqrt{m}}|m\right)  sn\left(  \frac{s-s_{0}}{\lambda\sqrt{m}
}|m\right)  dn\left(  \frac{s-s_{0}}{\lambda\sqrt{m}}|m\right)  \right]
_{-L/2}^{L/2}+C\int_{-L/2}^{L/2}\tilde{\psi}\mathcal{L}\tilde{\psi}ds \nonumber\\
&  =C\int_{-L/2}^{L/2}cn\left(  \frac{s-s_{0}}{\lambda\sqrt{m}}|m\right)
\mathcal{L}cn\left(  \frac{s-s_{0}}{\lambda\sqrt{m}}|m\right)
ds\nonumber\\
&  =C\frac{\left(  1-m\right)  }{\lambda^2m}\int_{-L/2}^{L/2}cn\left(
\frac{s-s_{0}}{\lambda\sqrt{m}}|m\right)  ^{2}ds<0 \; .
\end{align}
All solutions of case (a) are thus unstable.

\subsubsection*{Case (b)} 
We now consider a chain which is longer than the period, \textit{i.e.}, $s=-L/2$ to
$s=-L/2+L_{p}=s_{1}$ and a portion of the chain from $s=s_{1}$ to $s=L/2$. Choosing
\begin{equation}
\tilde{\psi}=dn\left(  \frac{s-s_{0}}{\lambda\sqrt{m}}|m\right)  \; ,
\end{equation}
which is a zero eigenmode of $\mathcal{L}$, we have
\begin{equation}
\delta^{2}E_{-L/2\rightarrow s_{1}}=-C\frac{\sqrt{m}}{\lambda}\left[
cn\left(  \frac{s-s_{0}}{\lambda\sqrt{m}}|m\right)  sn\left(  \frac{s-s_{0}
}{\lambda\sqrt{m}}|m\right)  dn\left(  \frac{s-s_{0}}{\lambda\sqrt{m}
}|m\right)  \right]  _{-L/2}^{s_{1}}+C\int_{-L/2}^{s_{1}}\tilde{\psi
}\mathcal{L}\tilde{\psi}ds=0 \; .
\end{equation}
Thus
\begin{align}
\delta^{2}E  &  =-C\frac{\sqrt{m}}{\lambda}\left[  cn\left(  \frac{s-s_{0}
}{\lambda\sqrt{m}}|m\right)  sn\left(  \frac{s-s_{0}}{\lambda\sqrt{m}
}|m\right)  dn\left(  \frac{s-s_{0}}{\lambda\sqrt{m}}|m\right)  \right]
_{s_{1}}^{\frac{L}{2}}+C\int_{s_{1}}^{\frac{L}{2}}\tilde{\psi}\mathcal{L}
\tilde{\psi}ds\nonumber\\
&  =2C\frac{\sqrt{m}}{\lambda}cn\left(  \frac{-L/2-s_{0}}{\lambda\sqrt{m}
}|m\right)  sn\left(  \frac{-L/2-s_{0}}{\lambda\sqrt{m}}|m\right)  dn\left(
\frac{-L/2-s_{0}}{\lambda\sqrt{m}}|m\right) \label{eq:deltaE2msup1caseb}\nonumber\\
&  =2C\frac{\sqrt{m}}{\lambda}\cos\left(  \psi\left(  -L/2\right)  \right)
\sin\left(  \psi\left(  -L/2\right)  \right)  \sqrt{1-m\sin^{2}\left(
\psi\left(  -\frac{L}{2}\right)  \right)  } \; ,
\end{align}
where we have used the fact that $\psi^{\prime}\left(  -L/2\right)  =dn\left(
\frac{-L/2-s_{0}}{\lambda\sqrt{m}}|m\right)  =\sqrt{1-m\sin^{2}\left(
\psi\left(  -\frac{L}{2}\right)  \right)  }=\omega_{3}>0$.

Since $\psi\left(  -L/2\right)  =\pm\arcsin\sqrt{\frac{1}{m}-\frac{1}{\mu}}$,
one finds that solutions with $\psi\left(  -L/2\right)  <0$ are unstable whereas
solutions with $\psi\left(  -L/2\right)  >0$ are stable.\


\end{document}